\documentclass[a4paper,12pt]{article}

\usepackage[dvips]{hyperref}
\usepackage[]{epsfig}
\usepackage[]{cite}
\usepackage[]{amssymb}
\newcommand{\Hosm}{\mbox{\ensuremath{\mathrm{H}_{\mathrm{SM}}}}}
\newcommand{\nn}{\mbox{\ensuremath{\nu \bar{\nu}}}}
\newcommand{\ee}{\mbox{${\mathrm{e}}^+ {\mathrm{e}}^-$}}
\newcommand{\epm}{\mbox{${\mathrm{e}}^+ {\mathrm{e}}^-$}}
\newcommand{\tautau}{\ensuremath{\tau^+\tau^-}}
\newcommand{\mm}{\mbox{$\mu^+\mu^-$}}

\newcommand{\glgl}{\mbox{${\mathrm{gg}}$}}

\newcommand{\mhmax}      {\mbox{$m_{\mathrm{h}}{\mathrm{-max}}$}}
\newcommand{\tevcs}      {\mbox{${\mathrm{TeV}}$}}

\newcommand{\qq}         {\ensuremath{\mathrm{q}\bar{\mathrm{q}}}}

\newcommand{\bb}         {\ensuremath{\mathrm{b}\bar{\mathrm{b}}}}
\newcommand{\cc}         {\mbox{$\mathrm{c}\bar{\mathrm{c}}$}}

\newcommand{\nunu}       {\mbox{$\nu\bar{\nu}$}}

\newcommand{\mZ}         {\mbox{$m_{\mathrm{Z}}$}}

\newcommand{\mH}         {\mbox{$m_{\mathrm{H}}$}}

\newcommand{\mHp}         {\mbox{$m_{\mathrm{H}^{\pm}}$}}
\newcommand{\mHone}         {\mbox{$m_{\mathrm{H}_{1}}$}}
\newcommand{\mHtwo}         {\mbox{$m_{\mathrm{H}_{2}}$}}

\newcommand{\mh}         {\mbox{$m_{\mathrm{h}}$}}
\newcommand{\mA}         {\mbox{$m_{\mathrm{A}}$}}
\newcommand{\gevcs}      {\mbox{${\mathrm{GeV}}$}}
\newcommand{\G}          {\mbox{${\mathrm{GeV}}$}}

\newcommand{\Hi}{\ensuremath{\mathrm{H}_i}}

\newcommand{\Hj}{\ensuremath{\mathrm{H}_j}}
\newcommand{\Hone}{\ensuremath{\mathrm{H}_1}}
\newcommand{\Htwo}{\ensuremath{\mathrm{H}_2}}
\newcommand{\Hthree}{\ensuremath{\mathrm{H}_3}}
\newcommand {\Ho}        {\ensuremath{\mathrm{H}}}
\newcommand {\Ao}        {\ensuremath{\mathrm{A}}}
\newcommand {\ho}        {\ensuremath{\mathrm{h}}}
\newcommand {\Zo}        {\ensuremath{\mathrm{Z}}}
\newcommand {\genH}      {\ensuremath{{\cal H}}}
\newcommand {\genmH}      {\ensuremath{m_{\cal H}}}
\newcommand {\genHone}      {\ensuremath{{{\cal H}_1}}}
\newcommand {\genHtwo}      {\ensuremath{{{\cal H}_2}}}
\newcommand {\genmHone}      {\ensuremath{m_{{\cal H}_1}}}
\newcommand {\genmHtwo}      {\ensuremath{m_{{\cal H}_2}}}
\newcommand {\Hsm}       {\ensuremath{\mathrm{H}_{\mathrm{SM}}}}

\newcommand{\mHpm}{\mbox{$m_{\mathrm{H}^{\pm}}$}}

\newcommand{\Zgs}        {\mbox{$\mathrm{(Z/\gamma)}^{*}$}}

\newcommand{\WW}         {\mbox{$\mathrm{W}^+\mathrm{W}^-$}}
\newcommand{\ZZ}         {\mbox{$\mathrm{Z}\mathrm{Z}$}}

\newcommand{\tanb}       {\mbox{$\tan\beta$}}
\newcommand{\msusy}       {\mbox{$m_{\mathrm{SUSY}}$}}
\newcommand{\mg}       {\mbox{$m_{\mathrm{\tilde g}}$}}

\newcommand{\ipb}         {\mbox{pb$^{-1}$}}

\newcommand{\Dbi}{\mbox{${\cal B}_i$}}

\newcommand{\ZH}     {\mbox{$\mathrm{H}\mathrm{Z}$}}
\newcommand{\Zh}     {\mbox{$\mathrm{h}\mathrm{Z}$}}

\newcommand{\Ah}     {\mbox{$\mathrm{h}\mathrm{A}$}}

\newcommand{\sqrts}     {\mbox{$\sqrt{s}$}}
\newcommand{\sprime}     {\mbox{$\sqrt{s^\prime}$}}
\def\etal{\mbox{{\it et al.}}}
\newcommand{\ra}        {\mbox{$\rightarrow$}}   
\begin{document}

%
%
%
%

\begin{titlepage}
\begin{center}{\large   EUROPEAN ORGANIZATION FOR NUCLEAR RESEARCH
}\end{center}\bigskip
\begin{flushright}
       CERN-PH-EP/2004-020  \\OPAL PR 399 \\ 26th March 2004\\ Revised 13 May 2004
\end{flushright}
\bigskip\bigskip\bigskip\bigskip\bigskip
\begin{center}{\huge\bf  Search for Neutral Higgs Bosons in\\ 
    CP-Conserving and CP-Violating\\[2mm] 
    MSSM Scenarios
}\end{center}\bigskip\bigskip
\begin{center}{\LARGE The OPAL Collaboration
    }\end{center}\bigskip\bigskip \bigskip\begin{center}{\large
    Abstract}\end{center} 

This report summarizes the final results from the OPAL collaboration
on searches for neutral Higgs bosons predicted by the Minimal
Supersymmetric Standard Model (MSSM).  CP-conserving and, for the
first time at LEP, CP-violating scenarios are studied.  New scenarios
are also included, which aim to set the stage for Higgs searches at
future colliders. The results are based on the data collected with the
OPAL detector at $\ee$ centre-of-mass energies up to 209~GeV.  The
data are consistent with the prediction of the Standard Model with no
Higgs boson produced. Model-independent limits are derived for the
cross-sections of a number of event topologies motivated by
predictions of the MSSM.  Limits on Higgs boson masses and other MSSM
parameters are obtained for a number of representative MSSM benchmark
scenarios.  For example, in the CP-conserving scenario \mhmax{} where
the MSSM parameters are adjusted to predict the largest range of
values for $\mh$ at each $\tanb$, and for a top quark mass of
174.3~GeV, the domain $0.7 < \tan\beta < 1.9$ is excluded at the 95\%
confidence level and Higgs boson mass limits of $\mh > 84.5$~GeV and
$\mA > 85.0$~GeV are obtained.  For the CP-violating benchmark
scenario CPX which, by construction, enhances the CP-violating effects
in the Higgs sector, the domain $\tanb<2.8$ is excluded but no
universal limit can be set on the Higgs boson masses.

\bigskip\bigskip\bigskip\bigskip


\bigskip\bigskip
\begin{center}{\large
(\emph{To be submitted to European Physical Journal {\bf \emph{C}}})
}\end{center}
\end{titlepage}
\begin{center}{\Large        The OPAL Collaboration
}\end{center}\bigskip
\begin{center}{
G.\thinspace Abbiendi$^{  2}$,
C.\thinspace Ainsley$^{  5}$,
P.F.\thinspace {\AA}kesson$^{  3,  y}$,
G.\thinspace Alexander$^{ 22}$,
J.\thinspace Allison$^{ 16}$,
P.\thinspace Amaral$^{  9}$, 
G.\thinspace Anagnostou$^{  1}$,
K.J.\thinspace Anderson$^{  9}$,
S.\thinspace Asai$^{ 23}$,
D.\thinspace Axen$^{ 27}$
% ...
}\end{center}\bigskip
\bigskip
$^{  1}$School of Physics and Astronomy, University of Birmingham,
Birmingham B15 2TT, UK
\newline
$^{  2}$Dipartimento di Fisica dell' Universit\`a di Bologna and INFN,
I-40126 Bologna, Italy
\newline
$^{  3}$Physikalisches Institut, Universit\"at Bonn,
D-53115 Bonn, Germany
\newline
% ...
$^{  5}$Cavendish Laboratory, Cambridge CB3 0HE, UK
\newline
% ...
$^{  9}$Enrico Fermi Institute and Department of Physics,
University of Chicago, Chicago IL 60637, USA
\newline
% ...
$^{ 16}$Department of Physics, Schuster Laboratory, The University,
Manchester M13 9PL, UK
\newline
% ...
$^{ 22}$Department of Physics and Astronomy, Tel Aviv University,
Tel Aviv 69978, Israel
\newline
$^{ 23}$International Centre for Elementary Particle Physics and
Department of Physics, University of Tokyo, Tokyo 113-0033, and
Kobe University, Kobe 657-8501, Japan
\newline
% ...
$^{ 27}$University of British Columbia, Department of Physics,
Vancouver BC V6T 1Z1, Canada
\newline
% ...
$^{  y}$ now at CERN
\newline
% ...

\clearpage

%
%
%
%

%
%
%
%
\newpage
\section{Introduction}

It is generally assumed that the Higgs mechanism~\cite{Higgs:ia} is
responsible for the breaking of electroweak symmetry and for the
generation of elementary particle masses.  In its simplest form,
implemented in the Standard Model (SM), this mechanism implies the
self-interactions of one doublet of complex scalar Higgs fields and
predicts the existence of one physical scalar particle, the SM Higgs
boson. The mass of this particle is not predicted by the model.
Despite an intense effort over the last few decades, the SM Higgs
boson has not been detected; a lower bound of 114.4~GeV has been set
on its mass~\cite{Barate:2003sz}.

The SM, while successfully describing all electroweak phenomena
investigated so far, provides no mechanism for stabilizing the
electroweak energy scale ($\approx M_\mathrm{W}$) in the presence of
quantum corrections (the ``scale hierarchy'' problem). Supersymmetry
(SUSY) is proposed as a possible solution to this problem since in
SUSY models fermionic quantum corrections are compensated by bosonic
corrections of similar size, and vice versa.

The Minimal Supersymmetric Standard Model (MSSM) is the SUSY extension
of the SM with minimal new particle content. It introduces two complex
Higgs field doublets which is the minimal Higgs structure required to
keep the theory free of anomalies and to give masses to all fermions.
The MSSM predicts five Higgs bosons: three neutral and two charged
particles.  At least one of the neutral Higgs bosons is predicted to
have its mass close to the electroweak energy scale; when radiative
corrections are included, this mass should be less than
135~GeV~\cite{Degrassi:2002fi}. This prediction provides a strong
motivation for searches at present and future colliders.

In the MSSM the Higgs potential is invariant under CP transformation
at tree level. However, it is possible to break CP symmetry in the
Higgs sector by radiative corrections, especially by contributions
from third generation scalar-quarks~\cite{Pilaftsis:1999qt}. Such a
scenario is theoretically attractive since it provides a possible
solution to the cosmic baryon asymmetry~\cite{Carena:2000id} while the
size of the CP-violating (CPV) effects occurring in the SM are far too
small to account for it.

The searches performed so far at LEP have been restricted to
CP-conserving (CPC) MSSM scenarios where all CPV phases in the soft
SUSY-breaking Lagrangian related to the Higgs sector are put to zero.
Under this assumption the three neutral Higgs bosons are CP
eigenstates: the $\ho$ and $\Ho$ bosons ($\ho$ is defined to be the
lighter of the two) are CP-even and the $\Ao$ boson is CP-odd.  Of
these, only the CP-even states couple to the Z boson in Higgsstrahlung
at tree level; thus, at LEP energies, these particles are mainly
produced through the Higgsstrahlung processes $\ee\ra\ho\Zo$ and
$\Ho\Zo$, and the pair production processes $\ee\ra\ho\Ao$ and
$\Ho\Ao$ ($\ee\ra\ho\Ho$ is forbidden by angular momentum and CP
conservation).  Production of $\ho$ and $\Ho$ via WW and ZZ fusion in
the $t$-channel, where a Higgs particle is produced in association
with a pair of neutrinos or electrons, is included and plays a role at
the kinematic limit of Higgs boson production at LEP energies.

In this paper CPV MSSM scenarios are also considered.  In such
scenarios the three neutral Higgs bosons, $\Hi$ ($i=1,2,3$) are
mixtures of the CP-even and CP-odd Higgs fields. Consequently, they
all couple to the Z boson by their CP-even field components and to
each other.  Therefore, the Higgsstrahlung processes $\ee\ra\Hi\Zo$
($i=1,2,3$) and pair production processes $\ee\ra\Hi\Hj$ ($i\neq j$)
may all occur, however with widely varying cross-sections.  In large
domains of the model parameters, the lightest neutral Higgs boson
$\Hone$ may escape detection, since its coupling to the Z boson may be
too weak. The other two Higgs boson masses may be out of reach on
kinematic grounds or may also have small production cross-sections.
As a result, the exclusion limits obtained previously for the CPC
scenario may be invalidated by CPV effects.

The decay properties of the Higgs bosons, while being quantitatively
different in the two scenarios, maintain a certain similarity. Since
Higgs bosons, in general, couple to mass, the largest branching ratios
are those to $\bb$ and $\tautau$ pairs. If kinematically allowed, the
cascade decays $\ho\ra\Ao\Ao$ (CPC scenario)~\cite{sven_hAA} or
$\Htwo\ra\Hone\Hone$ (CPV scenario) would occur and can even be the
dominant decays. The searches described below take this possibility
into account.

Since the topological searches are essentially insensitive to the CP
parity of the Higgs bosons, their results can be applied both to the
CPC and CPV MSSM scenarios. For example, the search for the
Higgsstrahlung signal topology $\ee\ra\Zh$ in CPC scenarios can be
interpreted as a search for the processes $\ee\ra\Hone\Zo$ and
$\Htwo\Zo$ in CPV scenarios; similarly, the search for the pair
production process $\ee\ra\Ah$ in CPC scenarios can be interpreted as
a search for the process $\ee\ra\Hone\Htwo$ in CPV scenarios.

The results presented in this paper are based on all data collected by
the OPAL collaboration, including LEP1 data taken at the Z resonance
and LEP2 data collected between 130 and 209~GeV. Many of the searches
are already described in earlier OPAL publications
\cite{OPALSMPAPER,pr285,bib:opalhiggsold1,bib:opalhiggsold2,bib:opalhiggsold3,bib:opalhiggsold4,bib:opalhiggsold5}.
In this paper we describe only the optimizations of the searches for
the data collected at the highest energies, between $\sqrts=192$ and
209~GeV.  The old and new searches are combined to produce the final
results presented here.

Recent searches for neutral Higgs bosons performed by other LEP
collaborations, limited to the CPC MSSM scenarios, are listed in
\cite{bib:otherlep} while results from the Tevatron collaborations are
reported in \cite{bib:tevatronhiggs}.

This paper is organised as follows. Section~\ref{sect:detector}
summarises the main features of the OPAL detector and the Monte Carlo
simulations. Section~\ref{sect:channels} describes the model
predictions for Higgs boson production.  The search topologies and the
experimental searches for the corresponding final states are described
in Section~\ref{sect:searches}, with emphasis on those searches which
have not been covered in earlier OPAL publications.  In
Section~\ref{sect:combmet} the statistical combination of individual
search channels and the derivation of exclusion limits in the MSSM
parameter space are described. In Section~\ref{sect:modelindep} the
search results are translated into model-independent limits on
topological cross-sections.  In Section~\ref{sect:benchmark} the MSSM
benchmark scenarios, including those recently suggested
in~\cite{lhc_benchmarks}, and the corresponding results are presented.
A short summary is given in Section~\ref{sect:conclusion}.


\section{OPAL detector and Monte Carlo samples}
\label{sect:detector}

The OPAL detector is described in \cite{detector}.  The tracking
detectors and calorimeters have nearly complete solid angle coverage.
The central tracking detector is placed in a uniform axial magnetic
field of 0.435~T.  The innermost part is occupied by a high-resolution
silicon strip (``microvertex'') detector~\cite{simvtx}. It surrounds
the beam pipe and covers the polar angle\footnote{ OPAL uses a
  right-handed coordinate system where the $+z$ direction is along the
  electron beam and where $+x$ points to the centre of the LEP ring.
  The polar angle $\theta$ is defined with respect to the $+z$
  direction and the azimuthal angle $\phi$ with respect to the $+x$
  direction.  The centre of the \ee\ collision region defines the
  origin of the coordinate system.} range $|\cos\theta|<0.93$, and is
the basic tool for an efficient b-tagging~\cite{bib:opalhiggsold1}.
This detector is followed by a high-precision vertex drift chamber, a
large-volume jet chamber, and chambers to measure the $z$ coordinates
along the particle trajectories.  A lead-glass electromagnetic
calorimeter with a presampler is located outside the magnet coil. The
iron return-yoke of the magnet is instrumented with streamer tubes and
thin-gap chambers for hadron calorimetry.  Finally, the detector is
completed by several layers of muon chambers.

A variety of Monte Carlo samples are generated to estimate the
detection efficiencies for the Higgs boson signal and to optimise the
rejection of the background.  The production cross-section and
kinematic properties of the Higgs boson signal vary rapidly with
energy near the kinematic limit.  For an accurate modelling, the
signal and background samples are generated at several centre-of-mass
energies, from 192~GeV to 210~GeV.  The main background processes are
generated with 15 to 25 times the statistics of the data.


Signal events for the Higgsstrahlung and pair production search
channels are generated using the HZHA03 \cite{Janot:1996hzha} program.
The pair production samples are described in this paper.  For
Higgsstrahlung, the Monte Carlo samples used are described in
\cite{OPALSMPAPER}. Typical sample sizes are 2000 to 5000 signal
events per Higgs mass point.  For the background processes the
following event generators are used: KK2f~\cite{kk2f} for
\Zgs\ra\qq($\gamma$), \mm$(\gamma)$ and \tautau$(\gamma)$,
grc4f~\cite{grc4f} for four-fermion processes (4f),
BHWIDE~\cite{bhwide} for \ee$(\gamma)$, and PHOJET~\cite{phojet},
HERWIG~\cite{herwig}, and Vermaseren~\cite{vermaseren} for hadronic
and leptonic two-photon processes ($\gamma\gamma$).
JETSET~\cite{pythia} is used as the principal model for fragmentation
and hadronisation.  The detector response is simulated in full
detail~\cite{gopal}.

%

\section{Higgs boson production processes}
\label{sect:channels}
In the MSSM, the Higgsstrahlung and pair production processes have
complementary cross-sections. Their relative rate is regulated by sum
rules which are different in the CPC and CPV scenarios and depend on
the precise choices of the MSSM parameters.

In the CPC scenario, the cross-sections for the processes
$\ee\ra\ho\Zo$, $\ee\ra\Ho\Zo$ and $\ee\ra\ho\Ao$ are given by
\begin{eqnarray}
  \ee\ra\ho\Zo\;:&&
  \sigma_{\mathrm{hZ}}=\sin^2(\beta -\alpha)~\sigma^{\mathrm{SM}}_{\mathrm{HZ}}(\mh),
  \label{equation:xsec_zh} \\
  \ee\ra\Ho\Zo\;:&&
  \sigma_{\mathrm{HZ}}=\cos^2(\beta -\alpha)~\sigma^{\mathrm{SM}}_{\mathrm{HZ}}(\mH),
  \label{equation:xsec_zH} \\
  \ee\ra\ho\Ao\;:&&
  \sigma_{\mathrm{hA}}=
  \cos^2(\beta-\alpha)~\bar{\lambda}~\sigma^{\mathrm{SM}}_{\mathrm{HZ}}(\mh),
  \label{equation:xsec_ah}
\end{eqnarray} 
where $\sigma^{\mathrm{SM}}_{\mathrm{HZ}}$ is the cross-section for
the SM Higgsstrahlung process $\ee\ra\Hosm\Zo$, $\tanb=v_2/v_1$ is the
ratio of the vacuum expectation values of the two Higgs field doublets
coupling to ``up'' ($v_2$) and ``down'' ($v_1$) type fermions, and
$\alpha$ is the mixing angle describing the combination of the two
CP-even weak eigenstates to produce the two CP-even Higgs mass
eigenstates. The symbol $\bar{\lambda}$ denotes the kinematic
phase-space factor
$$  {\bar\lambda}=\frac{\lambda_{{\rm A}{\rm h}}^{3/2}}{\lambda_{{\rm Z}{\rm h}}^{1/2}
  (12M_{\rm Z}^2/s + \lambda_{{\rm Z}{\rm h}})} $$
with 
$$\lambda_{ij} = \frac{1-(m_i+m_j)^2/s}{1-(m_i-m_j)^2/s}.$$
Due to the
complementarity of the Higgsstrahlung and pair production processes,
expressed in Equations (\ref{equation:xsec_zh}) and
(\ref{equation:xsec_ah}), the searches have to include both of them in
order to maintain a high sensitivity over the whole MSSM parameter
space.
 
Similar, but more complex, sum rules regulate the relative rates in
the CPV scenario. The cross-sections for the processes $\ee\ra\Hi\Zo$
and $\ee\ra\Hi\Hj$ are given by~\cite{Carena:2000ks}
\begin{eqnarray}
  \ee\ra\Hi\Zo\;:&&
  \sigma_{\Hi\Zo}=g^{2}_{\Hi\Zo\Zo}~\sigma^{\mathrm{SM}}_{\mathrm{HZ}}(m_{\Hi}),
  \label{equation:xsec_zh_cp} \\
  \ee\ra\Hi\Hj\;:&&
  \sigma_{\Hi\Hj}=
  g^{2}_{\Hi\Hj\Zo}~\bar{\lambda}~\sigma^{\mathrm{SM}}_{\mathrm{HZ}}(m_{\Hi}),
  \label{equation:xsec_ah_cp}
\end{eqnarray} 
where the couplings 
\begin{eqnarray}
g_{\Hi\Zo\Zo} & = & \cos\beta\,O_{1i}+\sin\beta\,O_{2i}\nonumber\\
g_{\Hi\Hj\Zo} & = & O_{3i}\,(\cos\beta\,O_{2j}-\sin\beta\,O_{1j})
                        - O_{3j}\,(\cos\beta\,O_{2i}-\sin\beta\,O_{1i})\nonumber
\end{eqnarray}   
obey the sum rules
\begin{eqnarray}
\sum_{i=1}^{3} g^{2}_{\Hi\Zo\Zo} & = & 1\nonumber\\
g_{\mathrm{H}_k\Zo\Zo} & = & \frac{1}{2}\sum_{i,j=1}^{3} \varepsilon_{ijk} g_{\Hi\Hj\Zo}.\nonumber
\end{eqnarray}   
The orthogonal matrix $O_{ij}$ ($i,j=1,2,3$) relating the weak eigenstates
to the mass eigenstates has non-zero off-diagonal elements only in a CPV scenario. 

%

\section{Signal topologies and experimental searches}
\label{sect:searches}

The topological searches are devised to detect Higgs boson production
via the Higgsstrahlung and pair production processes. Many of these
searches are described in earlier publications. Here we emphasize the
most recent searches, optimized at the highest LEP energies and some
modifications of the earlier searches to increase the sensitivity in
parts of the MSSM parameter space which were not sufficiently covered
before.  The topological searches for the Higgsstrahlung process are
listed in Table~\ref{tab:channel_intro_hz}, and those for the pair
production process in Table~\ref{tab:channel_intro_ah}.

In the following, the notation $\ee\ra\genH\Zo$ and
$\ee\ra\genHone\genHtwo$ is used as a generic notation designating
both the CPC processes $\ee\ra\Zh$, $\ee\ra\ZH$ and $\ee\ra\Ah$, and
the CPV processes $\ee\ra\Hone\Zo$, $\ee\ra\Htwo\Zo$ and
$\ee\ra\Hone\Htwo$.  Since in $\ee$ collisions the kinematic
properties of the CPC and CPV signal processes are expected to be very
similar, the topological searches described in the following can be
applied in both scenarios.  In both CPC and CPV models, the production
process in Higgsstrahlung only contains couplings between the $\Zo$
boson and a CP-even Higgs state, and the pair production always
involves a CP-even and a CP-odd state.  The production angle
distributions are therefore expected to be the same. In CPV models,
the Higgs mass eigenstate is a mixture of CP eigenstates and is able
to change from a CP-even to a CP-odd state. Even in this case, the
Higgs decay angle distributions will stay the same since both the
CP-odd and the CP-even states are spin~0 bosons. Small differences not
measurable at LEP may arise from different spin correlations in the
decay products of the Higgs boson.


\subsection{Searches for Higgsstrahlung processes}
\label{sec:higgsstrahlung}

\begin{enumerate}
\item[(a)] The most efficient search for the Higgsstrahlung processes
  is the one which is designed to search for the Standard Model Higgs
  boson. This search~\cite{OPALSMPAPER} is interpreted here as a
  generic search for the corresponding MSSM process $\ee\ra\genH\Zo$.
  It takes advantage of the preferential decay of Higgs bosons into
  $\bb$ and $\tautau$ pairs and addresses the following $\Zo$ boson
  decays: $\Zo\,\ra\,\qq,\nunu,\ee,\mm,\tautau$.  Moreover, this
  search is also sensitive to contributions to the signal from the
  $\WW$ and $\ZZ$ fusion processes $\ee\ra\genH\nunu$ and $\genH\ee$,
  which may become important at the kinematic limit of the
  Higgsstrahlung process.
\item[(b)] The Higgs cascade decay $\genHtwo\ra\genHone\genHone$ may
  play an important role in regions of the MSSM parameter space where
  it is kinematically accessible. In order to increase the sensitivity
  to cascade decays, the search described in \cite{OPALSMPAPER} is
  adapted, in those parts which deal with the ``four-jet'' final state
  $\ee\ra(\genH\ra\bb)(\Zo\ra\qq)$ and the ``missing energy'' final
  state $\ee\ra(\genH\ra\bb)(\Zo\ra\nunu)$.  These searches modified
  for $\ee\ra(\genHtwo\ra\genHone\genHone)\Zo$ are described below in
  Sections \ref{sect:fourjet} and \ref{sect:missing_energy}.
\item[(c)] The search for Higgs cascade decays is complemented by an
  earlier search for
  $\ee\ra(\genHtwo\ra\genHone\genHone)\Zo$~\cite{lowma}, which is
  specifically designed to be efficient in the domain $\genmHone <
  10$~GeV.
\item[(d)] For Higgs bosons produced in Higgsstrahlung
  $\ee\ra\genH\Zo$ and decaying into particles other than b-quarks or
  $\tau$ leptons, a flavour-independent search for
  $\ee\ra(\genH\ra\mathrm{hadrons})\,\Zo$
  \cite{Abbiendi:2000ug,2HDMFINAL} is used.
\end{enumerate}

%
%

\subsubsection{Modification of the search in the four-jet channel}
\label{sect:fourjet}

The search for the SM Higgs boson in the channel
$\ee\ra\Hosm\Zo\ra\bb\qq$~\cite{OPALSMPAPER} is modified to be
sensitive to the cascade decay $\genHtwo\ra\genHone\genHone$.  The
event selection is identical to the SM search, and thus the same
candidate events are observed with the same expected background.  The
whole event is forced into four jets using the Durham jet
finder~\cite{durham_jetfinder}.  If $\genHone$ is not too heavy, the
two jets from $\genHone\ra\bb$ are often joined into one jet.  The SM
four-jet search is therefore also efficient for this decay and the
expected signal rates from $\ee\ra(\genHtwo\ra\bb)\Zo$ and
$\ee\ra(\genHtwo\ra\genHone\genHone\ra\bb\bb)\Zo$ can simply be added.
The efficiencies for various combinations of $(\genmHone,\genmHtwo)$
are given in Table~\ref{tab:effbbbbqq}.  The shape of the distribution
of the discriminating variable $\cal D$~\cite{OPALSMPAPER}, however,
differs for the two decay modes. $\cal D$ is a product of a mass
independent and a mass dependent likelihood. Depending on $\genmHone$,
the mass reconstruction is diluted by wrong jet pairings inside one
jet, and thus the likelihood distributions are broadened.  The signal
distribution of $\cal D$ is therefore constructed at each point of the
MSSM parameter space, taking into account the changing relative
contributions from the two decays by first adding the relative
contributions of $\genHtwo\ra\bb$ and $\genHtwo\ra\bb\bb$ in the two
likelihoods and then calculating the product.  The systematic
uncertainties are essentially the same as for the SM channel
$\ee\ra\Ho_{\mathrm{SM}}\Zo\ra\bb\qq$ \cite{OPALSMPAPER}.

%
%

\subsubsection{Modification of the search in the missing energy channel}
\label{sect:missing_energy}

For the data taken at $\sqrts=199$ to $209$~GeV, the Artificial Neural
Network (ANN) analysis for $\ee\ra\Hsm\Zo\ra\bb\nn$~\cite{OPALSMPAPER}
is reoptimized for $100 < \genmHtwo < 110$~GeV and modified to be
sensitive to $\genHtwo\ra\bb$ and
$\genHtwo\ra\genHone\genHone\ra\bb\bb$ decays simultaneously. In this
mass range, the $\genHtwo\ra\genHone\genHone$ decay is crucial
especially in the CPV scenario.

In the preselection, the event sample is split into two subsamples,
according to the 2-to-3 jet resolution parameter $y_{32}$ of the
Durham jet finder.  Subsample~A contains events with $y_{32} < 0.05$
($\genHtwo\ra\bb$ events and most of the
$\genHtwo\ra\genHone\genHone\ra\bb\bb$ with light $\genHone$) and
subsample~B events with $y_{32} \ge 0.05$ (most of the
$\genHtwo\ra\genHone\genHone\ra\bb\bb$ events with heavier
$\genHone$).  The separating value of 0.05 is chosen so that the
efficiency for $\genHtwo\ra\bb$ events in subsample A is approximately
the same as in the SM search.  Separate neural networks,
ANN$_{\mathrm{A}}$ and ANN$_{\mathrm{B}}$ are then trained for events
belonging to the two subsamples.  For ANN$_{\mathrm{A}}$ the training
is based on $\genHtwo\ra\genHone\genHone\ra\bb\bb$ signal events with
$\genmHone=12$~GeV and $100\leq\genmHtwo\leq110$~GeV, while for
ANN$_{\mathrm{B}}$ $\genmHone=40$~GeV and
$100\leq\genmHtwo\leq110$~GeV is used. The analysis is slightly
different from the SM missing energy analysis~\cite{OPALSMPAPER}.

The preselection cuts 1 to 5 are identical to those described in
\cite{OPALSMPAPER}.  They are designed to remove accelerator-related
backgrounds (such as beam-gas interactions and instrumental noise),
dilepton final states, two-photon processes and radiative \qq\ events,
and to select events with a significant amount of missing energy.  The
following new preselection cuts are applied:
\begin{itemize} 
\item[6.]  The tracks and clusters in the event are grouped into jets
  using the Durham algorithm. Depending on $y_{32}$, the event is
  either grouped into two jets ($y_{32}<0.05$) in subsample A or into
  four jets ($y_{32}\geq0.05$) in subsample B.  Each event in
  subsample B is required to have at least one track per jet.
\end{itemize} 
Additional requirements are imposed for subsample A ($y_{32}<0.05$):
\begin{itemize} 
\item[7.] The acoplanarity angle ($180^\circ$ minus the angle between
  the two jets when projected into the $xy$ plane) must be between
  $3^\circ$ and $100^\circ$ to reject $\Zo/\gamma\ra\qq$ events with
  back-to-back jets.
\item[8.]  To reduce the background from \WW\ events, the event must
  not have an identified isolated lepton~\cite{bib:opalhiggsold2}.
\end{itemize} 

\begin{figure}[pt]
  \begin{center}
    \epsfig{file=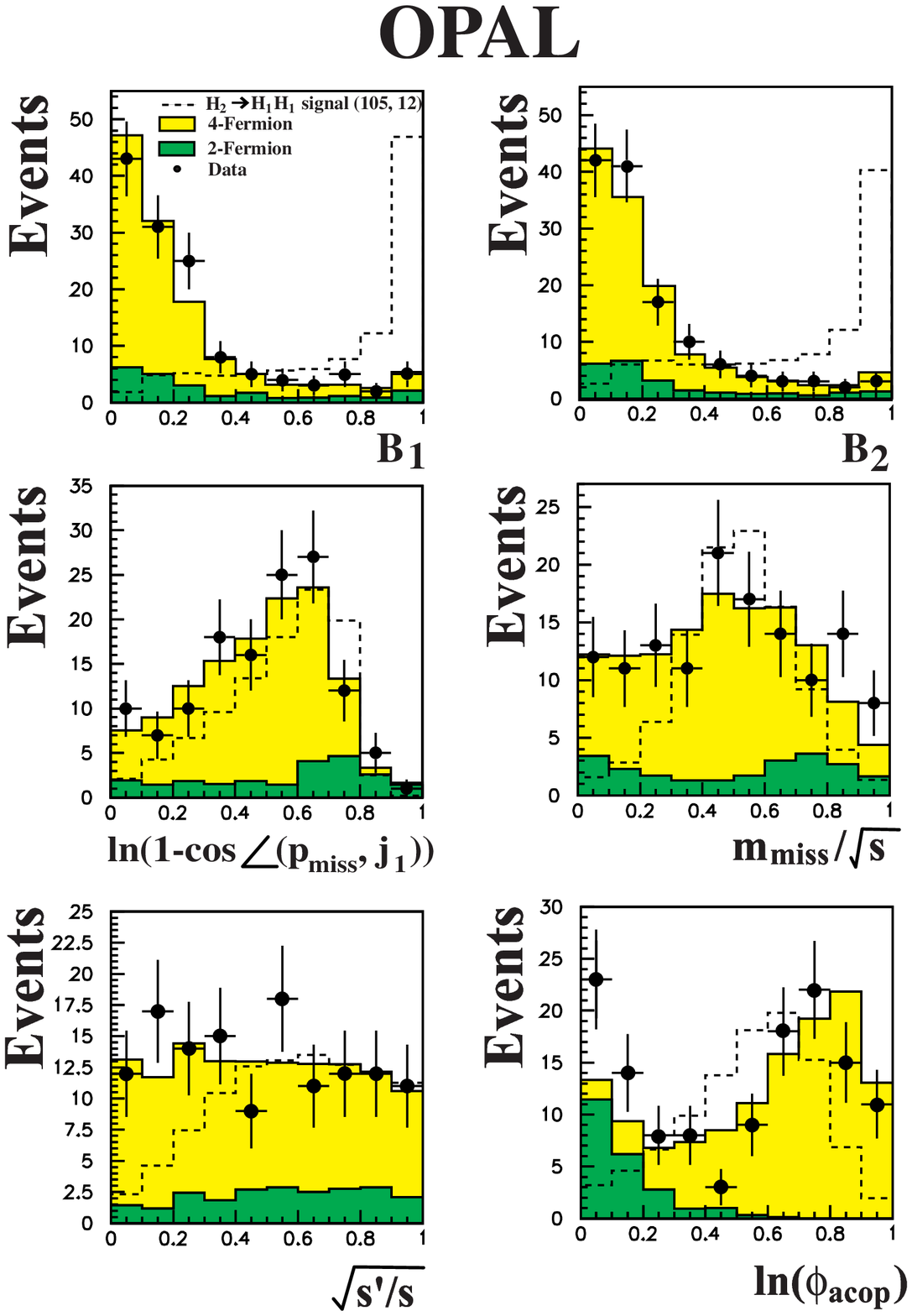, width=0.8\textwidth}
  \end{center}
  \caption{\sl{Distributions of some ANN input variables
      for subsample A
      for $\ee\ra\genHtwo\Zo\ra\bb\nn$ and
      $\ee\ra\genHone\genHone\Zo\ra\bb\bb\nn$ (see Section~\ref{sect:missing_energy}) in the region 
      100~GeV~$< \genmHtwo < $~110 GeV for the data taken at $\sqrts=199-209$~GeV.  
      Light grey (yellow)
      is the four-fermion  background, dark grey (green) the 
      contribution from the $\qq(\gamma)$ background.
      The dashed line shows the arbitrarily scaled signal expectation for $\genmHone=12$~GeV and 
      $\genmHtwo=105$~GeV, as given in the legend.
      The points with error bars are the data.}}\label{fig:emis_net2_1}
\end{figure}

\begin{figure}[pt]
  \begin{center}
    \epsfig{file=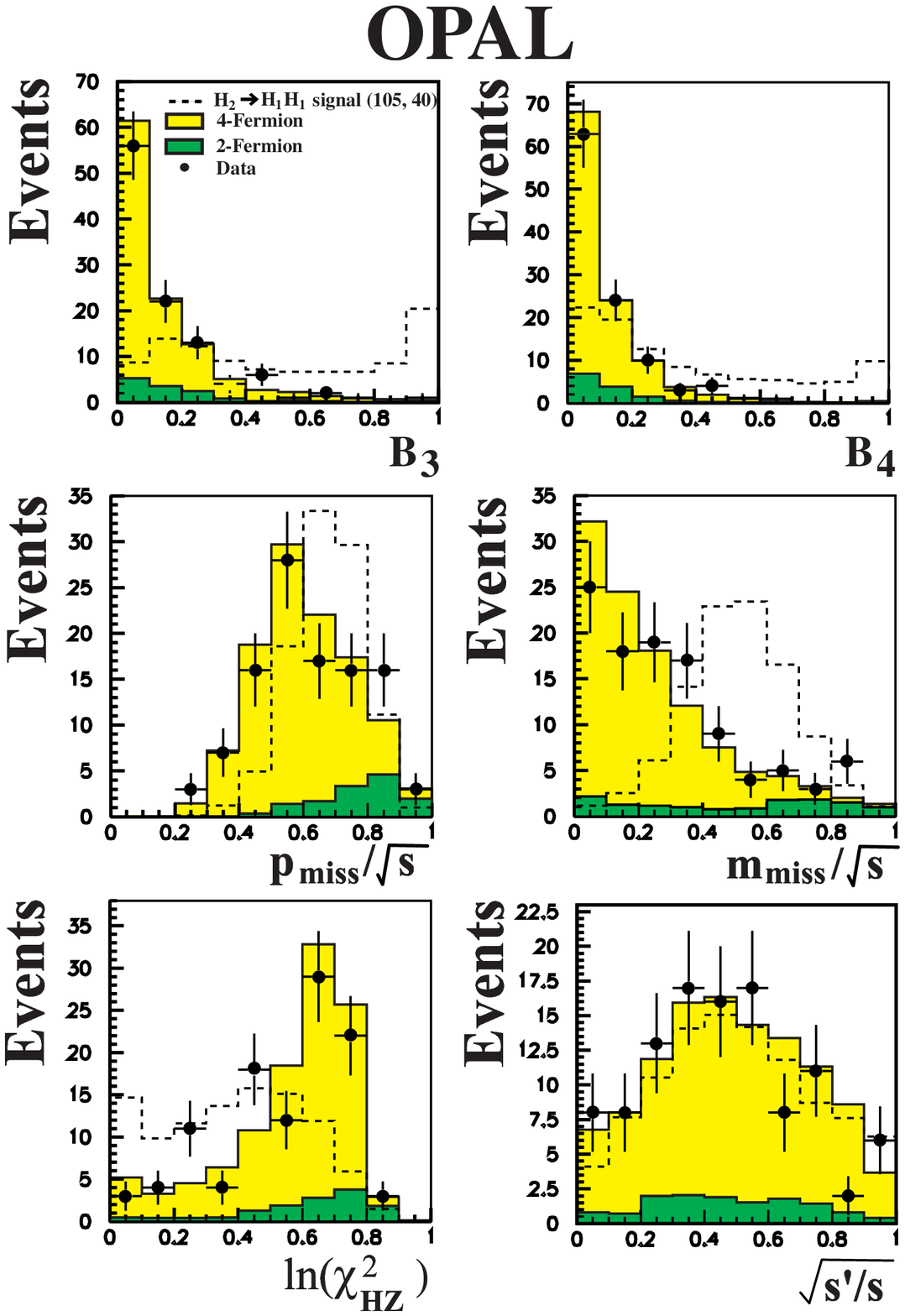, width=0.8\textwidth}
  \end{center}
  \caption{\sl{Distributions of some ANN input variables for subsample B
      for $\ee\ra\genHtwo\Zo\ra\bb\nn$ and
      $\ee\ra\genHone\genHone\Zo\ra\bb\bb\nn$ (see Section~\ref{sect:missing_energy}) in the region 
      100~GeV~$< \genmHtwo < $~110 GeV.  
      }}\label{fig:emis_net4_1}
\end{figure}

The effects of the cuts on the data, the simulated signal and
background samples are given in Table~\ref{tab:cutbbbbnn}.  The 12
(11) variables used as inputs to ANN$_{\mathrm{A}}$
(ANN$_{\mathrm{B}}$) are listed below.  All variables are scaled to
values between zero and one, and in some cases the logarithm of the
variable is used, which gives less peaked distributions and is
therefore better suited to an ANN analysis.  The distributions of some
of these input variables are shown in Figs.~\ref{fig:emis_net2_1} and
\ref{fig:emis_net4_1}. The following 9 variables are common to both
ANN$_{\mathrm{A}}$ and ANN$_{\mathrm{B}}$.
\begin{itemize} 
\item[1.] The scaled effective centre-of-mass energy~\cite{l2mh}
  $\sprime/\sqrt{s}$,
\item[2.] The scaled missing mass $m_{\mathrm{miss}}/\sqrt{s}$,
\item[3.] The polar angle of the missing momentum vector
  $|\cos\theta_{\mathrm{miss}}|$,
\item[4.] The b-tag likelihood output ${\cal B}_1$~\cite{pr285} of the
  first (highest energy) jet,
\item[5.] The b-tag likelihood output ${\cal B}_2$ of the second
  highest energy jet,
\item[6.] The angle between the first jet and the missing momentum
  vector, \\ $\ln(1-\cos\angle{(j_1,p_{\mathrm{miss}})})$,
\item[7.] The angle between the second jet and the
  missing momentum vector, \\
  $\cos\angle{(j_2,p_{\mathrm{miss}})}$,
\item[8.] The $\chi^2$ of the one-constraint kinematic fit,
  $\ln(\chi^2_{\mathrm{HZ}}$), where the missing mass
  $m_{\mathrm{miss}}$ is forced to $\mZ$,
\item[9.] The scaled missing momentum $p_{\mathrm{miss}}/\sqrt{s}$.
\end{itemize} 
The following three variables are used in the construction of
ANN$_{\mathrm{A}}$
\begin{itemize} 
\item[10.] The polar angle of the thrust axis
  $|\cos\theta_{\mathrm{thr}}|$,
\item[11.] The acoplanarity angle of the jets
  $\ln{(\phi_{\mathrm{acop}})}$,
\item[12.] The logarithm of the energy difference between the two jets
  $\ln|E_1-E_2|$.
\end{itemize} 
The following two variables are used in the construction of
ANN$_{\mathrm{B}}$
\begin{itemize} 
\item[10.] The b-tag likelihood output ${\cal B}_3$ of the third jet,
\item[11.] The b-tag likelihood output ${\cal B}_4$ of the fourth jet.
\end{itemize} 

\begin{figure}[pt]
  \begin{center}
    \epsfig{file=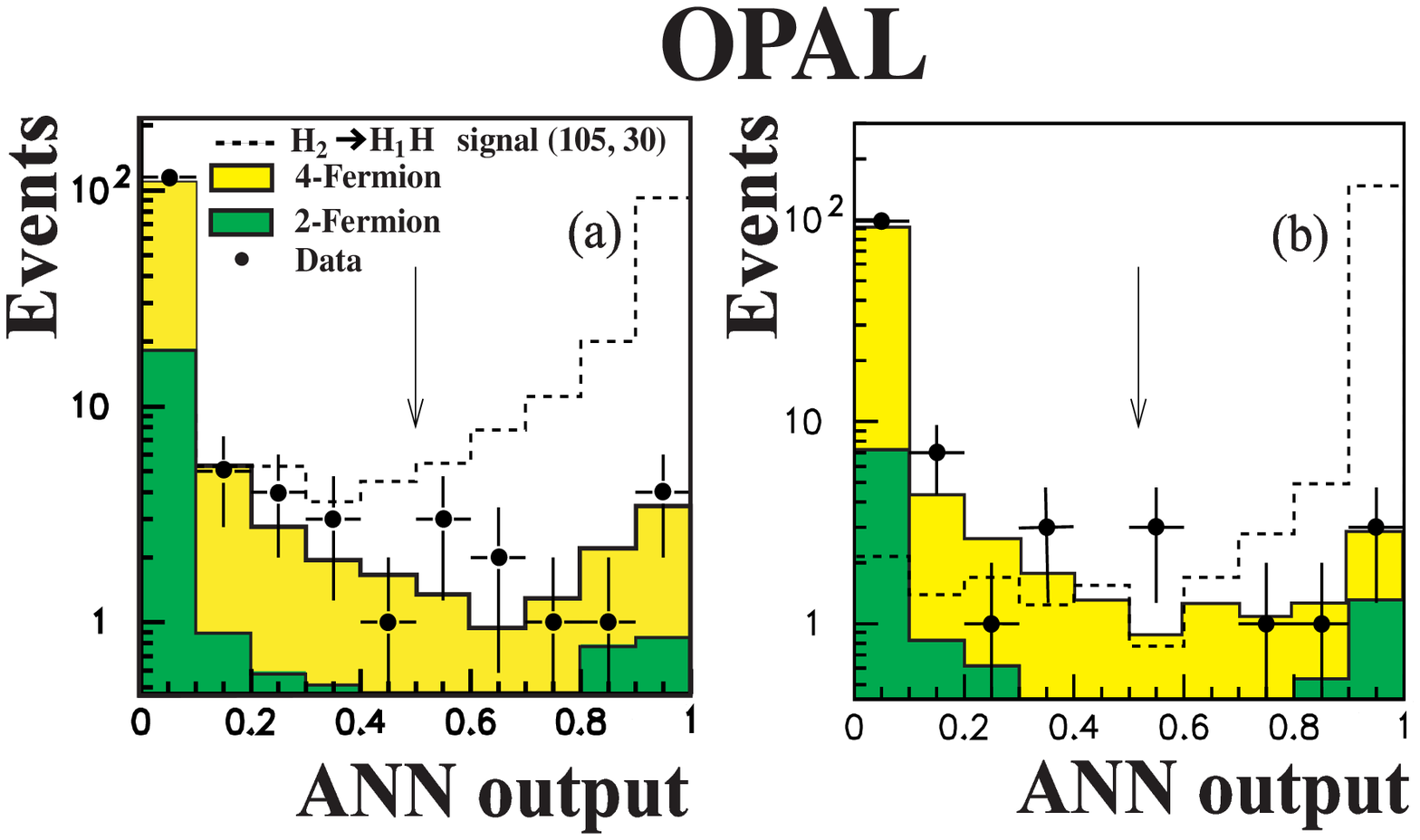, width=0.9\textwidth}
  \end{center}
  \caption{\sl{Distributions of ANN$_{\mathrm{A}}$ (a) and ANN$_{\mathrm{B}}$ (b)
      in the dedicated selection for $\ee\ra\genHtwo\Zo\ra\bb\nn$ and
      $\ee\ra\genHone\genHone\Zo\ra\bb\bb\nn$ (see Section~\ref{sect:missing_energy}) in the region 
      100~GeV~$< \genmHtwo < $~110 GeV.  The arrows indicate the cut on the
      ANN output value.
      }}\label{fig:abbbbnn}
\end{figure}

\begin{figure}[pt]
  \begin{center}
    \epsfig{file=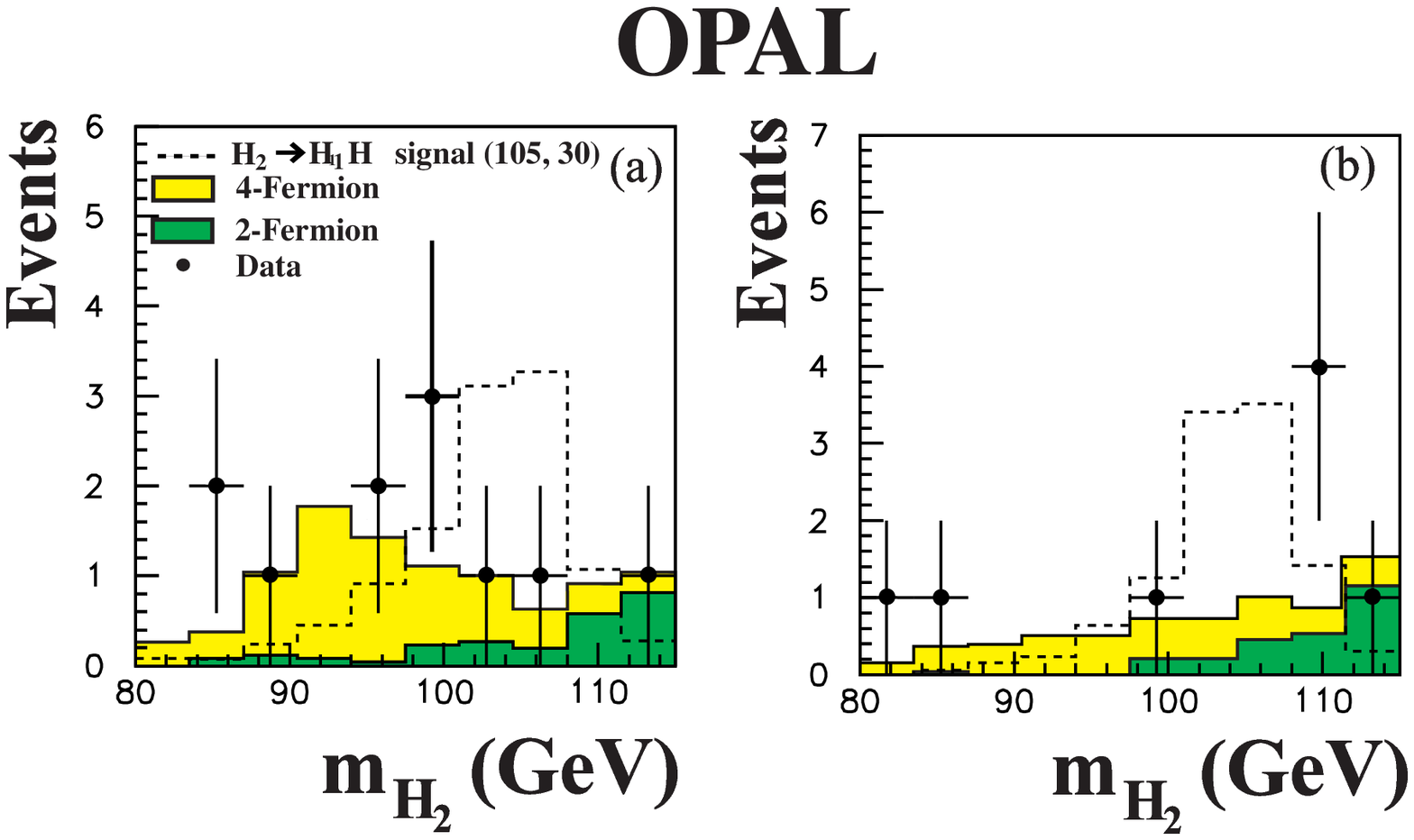, width=0.9\textwidth}
  \end{center}
  \caption{\sl{Distributions of the reconstructed masses in the selection $A$ and $B$ of the
      missing energy channel for data taken at $\sqrts=199$ to $209$~GeV in the region 
      100~GeV~$< \genmHtwo < $~110 GeV (see Section~\ref{sect:missing_energy}).  
      }}\label{fig:mbbbbnn}
\end{figure}

The Higgs mass is reconstructed using the di-jet invariant mass after
the 1 constraint kinematic fit.  The distributions of the
ANN$_{\mathrm{A}}$ and ANN$_{\mathrm{B}}$ output variables are shown
in Fig.~\ref{fig:abbbbnn}, and the distributions of the reconstructed
mass in Fig.~\ref{fig:mbbbbnn}.  Candidate events are selected if
ANN$_{\mathrm{A}} > 0.5 $ for events in subsample A and
ANN$_{\mathrm{B}} > 0.5$ in subsample B.  The efficiencies for signal
events for both \genHtwo\ra\bb{} and
\genHtwo\ra\genHone\genHone\ra\bb\bb{} are determined for both
selections.  In each point of the MSSM parameter space, the expected
signal distributions in the output variables are added from both
signal sources according to their expected rates. The number of
candidate events in subsample A(B) is 11(8) with 10.0 (7.2) events
expected from background (see Table~\ref{tab:cutbbbbnn}). The signal
efficiencies are shown in Table~\ref{tab:zhqa_eff} for various values
of (\genmHtwo, \genmHone).  The reconstructed masses and the ANN
outputs are used to construct the discriminating variable ${\cal D}$,
which is used in the statistical combination with other search
channels.

The systematic uncertainties for this channel 
are evaluated by analogy to the SM missing energy
search~\cite{OPALSMPAPER}.  They amount to 2.2\% (9.0\%) in the signal
(background) for subsample A and to 2.5\% (19.3\%) in subsample B. The
strongest contributions arise from the uncertainty of the B hadron
fragmentation and from Monte Carlo statistics.  Systematic
uncertainties of 9\% for the background and 1\% for the signal are
added in quadrature to account for the uncertainty introduced when
simulating the separation of the events into the two subsamples. These
are estimated by shifting the value of $y_{32}$ of each event by the
difference in the mean values of the background and the data
distributions of $y_{32}$ and repeating the selection with the
modified value of $y_{32}$.

%
%

\subsection{Searches for pair production processes}
\label{fourb}

\begin{enumerate}
\item[(a)] The search for the four-b final state
  $\ee\ra(\genHone\ra\bb)(\genHtwo\ra\bb)$ provides the highest
  sensitivity. While in the CPC scenario the pair production process
  is dominant only for $\genmHone\approx\genmHtwo$, this is not the
  case in the CPV scenario. The search in this channel is therefore
  optimized separately for small $\genmHone$ and large $\genmHone$.
  These are described below in Sections \ref{sect:ahbb_eq} and
  \ref{sect:ahbb_gg}.
\item[(b)] For the Higgs cascade decay
  $\ee\ra(\genHone\ra\bb)(\genHtwo\ra\genHone\genHone\ra\bb\bb)$ with
  6 b-quarks in the final state, the search in the four-b final state
  for similar masses described in Section~\ref{sect:ahbb_eq} is used
  because it has a reasonably good efficiency.  Even for large mass
  differences and thus small $\genmHone$ this search is more efficient
  than the one described in Section~\ref{sect:ahbb_gg}, due to the
  highly spherical shape of the six-b events. This search is described
  in Section~\ref{sect:ah6b}
\item[(c)] The search for the final states
  $\ee\ra(\genHone\ra\bb)(\genHtwo\ra\tautau)$ and
  $\ee\ra(\genHone\ra\tautau)(\genHtwo\ra\bb)$ follow the technique
  described in \cite{OPALSMPAPER} for the corresponding Standard Model
  channels.  The final likelihood selection and its optimization for
  the MSSM case is described in Section~\ref{btau}.
\end{enumerate}

%
%
\boldmath
\subsubsection{Search for $\ee\ra\genHone\genHtwo\ra\bb\bb$ optimized for high $\genmHone$}
\unboldmath
\label{sect:ahbb_eq}

Events from the process \genHone\genHtwo\ra\bb\bb\ with high
$\genmHone$ have four energetic b-jets and a total visible energy
close to the centre-of-mass energy.  The dominant backgrounds arise
from the four-fermion processes $\ee\ra\ZZ$ and $\ee\ra\WW$ and from
two-fermion processes $\ee\ra\qq(\gamma)$.  The events are forced into
four jets using the Durham jet finding algorithm and the following
preselection is applied.

\begin{enumerate}
\item The event must qualify as a multi-hadronic final state according
  to \cite{l2mh},
\item The effective centre-of-mass energy $\sqrt{s'}$ is required to
  be higher than $0.794\,\sqrts$,
\item The 3-to-4 jet resolution parameter
  $y_{43}$~\cite{durham_jetfinder} is required to be larger than
  0.003,
\item The $C$-Parameter \cite{Parisi:1978eg}, which is a measure of
  the spherical shape of the event, is required to be larger than
  0.45,
\item The sum of the number of reconstructed tracks and
  electromagnetic clusters not associated to tracks~\cite{opal_eflow}
  belonging to each jet has to be larger than six,
\item To discriminate against poorly reconstructed events, a
  4-constraint kinematic fit is applied, using energy and momentum
  conservation; this fit is required to converge and the $\chi^2$
  probability is required to be larger than $10^{-5}$.
\end{enumerate}

\begin{figure}[htbp]
  \vspace*{-1.0cm}
  \centerline{\hbox{\epsfig{file=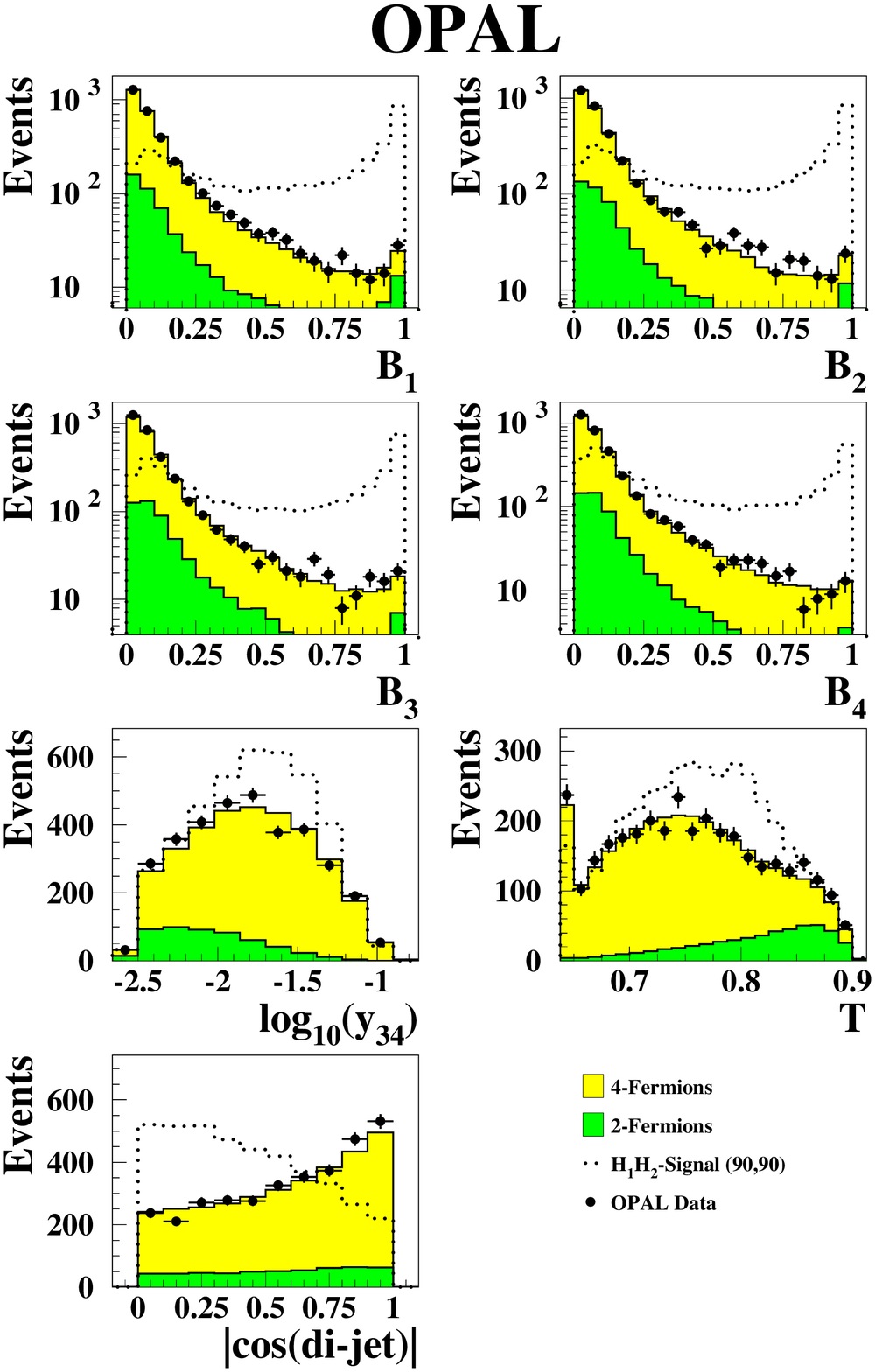, width=0.8\textwidth}}}
  \caption[]{\label{fig:ahbbbb192-209_lin}\sl
    Distributions of the likelihood input variables for the $\ee\ra\genHone\genHtwo\ra\bb\bb$ searches for high $\genmHone$ at 
    192--209 GeV (see Section~\ref{sect:ahbb_eq}).       
    }
\end{figure}

For the events passing the preselection, a likelihood function is
constructed from seven input variables, allowing the events to be
classified as signal or background.  The likelihood variables are:
\begin{enumerate} 
\item The four b-tagging discriminants $\Dbi$~\cite{pr285} for each of
  the four jets, ordered by energy,
\item The logarithm of the jet resolution parameter $y_{43}$,
\item The event thrust value $T$,
\item The estimate of the $\genHone\genHtwo$ production angle,
  $|\cos{\theta}_{\mathrm{dijet}}|$, which is defined as follows. For
  the jet pairing that yields the smallest difference between the two
  dijet-masses, $|\cos{\theta}_{\mathrm{dijet}}|$ is the absolute
  value of the cosine of the dijet polar angle.
\end{enumerate}
The signal reference histograms are obtained using Monte Carlo samples
with $\genmHone\ge 60$~GeV and $\genmHtwo\ge 60$~GeV. For the
background, reference histograms are formed from $\ee\ra\qq(\gamma)$
and $\ee\ra\qq\qq$ events.  The distributions of these variables are
shown in Figure ~\ref{fig:ahbbbb192-209_lin}.  Events are selected if
they satisfy ${\cal L}>0.95$, which provides the best sensitivity
measured in terms of $s/\sqrt{b+2}$ for $\genmHone=\genmHtwo=90$~GeV.

\begin{figure}[htbp]
  \vspace*{-1.0cm}
  \centerline{\hbox{\epsfig{file=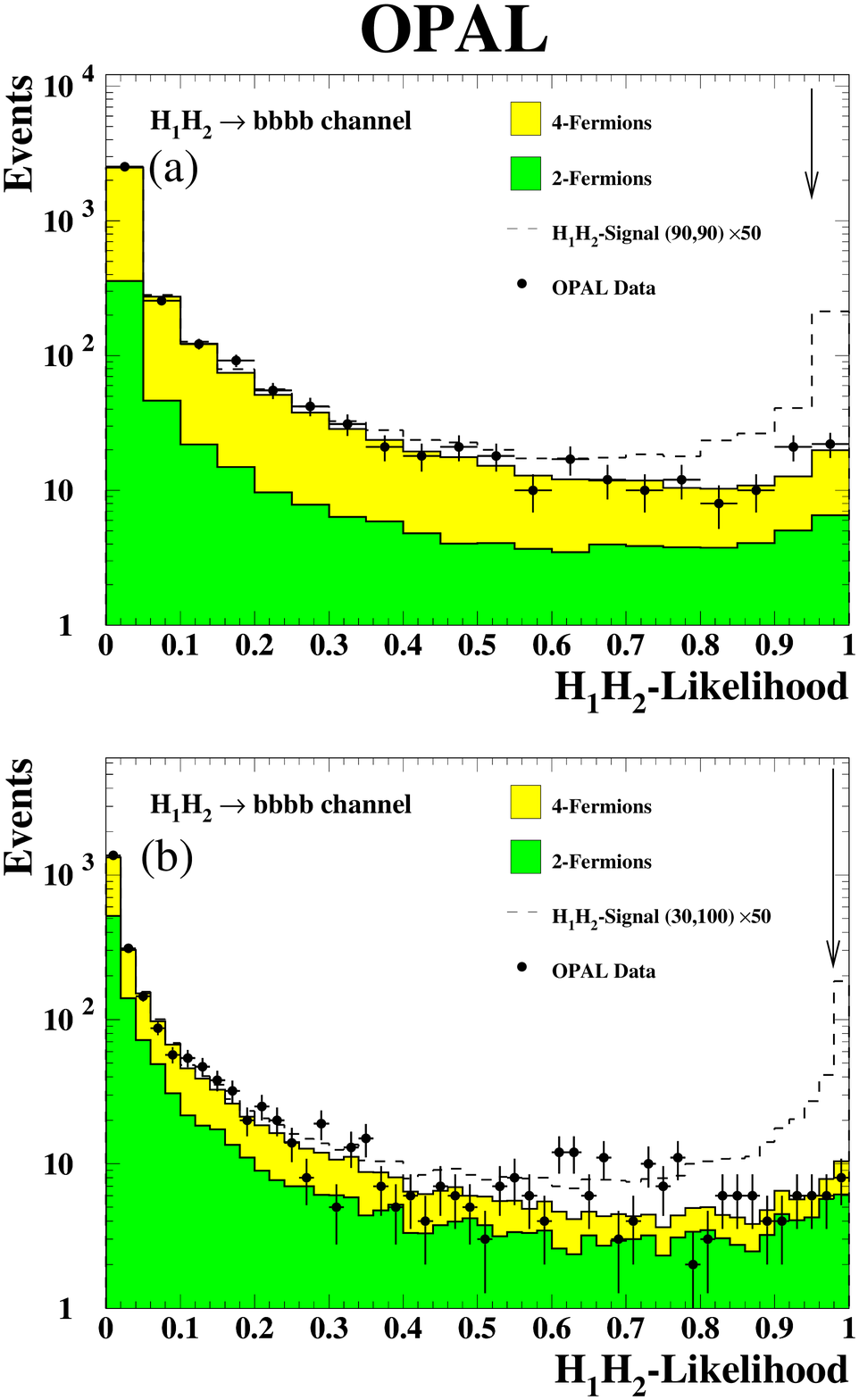, width=0.7\textwidth}}}
  \caption[]{\label{fig:ahbb192-209_lhout}\sl
    Searches for $\ee\ra\genHone\genHtwo\ra\bb\bb$ with high $\genmHone$ (see Section~\ref{sect:ahbb_eq}, \ref{sect:ahbb_gg}) at 192--209 GeV.
    Likelihood outputs for a) high $\genmHone$, and b) low $\genmHone$.
    The arrow indicates the cut position. The signal is 
    scaled with a factor of 50 with respect to a $\ee\ra\genHone\genHtwo\ra\bb\bb$ signal for $\cos^2(\beta-\alpha)=1$.
    }
\end{figure}

The numbers of observed and expected background events after each
preselection cut and the final likelihood cut are shown in
Table~\ref{tab:ah_cutflow} for data taken at $\sqrts=192$ to 209~GeV.
The distribution of the likelihood output is shown in
Figure~\ref{fig:ahbb192-209_lhout}~(a). The efficiencies for various
combinations of $(\genmHone,\genmHtwo)$ are shown in
Table~\ref{tab:ahbb_eff}.

\begin{figure}[htbp]
  \centerline{\hbox{\epsfig{file=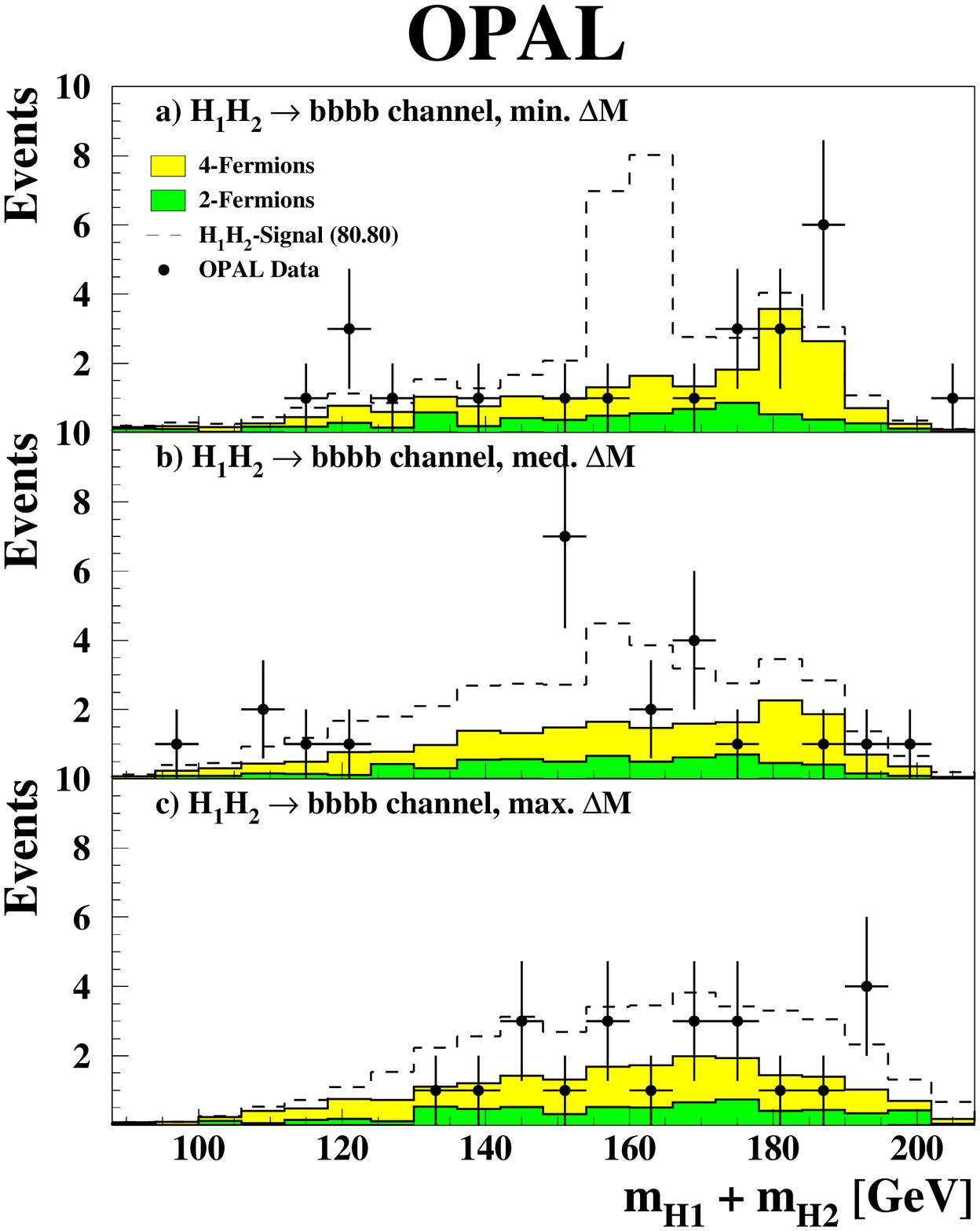, width=0.8\textwidth}}}
  \caption[]{\label{fig:ahbb192-209_mass}\sl
    Searches for $\ee\ra\genHone\genHtwo$ with high $\genmHone$ at 192--209 GeV.
    The sum of the reconstructed Higgs boson masses, 
    $\genmHone^{\mathrm {rec}} + \genmHtwo^{\mathrm {rec}}$, for
    (a) the $\genHone\genHtwo\rightarrow$4b channel with the di-jet pairing combination 
    which yields the smallest mass difference,
    $\Delta M\equiv |\genmHtwo^{\mathrm {rec}}-\genmHone^{\mathrm {rec}}|$,
    (b) the $\genHone\genHtwo\rightarrow$4b channel with the intermediate $\Delta M$ combination,
    (c) the $\genHone\genHtwo\rightarrow$4b channel with the largest $\Delta M$ combination.
    OPAL data are
    indicated by points with error bars, the four-fermion background
    by the light grey (yellow) histograms, and the two-fermion background by the
    darker grey (green) histograms. 
    Shown as dashed histograms are the contributions expected from
    a Higgs boson signal with full strength at $\genmHone=\genmHtwo=80$~GeV 
    for a luminosity of $207\,\ipb$ taken at $\sqrts=199-209$~GeV, added
    to the background expectation for a luminosity of $424\,\ipb$ and the cross-section
    for $\cos^2(\beta-\alpha)=1$.
    }
\end{figure}

The mass of the Higgs boson candidates is reconstructed using a
constrained fit requiring energy and momentum conservation.
Figures~\ref{fig:ahbb192-209_mass}(a)--(c) show the distributions of
the sum of the reconstructed Higgs boson masses, $M_{\mathrm{sum}} =
m_{{\cal H}_1}^{\mathrm{rec}}+m_{{\cal H}_2}^{\mathrm{rec}}$, for the
jet combination with the largest, second largest and smallest value
for $|m_{{\cal H}_1}^{\mathrm{rec}}-m_{{\cal H}_2}^{\mathrm{rec}}|$.
No significant excess over the expected background is observed.  The
discriminating variable ${\cal D}$ is a two-dimensional array of
reconstructed masses
$\genmHtwo^{\mathrm{rec}}+\genmHone^{\mathrm{rec}}$ and
$\genmHtwo^{\mathrm{rec}}-\genmHone^{\mathrm{rec}}$.

The systematic uncertainties on the signal efficiencies and background
expectation for the $\genHone\genHtwo\ra\bb\bb$ search are given in
Table~\ref{ah:allsyst}.  They are evaluated as for the SM searches
in~\cite{OPALSMPAPER} and include Monte Carlo statistics (uncorrelated
between channels, energies and signal and background), detector
modelling, such as tracking resolution in $r\phi$ and $z$, hit
matching efficiency in the silicon microvertex detector for $r\phi$
and $z$, B-hadron decay multiplicity and momentum spectrum, c-hadron
momentum spectrum, comparison between different SM Monte Carlo
generators, uncertainties of the four-fermion cross-section (all taken
to be fully correlated between channels and energies) and
uncertainties in the modelling of the likelihood variables (taken to
be uncorrelated between channels and fully correlated between energies
of the same channel).  The systematic uncertainty amounts to $3.1\%$
for the signal and $10.3\%$ for the background expectation.

%
%
\boldmath
\subsubsection{Search for $\ee\ra\genHone\genHtwo\ra\bb\bb$ for low $\genmHone$}
\unboldmath
\label{sect:ahbb_gg}

The region 12~GeV$\,<m_{\genHone}<30$~GeV and $\genmHtwo>90$~GeV is of
particular interest in the CPV scenario. The following selection is
optimized for that kinematic region and replaces the selection of
Section~\ref{sect:ahbb_eq}.  Events with large
$m_{\genHtwo}-m_{\genHone}$ look like asymmetrically boosted three-jet
events.

The preselection is identical to the one of Section~\ref{sect:ahbb_eq}
except for cuts (3) and (4). In (3) the cut on $y_{43}$ is relaxed to
0.0003. In order to compensate for the increased
$\qq\gamma$-background an additional requirement is introduced in cut
(3): the sum of the two smallest angles between any jets, $J_2$, has
to satisfy the requirement $30^{\circ}<J_2<175^{\circ}$ and the sum of
the four smallest angles between jets, $J_4$, has to satisfy
$220^{\circ}<J_4<400^{\circ}$. Cut (4) is relaxed to $C>0.2$. This
increases the acceptance for asymmetric three-jet-like events. The
number of selected events after each cut, along with the expected
background, is shown in Table~\ref{tab:ahflow200-209_offdiag}.

\begin{figure}[htbp]
  \vspace*{-1.0cm}
  \centerline{\hbox{\epsfig{file=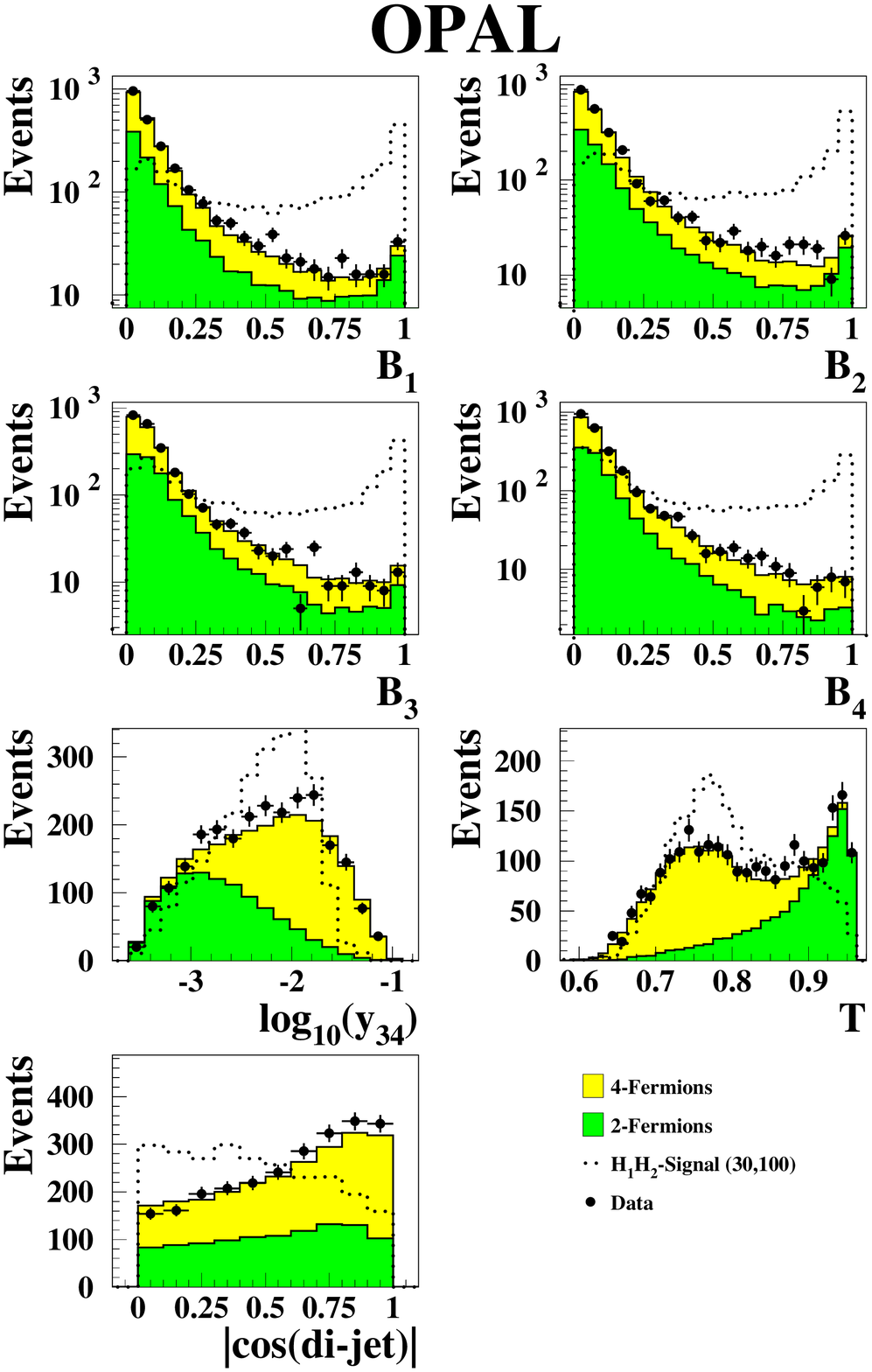, width=0.8\textwidth}}}
  \caption[]{\label{fig:ahbbbb192-209_lin_offdiag}\sl
    Distributions of the input variables for the $\ee\ra\genHone\genHtwo\ra\bb\bb$ searches for low $\genmHone$ at 
    199--209 GeV (see Section~\ref{sect:ahbb_gg}).       
    }
\end{figure}

After the preselection, a likelihood function is constructed from the
seven variables described in Section~\ref{sect:ahbb_eq}.  Signal
reference histograms are formed from Monte Carlo samples with
$12<m_{\genHone}<30$~GeV and $90<m_{\genHtwo}<110$~GeV.  The
distributions of the input variables are shown in
Fig.~\ref{fig:ahbbbb192-209_lin_offdiag}. The resulting likelihood
distribution is shown in Fig.~\ref{fig:ahbb192-209_lhout}(b).  The cut
${\cal L}>0.98$ is applied, which is optimal for $\genmHone=30$~GeV
and $\genmHtwo=100$~GeV.

\begin{figure}[htbp]
  \centerline{\hbox{\epsfig{file=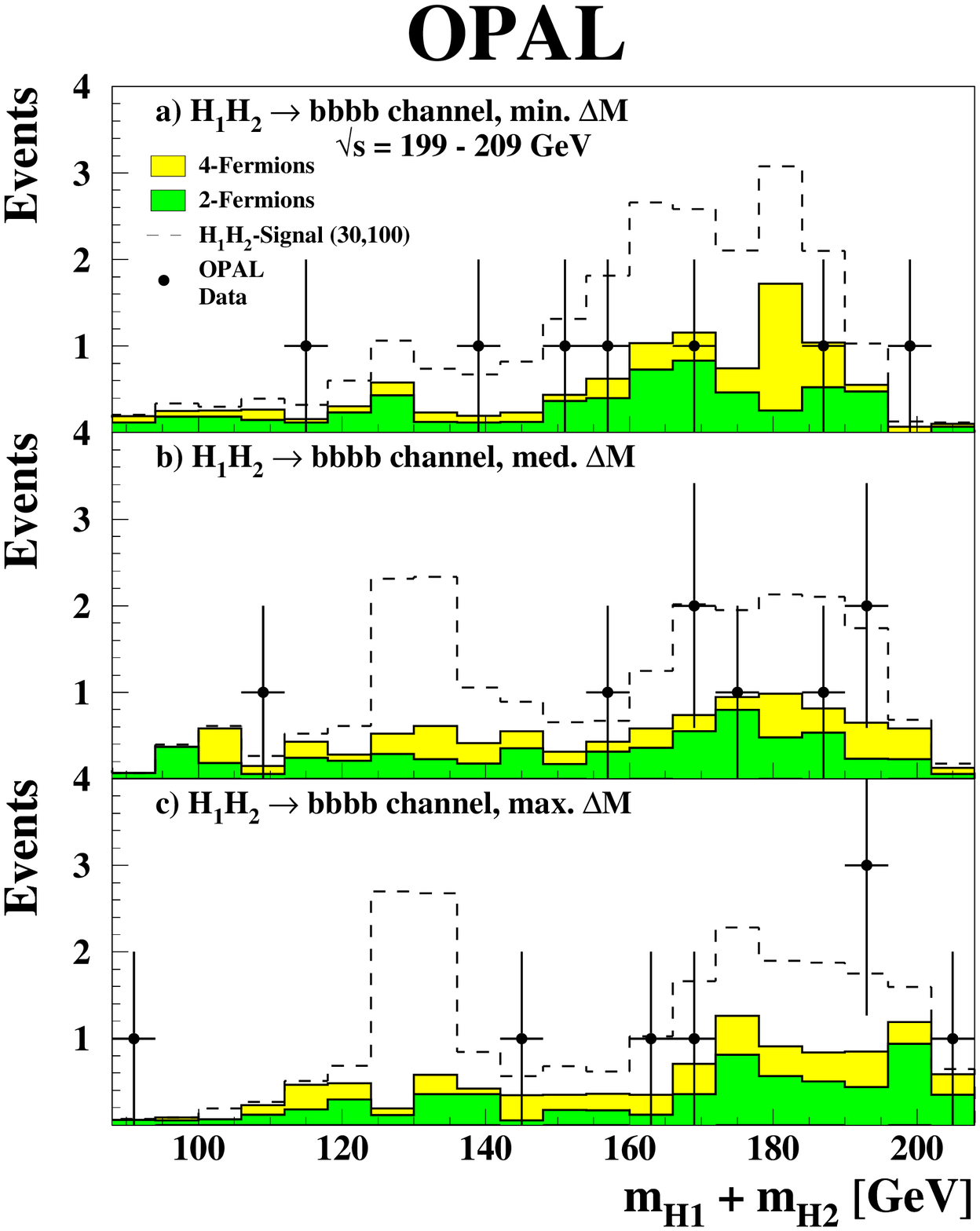, width=0.8\textwidth}}}
  \caption[]{\label{fig:ahbb192-209_mass_offdiag}\sl
    Searches for $\ee\ra\genHone\genHtwo$ with low $\genmHone$ at 199--209 GeV.
    The sum of the reconstructed Higgs boson masses, 
    $\genmHone^{\mathrm {rec}} + \genmHtwo^{\mathrm {rec}}$, for
    (a) the $\genHone\genHtwo\rightarrow$4b channel with the di-jet pairing combination 
    which yields the smallest mass difference,
    $\Delta M\equiv |\genmHtwo^{\mathrm {rec}}-\genmHone^{\mathrm {rec}}|$,
    (b) the $\genHone\genHtwo\rightarrow$4b channel with the medium $\Delta M$ combination,
    (c) the $\genHone\genHtwo\rightarrow$4b channel with the maximum $\Delta M$ combination.
    OPAL data are
    indicated by points with error bars, the four-fermion background
    by the light grey (yellow) histograms, and the two-fermion background by the
    darker grey (green) histograms. 
    Shown as dashed histograms are the contributions expected from
    a Higgs boson signal with $\genmHone=30$~GeV, $\genmHtwo=100$~GeV, added
    to the background expectation for a luminosity of $207\,\ipb$ and the cross-section
    for $\cos^2(\beta-\alpha)=1$. 
    }
\end{figure}

The efficiencies for various combinations of $(\genmHone,\genmHtwo)$
are shown in Table~\ref{tab:ahbb_eff_offdiag}.  The distribution of
the reconstructed mass sum $M_{\mathrm{sum}}$ is shown in
Fig.~\ref{fig:ahbb192-209_mass_offdiag}. No significant excess over
the background is observed.  The discriminating variable ${\cal D}$ is
a two-dimensional array of reconstructed masses
$\genmHtwo^{\mathrm{rec}}+\genmHone^{\mathrm{rec}}$ and
$\genmHtwo^{\mathrm{rec}}-\genmHone^{\mathrm{rec}}$.  The systematic
uncertainties are listed in Table~\ref{ah:allsyst} and are derived in
the same way as for the search described in
Section~\ref{sect:ahbb_eq}.  They amount to $4.7\%$ for the signal and
$10.5\%$ for the background expectation.

%
%
\boldmath
\subsubsection{Search for $\ee\ra\genHone\genHtwo\ra\genHone\genHone\genHone\ra\bb\bb\bb$}
\unboldmath
\label{sect:ah6b}
The search channel for $\genHone\genHtwo\ra\bb\bb$ optimized for high
$\genmHone$ is also used to search for events of the type
$\ee\ra\genHone\genHtwo\ra\genHone\genHone\genHone\ra\bb\bb\bb$.
Despite the large mass difference $\genmHtwo-\genmHone$ and generally
relatively low $\genmHone$, the selection for high $\genmHone$
(Section~\ref{sect:ahbb_eq}) is more efficient than the selection for
low $\genmHone$ (Section~\ref{sect:ahbb_gg}) due to the spherical
shape of the $\bb\bb\bb$ events.

The expected signal distribution is added to the one from the
4b-channel in the same way as described in Section~\ref{sect:fourjet}.
The efficiency of this search is shown in Table~\ref{tab:effbbbbbb}
for various $(\genmHone,\genmHtwo)$.  The systematic error of the
signal of this channel is taken to be the same as for the search
optimized for high $\genmHone$.

%
%

\boldmath
\subsubsection{Search for the process $\ee\ra\genHone\genHtwo\ra\bb\,\tautau$ and $\tautau\bb$}
\unboldmath
\label{btau}

\begin{figure}[htbp]
  \centerline{\hbox{\epsfig{file=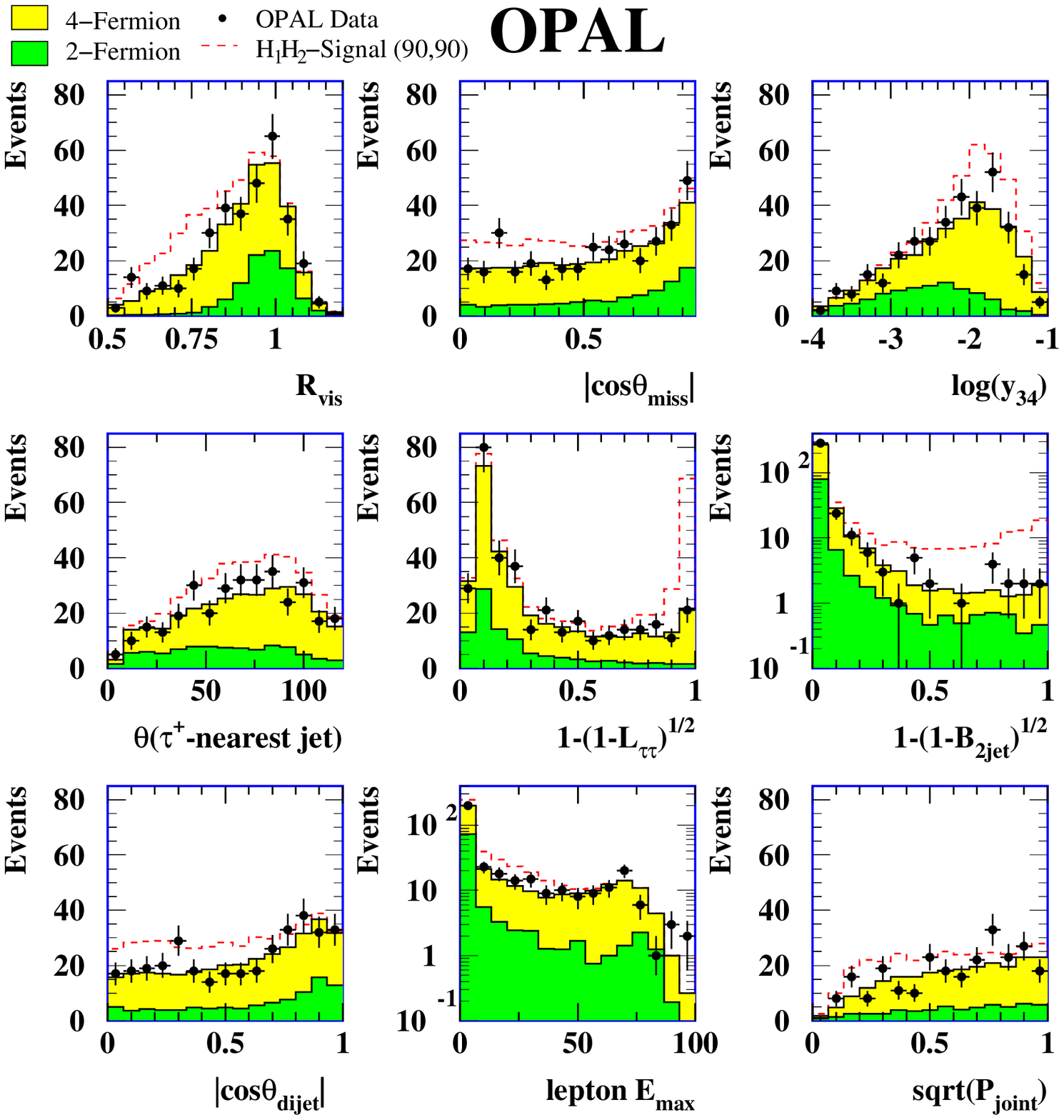, width=1\textwidth}}}
  \caption[]{\label{fig:ahbbtt192-209_lin}\sl
    Input variables~\cite{OPALSMPAPER} for the $\ee\ra\genHone\genHtwo\ra\bb\tautau$ searches at 
    192--209 GeV. 
    }
\end{figure}

\begin{figure}[htbp]
  \centerline{\hbox{\epsfig{file=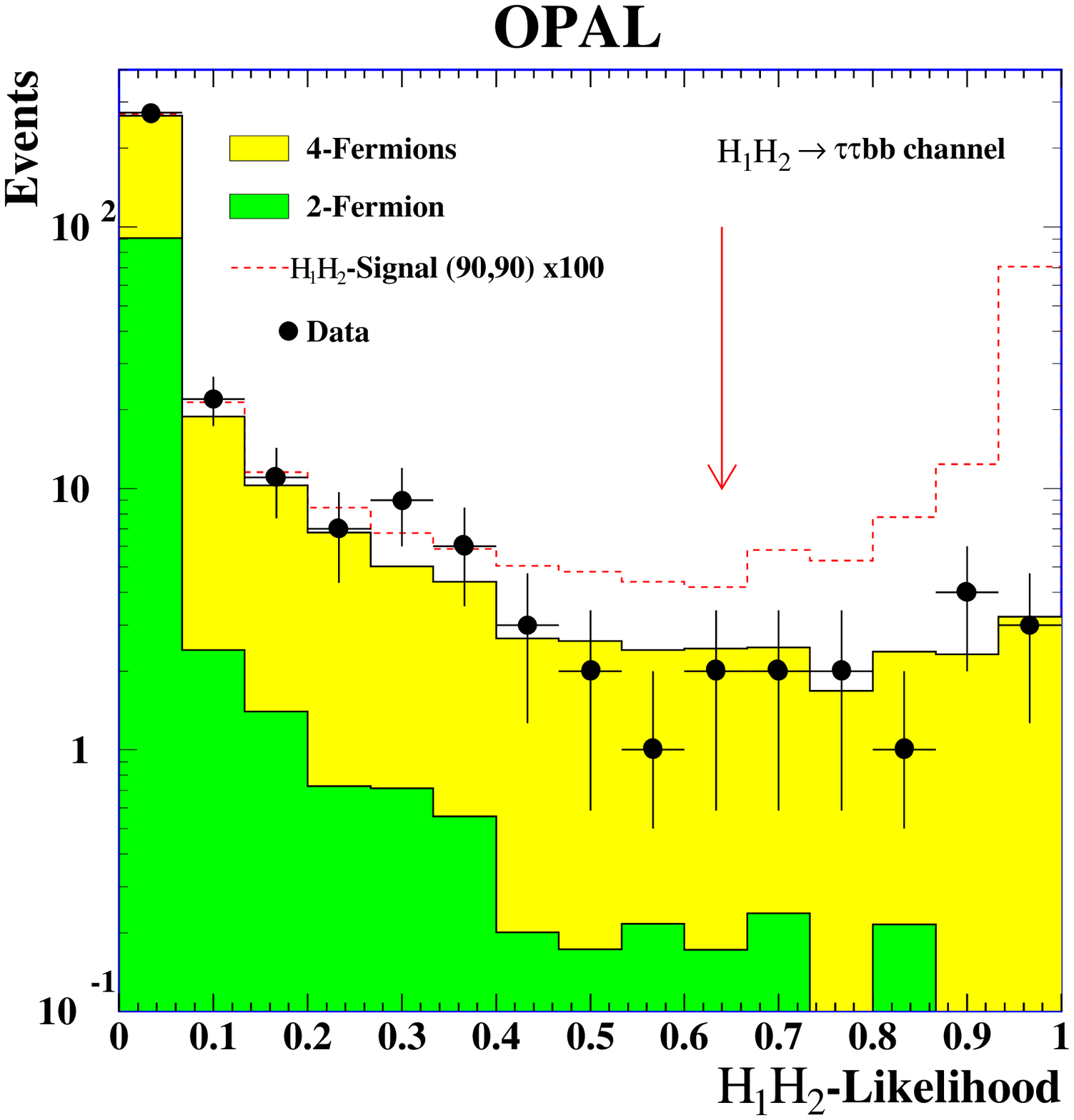, width=0.6\textwidth}}}
  \caption[]{\label{fig:ahbt192-209_lhout}\sl
    Likelihood output of the $\ee\ra\genHone\genHtwo\ra\bb\tautau$ search at 192--209 GeV.
    The arrow indicates the cut position. The signal is 
    scaled with a factor of 100 with respect to a $\ee\ra\genHone\genHtwo\ra\bb\tautau$ signal for $\cos^2(\beta-\alpha)=1$.}
\end{figure}

The search for the \genHone\genHtwo\ra\bb\tautau\ final state, where
either \genHone\ or \genHtwo\ decays into a tau pair, uses the same
techniques as the SM tau search ~\cite{OPALSMPAPER}.  The final
likelihood selection including the b-tagging is optimized for the MSSM
process.  The likelihood ${\cal L}_{\genHone\genHtwo}$ differs from
the SM likelihood in the following way.
\begin{enumerate}
\item The signal reference histograms are constructed from simulated
  $\genHone\genHtwo$ events, using a large range of $\genmHone$ and
  $\genmHtwo$ values,
\item The $\chi^2$ probability of the 3-constraint fit, constraining
  the invariant mass of the $\tautau$ pair to $\mZ$, is dropped,
\item The variable $|\cos{\theta}_{\mathrm{dijet}}|$ (cf.
  Section~\ref{sect:ahbb_eq}) is introduced.  Here the dijet pairing
  is defined by the pair of b-tagged jets and the pair of tau jets.
  The mean value of the cosines of the two systems is taken.
\end{enumerate}
The distributions of the likelihood input variables are shown in
Fig.~\ref{fig:ahbbtt192-209_lin}.  The distribution of the likelihood
${\cal L}$ is shown in Fig.~\ref{fig:ahbt192-209_lhout}, and the
number of selected events after each cut along with the expected
background is shown in Table~\ref{tab:ah_cutflow}. The discriminating
variable ${\cal D}$ is a two-dimensional array of reconstructed masses
$\genmHtwo^{\mathrm{rec}}+\genmHone^{\mathrm{rec}}$ and
$\genmHtwo^{\mathrm{rec}}-\genmHone^{\mathrm{rec}}$.  The cut ${\cal
  L}>0.64$ is applied.  The efficiencies of the selection are given in
Table~\ref{tab:bbtautaueff}.  Systematic uncertainties are listed in
Table~\ref{ah:allsyst}. They are derived in the same way as in
Section~\ref{sect:ahbb_eq}. Additionally the systematic error on the
tau identification is evaluated using event mixing~\cite{OPALSMPAPER}.
They amount to $2.6\%$ for the signal and $15.8\%$ for the background
expectation.

%
%

\subsection{Comparison of Data and Expected Background Monte Carlo}

In Table~\ref{tab:channel_sum} the results of the new searches, which
are described in detail in this section, are summarized in terms of
total background and data events. There is good agreement between the
selected data events and the expected SM background.  The numbers in
this table are obtained by choosing for illustration
$\genmHone=39$~GeV and $\genmHtwo=105$~GeV in those searches where the
selections depend explicitly on the hypothetical Higgs boson masses.

No significant excess of candidate events over the expected background
is found in any of the old and new search channels.

%
%

\section{Combination of search channels and hypothesis testing}\label{sect:combmet}

%
%

\subsection{Combined confidence levels}

The sensitivity of the searches for hypothetical Higgs bosons is
increased by combining the results of the various topological
searches. This is done following the statistical method described in
\cite{OPALSMPAPER}.

In order to compute confidence levels, a test statistic $Q$ is defined
that can be used to quantify the compatibility of the data with two
hypotheses: the background hypothesis and the signal+background
hypothesis.  The confidence levels are computed from a comparison of
the observed test statistic and its probability distributions for a
large number of simulated experiments for these two hypotheses.  In
this paper, the ratio $Q={\cal L}_{s+b}/{\cal L}_{b}$ of the
likelihoods for the two hypotheses is chosen as the test statistic.
The results of all search channels are expressed in fine bins of
discriminating variables ${\cal D}$, as defined in the descriptions of
the individual searches.
For each bin $i$ of ${\cal D}$ three numbers are calculated: $s_i$,
the number of expected signal events for a given set of model
parameters (Higgs masses etc.), $b_i$, the number of expected
background events, and $n_i$, the number of observed events.  Each bin
is considered to be a statistically independent counting experiment
obeying Poisson statistics.  The test statistic can then be
computed~\cite{OPALSMPAPER} as
$$-2 \ln Q=2\sum_i s_i -2\sum_i n_i\ln(1+s_i/b_i).$$
The confidence
level for the background hypothesis, ${\mathrm{CL}}_b$, is defined as
the probability to obtain values of $Q$ no larger than the observed
value $Q_{\mathrm{obs}}$, given a large number of hypothetical
experiments with background processes only,
$$\mathrm{CL}_b=P(Q\le Q_{\mathrm{obs}}|{\mathrm{background}}).$$
Similarly, the confidence level for the signal+background hypothesis,
${\mathrm{CL}}_{s+b}$, is defined as the probability to obtain values
of $Q$ smaller than observed, given a large number of hypothetical
experiments with signal and background processes,
$${\mathrm{CL}}_{s+b}=P(Q\le
Q_{\mathrm{obs}}|{\mathrm{signal+background}}).$$
In principle,
${\mathrm{CL}}_{s+b}$ can be used to exclude the signal+background
hypothesis, given a model for Higgs boson production.  However, this
procedure may lead to the undesired possibility that a downward
fluctuation of the background would allow hypotheses to be excluded
for which the experiment has no sensitivity due to the small expected
signal rate. This problem is avoided by introducing the ratio
$${\mathrm{CL}}_s={\mathrm{CL}}_{s+b}/{\mathrm{CL}}_b.$$
Since
${\mathrm{CL}}_b$ is a positive number less than one,
${\mathrm{CL}}_s$ will always be greater than ${\mathrm{CL}}_{s+b}$
and the limit obtained in this way will therefore be conservative. We
adopt this quantity for setting exclusion limits and consider a
hypothesis to be excluded at the $95\,\%$ confidence level if the
corresponding value of ${\mathrm{CL}}_s$ is less than 0.05.

The expected confidence levels are obtained by replacing the observed
data configuration by a large number of simulated event configurations
for the two hypotheses background only or signal+background.  These
can be used to estimate the expected sensitivity of a search and to
compare the observed exclusion with the one expected with no signal
present.

The effect of systematic uncertainties of the individual channels is
calculated using a Monte Carlo technique.  The signal and background
estimations are varied within the bounds of the systematic
uncertainties, assuming Gaussian distributions of the uncertainties.
Correlations are taken into account.  These variations are convoluted
with the Poisson statistical variations of the assumed signal and
background rates in the confidence level calculation.  The effect of
systematic uncertainties on the exclusion limits turns out to be
generally small.

In case of overlapping channels, i.e. channels sharing a fraction of
events, the approach described above is modified. Such a situation
occurs for example for the Higgsstrahlung searches with and without
b-tagging~\cite{OPALSMPAPER,2HDMFINAL} or in the case of the
Higgsstrahlung four-jet channel and the pair production four-b
channel. We calculate the expected CL$_s$ for each of the overlapping
channels in turn, and retain only the channel that yields the smaller
expected CL$_s$.  This procedure is repeated for each signal
hypothesis. For different Higgs boson masses therefore different
search channels give the exclusion.

The same procedure is applied if two signal processes, for example
$\genHone\Zo$ and $\genHtwo\Zo$, can contribute to the same event
topology, but at different mass values. In the case of the four-jet
channel the selection procedure and discriminant variable ${\cal D}$
depend on the Higgs mass hypothesis (test mass). Two different test
masses have not only different signal distributions $s_i$ but also
different background and data distributions $b_i$ and $n_i$.  The
selected events in searches for $\genHone\Zo$ and for $\genHtwo\Zo$
cannot be combined
since the inconsistent background and data distributions for the two
hypotheses in general contain an overlapping sample of data events.
Therefore only the hypothesis that yields the lower expected CL$_s$ is
retained.

%
%


\subsection{Additional experimental constraints}
\label{sect:add_exp_constr}

If a given model for Higgs production is not excluded by using the
search channels described above, the following additional constraints
are considered:
\begin{enumerate}
\item[(a)] The constraint from the measured $\Zo$ boson decay width
  $\Gamma_{\Zo}$: a model is regarded as excluded if the condition
  $$
  \sum_i \sigma_{{\cal H}_i\Zo}(91.4\,\mathrm{GeV})+ \sum_{i,j}
  \sigma_{{\cal H}_i{\cal H}_j}(91.4\,\mathrm{GeV}) >
  \sigma_{\Zo}(91.4\,\mathrm{GeV})\,
  \frac{\Delta\Gamma_{\Zo}}{\Gamma_{\Zo}}$$
  is satisfied using results
  from \cite{Zwidth:2000se}.  The nominal LEP1 centre-of-mass energy
  of $\sqrts=91.4$~GeV is used.  $\Gamma_{\Zo}$ is the total $\Zo$
  width, $\sigma_{\Zo}$ is the total $\Zo$ cross-section and
  $\Delta\Gamma_{\Zo}=6.5$~MeV is the maximum additional width that is
  compatible with the measured width, given the SM hypothesis
  (obtained from ZFITTER \cite{Bardin:1999yd}) at the 95\%~CL.
  
\item[(b)] The constraint from the decay mode independent search for
  $\ee\ra\mathrm{H}\Zo$ \cite{DECAYMODEINDEP}: a model is regarded as
  excluded if
  $$
  \sigma_{{\cal H}_i\Zo} > k(m_{{\cal
      H}_i})\,\sigma_{\mathrm{H}\Zo}^{\mathrm{SM}}(m_{{\cal
      H}_i})\qquad\mathrm{with}\qquad
  m_{{\cal H}_i} = m_{\mathrm{H}}$$
  is fulfilled, where $k(m_{{\cal
      H}_i})$ is the smallest scale factor for the SM Higgs production
  cross-section that is excluded at the 95\% CL by this search. This
  criterion is used for ${\cal H}_i=\genHone$ and ${\cal
    H}_i=\genHtwo$ and at $\sqrt{s}=91.4,\,183$ and 206~GeV.  The use
  of $\Zo$ width constraints and decay mode independent analyses is
  especially helpful for the range $m_{\genHone}<6$~GeV. In the CPV
  scenarios all points excluded by the $\Zo$-width constraint turn out
  to be also excluded by the decay mode independent search.
  
\item[(c)] The constraint from a search for Yukawa production of a
  light Higgs boson~\cite{bib:yukawa}: a model is regarded as excluded
  if the predicted value of the Yukawa enhancement factor $\xi$ for
  $\genHone$, multiplied with the branching fraction
  BR($\genHone\ra\tautau$), is larger than the smallest value excluded
  in \cite{bib:yukawa}. In the case of the CPV scan, where $\genHone$
  is composed of CP-odd and CP-even parts, the weaker of the two
  limits calculated for the Yukawa production of a CP-even or a CP-odd
  Higgs boson is used in the comparison. For CP-even Higgs bosons,
  $\xi=-\sin\alpha/\cos\beta$, while for CP-odd Higgs bosons
  $\xi=\tanb$ holds.  This constraint is helpful in excluding models
  with large $\tanb$, $2\,m_{\tau}<\genmHone<2\,m_{\mathrm{b}}$ and
  vanishing $\ee\ra\genHone\Zo$ cross-section.
  
\item[(d)] As an overlay, the constraint from measurements of
  inclusive decays of a b quark into an s quark and a photon
  $\mathrm{BR}(\mathrm{b}\ra\mathrm{s}\gamma)$ is shown.  A model is
  shown as excluded by this constraint if the corresponding branching
  ratio, calculated using \cite{bsg_calc}, falls outside the bounds
  $2.33\times10^{-4}<\mathrm{BR}(\mathrm{b}\ra\mathrm{s}\gamma)<4.15\times10^{-4}$
  (95\,\% CL) \cite{bsg_measurement}. This limit is used to constrain
  the CPC scenarios.
\end{enumerate}

%
%

\section{Model-independent limits on topological cross-sections}\label{sect:modelindep}

For the model-independent interpretation of the OPAL Higgs searches
the scaling factor
$$s_{95}=\frac{\sigma_{\mathrm{max}}}{\sigma_{\mathrm{ref}}}$$
is
computed, where $\sigma_{\mathrm{max}}$ is the largest production
cross-section allowed at 95\,\% CL and $\sigma_{\mathrm{ref}}$ is a
reference cross-section. For Higgsstrahlung the SM cross-section
$\sigma_{\mathrm{SM}}$ is used as $\sigma_{\mathrm{ref}}$; for pair
production the cross-section of equation (\ref{equation:xsec_ah}) with
$\cos^2(\beta-\alpha)=1$ is used. Initial-state radiation is included
according to \cite{bib:kleissgen}.

Cross-section limits on the SM-like production and decay can be found
in~\cite{OPALSMPAPER} and for flavour independent $\genH\ra\qq$ decays
in~\cite{2HDMFINAL}.

\begin{figure}[p]
  \begin{center}
    \epsfig{file=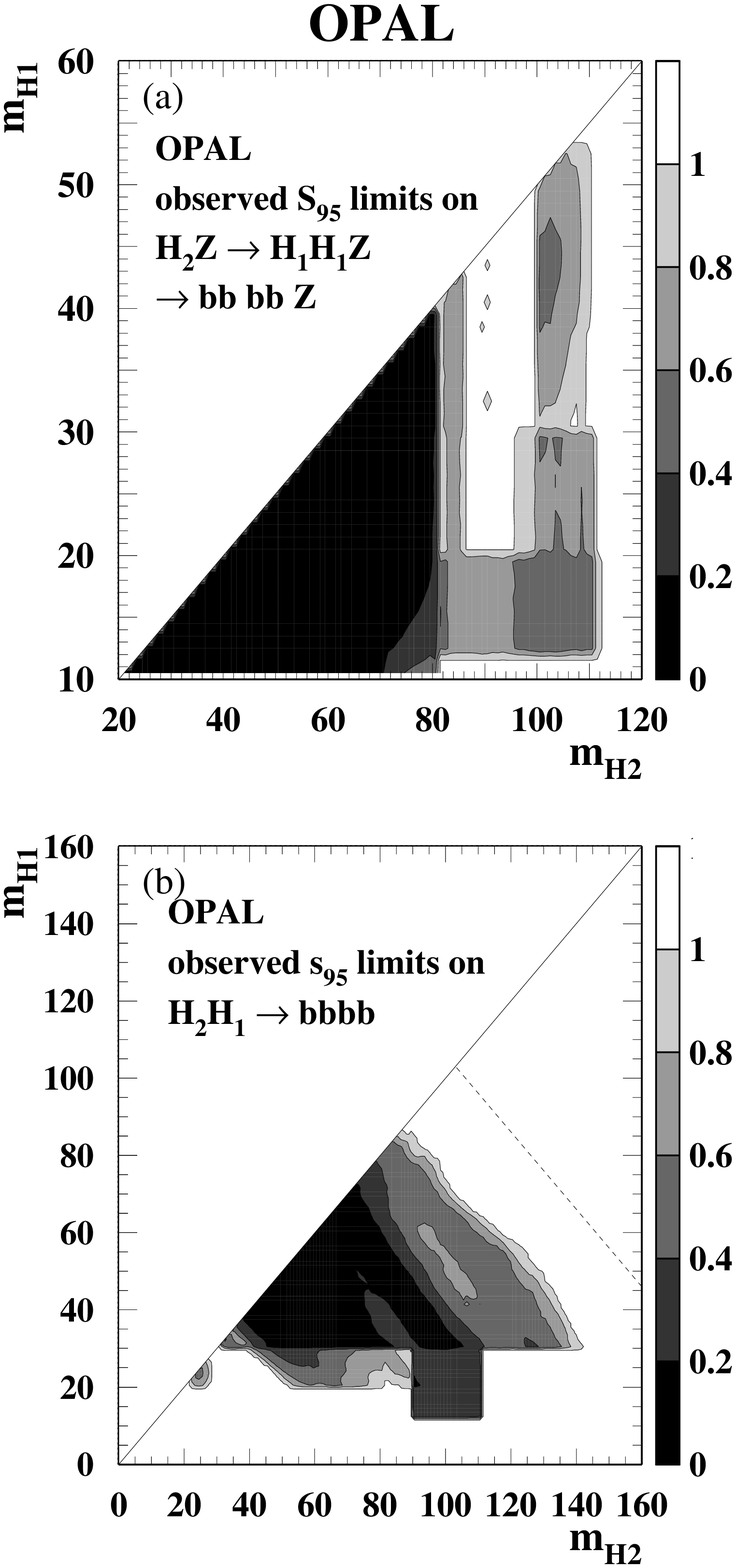,width=0.7\textwidth}\vspace{-0.8cm}\\
  \end{center}
  \caption{{\sl Model-independent upper bounds on $\sigma\times\mathrm{BR}$ for
      (a)  the $\ee\ra\genHtwo\Zo\ra\genHone\genHone\Zo\ra\bb\bb\Zo$ channel and (b) the $\ee\ra\genHone\genHtwo\ra\bb\bb$ channel. 
      For (a), the SM cross-section for $\Hosm\Zo$ production
      is taken as normalization. 
      For (b), The MSSM cross-section for $\genHone\genHtwo$ production with $\cos^2(\beta-\alpha)=1$
      is taken as normalization. The dashed line indicates the kinematic limit for $\sqrts=206$~GeV.
      }}\label{fig:hz_aaz_limits}
\end{figure}

Fig.~\ref{fig:hz_aaz_limits}~(a) shows $s_{95}$ for the production
process $\ee\ra\genHtwo\Zo\ra\genHone\genHone\Zo\ra\bb\bb\Zo$.
$\mathrm{BR}(\genHtwo\ra\genHone\genHone)=1$ and
$\mathrm{BR}(\genHone\ra\bb)=1$ is assumed.  The observed borders and
discontinuities stem from a number of different searches contributing
and being sensitive in different mass ranges.  For $\genmHtwo<80$~GeV,
specific searches for this final state at 183~GeV provide a strong
exclusion. For $80<\genmHtwo<100$~GeV, only the $\Zo\ra\qq$ final
state is used, giving a weaker exclusion. For $100<\genmHtwo<110$~GeV,
the $\Zo\ra\nunu$ final state is also employed.  The limits are
calculated for $\genmHone>10.5$~GeV only where the decay
$\genHone\ra\bb$ becomes kinematically possible.

Fig.~\ref{fig:hz_aaz_limits}~(b) shows $s_{95}$ for the process
$\ee\ra\genHone\genHtwo\ra\bb\bb$.
$\mathrm{BR}(\genHone\ra\bb)=\mathrm{BR}(\genHtwo\ra\bb)=1$ is
assumed.  The kinematic limit for $\sqrts=206$~GeV is indicated as a
dashed line.  Most searches apply only for $\genmHone>30$~GeV.  Below
$\genmHone=30$~GeV, only searches for pair production at
$\sqrts=183$~GeV or lower contribute. Additionally, the area of
$\genmHone>12$~GeV and $90<\genmHtwo<110$~GeV is studied in the data
at $\sqrts=199$ to 209~GeV.

\begin{figure}[p]
  \begin{center}
    \epsfig{file=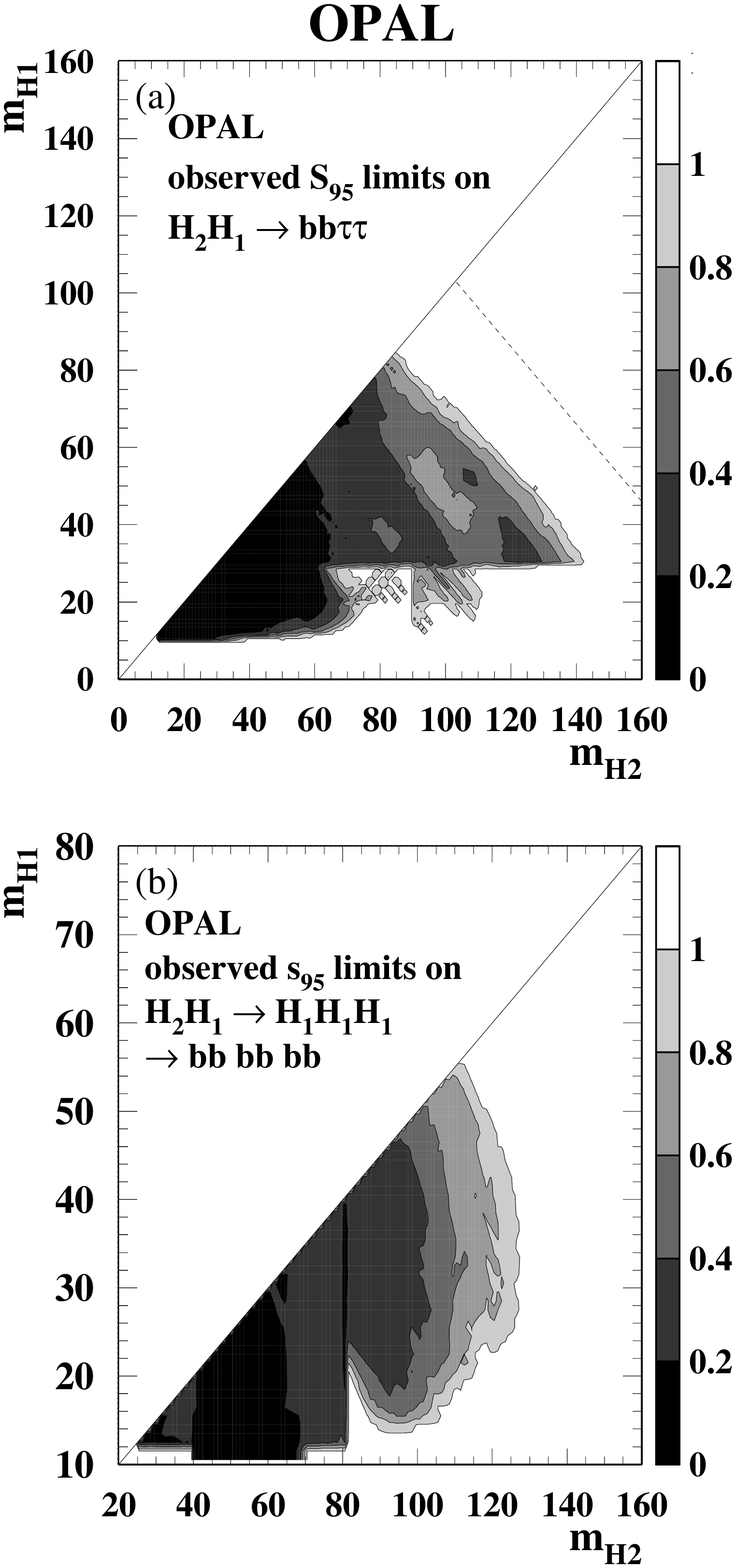,width=0.7\textwidth}\vspace{-0.8cm}\\
  \end{center}
  \caption{{\sl Model-independent upper bounds on $\sigma\times\mathrm{BR}$ for
      (a) the  $\ee\ra\genHone\genHtwo\ra\bb\tautau$ channel and (b) the 
      $\ee\ra\genHone\genHtwo\ra\genHone\genHone\genHone\ra\bb\bb\bb$ channel.  
      The MSSM cross-section for $\genHone\genHtwo$ production with $\cos^2(\beta-\alpha)=1$
      is taken as normalisation. The dashed line indicates the kinematic limit for $\sqrts=206$~GeV.
      }}\label{fig:hh_bbtautau_limits}
\end{figure}

In Fig.~\ref{fig:hh_bbtautau_limits}~(a) $s_{95}$ for the process
$\ee\ra\genHone\genHtwo\ra\bb\tautau$ is shown. The branching ratios
are set to
$\mathrm{BR}(\genHone\ra\bb)=\mathrm{BR}(\genHtwo\ra\bb)=0.5$ and
$\mathrm{BR}(\genHone\ra\tautau)=\mathrm{BR}(\genHtwo\ra\tautau)=0.5$.
The kinematic limit for $\sqrts=206$~GeV is indicated as a dashed
line.  The domain below $\genmHone=30$~GeV is covered only by data
collected at $\sqrts=183$~GeV or lower.

Fig.~\ref{fig:hh_bbtautau_limits}~(b) shows the exclusion region for
the process
$\ee\ra\genHone\genHtwo\ra\genHone\genHone\genHone\ra\bb\bb\bb$.
$\mathrm{BR}(\genHtwo\ra\genHone\genHone)=1$ and
$\mathrm{BR}(\genHone\ra\bb)=1$ are assumed.  At $\genmHtwo<80$~GeV
the exclusion is stronger than for higher $\genmHtwo$ due to dedicated
searches at $\sqrts$ up to 189~GeV.  Above $\genmHtwo=80$~GeV, only
searches using data recorded with $\sqrts$ of 199 to 209~GeV are
available.

%
%

\section{Interpretation of the search results in the MSSM}\label{sect:benchmark}

The presence of neutral Higgs bosons is tested in a constrained MSSM
with seven parameters. Two of these parameters are sufficient to
describe the Higgs sector at tree level. A convenient choice is
$\tanb$ (the ratio of the vacuum expectation values of the Higgs
fields) and one Higgs mass; $\mA$ is chosen in the case of the CPC
scenario and $\mHp$ in the CPV scenario. Additional parameters appear
at the level of radiative corrections; these are: \msusy, $M_2$,
$\mu$, $A$, and \mg.  All soft SUSY-breaking parameters in the
sfermion sector are set to \msusy\ at the electroweak scale.  $M_2$ is
the SU(2) gaugino mass parameter at the electroweak scale and $M_1$,
the U(1) gaugino mass parameter, is derived from $M_2$ using the GUT
relation $M_1=M_2(5\sin^2\!\theta_W/3\cos^2\!\theta_W)$, where
$\theta_W$ is the weak mixing angle\footnote{$M_3$, $M_2$ and $M_1$
  are the mass parameters associated with the SU(3), SU(2) and U(1)
  subgroups of the Standard Model. $M_3$ enters only via loop
  corrections sensitive to the gluino mass.}.  The supersymmetric
Higgs mass parameter is denoted $\mu$. The parameter
$A=A_{\mathrm{t}}=A_{\mathrm{b}}$ is the common trilinear Higgs-squark
coupling for up-type and down-type squarks.  The stop and sbottom
mixing parameters are defined as
$X_{\mathrm{t}}=A_{\mathrm{t}}-\mu\cot\beta$ and
$X_{\mathrm{b}}=A_{\mathrm{b}}-\mu\tan\beta$.  The parameter $\mg$ is
the gluino mass.  For the CPV scenario the complex phases related to
$A_{\mathrm{t,b}}$ and $\mg$ are additional parameters. The phase
related to $A_{\mathrm{t,b}}$ enters at one-loop level while the one
related to $\mg$ enters as a second-order correction to stop and
sbottom loops.  Large radiative corrections to the predicted mass
$\genmHone$ arise from scalar top loops, while the contributions from
scalar bottom loops are smaller.

The precise mass of the top quark has a strong impact on $\genmHone$;
it is taken to be $m_{\mathrm{top}}=174.3$~GeV, the current average of
the Tevatron measurements~\cite{RPP2000}.  To account for the current
experimental uncertainty, all MSSM interpretations are also done for
$m_{\mathrm{top}}=169$~GeV and $m_{\mathrm{top}}=179$~GeV.

Rather than varying all of the above MSSM parameters independently, we
consider only a certain number of ``benchmark sets'' where the tree
level parameters $\tanb$ and $\mA$ (CPC scenario) or $\mHpm$ (CPV
scenario) are scanned while all other parameters are fixed.  Results
are presented for eight benchmark
sets~\cite{newbenchmarks,lhc_benchmarks} in the CPC scenario and nine
in the CPV scenario\cite{Carena:2000ks}.  Each scan point within a
given benchmark set defines an independent realization of the MSSM (a
model), which is tested by comparing its predicted observables
(masses, cross-sections and decay branching ratios) with the
experimental data. The parameters of the scans are summarized in
Table~\ref{tab:benchmark_sum}.

For a given scan point the observables in the Higgs sector are
calculated using two theoretical approaches. The FEYNHIGGS
program~\cite{feynhiggs,Heinemeyer:CFeyn} is based on a two-loop
diagrammatic approach~\cite{MSSMMHBOUND7,weigheiholl} and uses the OS
renormalization scheme, while SUBHPOLE and its CPV variant
CPH~\cite{Carena:2000ks} are based on a one-loop renormalization group
improved
calculation~\cite{MSSMMHBOUND5,carenamrennawagner,reconciliation,espinosareconciliation}
and uses the $\overline{\mathrm{MS}}$ scheme.  Both calculations give
consistent results although small differences naturally exist.
Numerical values for parameters in this paper are given in the
$\overline{\mathrm{MS}}$ scheme.

In the CPC case, the FEYNHIGGS calculation is retained for the
presentation of the results since it yields slightly more conservative
results (the theoretically allowed parameter space is wider) than
SUBHPOLE.  Also, FEYNHIGGS is preferred on theoretical grounds since
its radiative corrections are more detailed than those of SUBHPOLE.

In the CPV case, neither of the two existing calculations is preferred
a priori on theoretical grounds. While FEYNHIGGS contains more
advanced one-loop corrections, CPH is more precise at the two-loop
level.  We therefore opted for a solution where, in each scan point,
the calculation yielding the more conservative result (less
significant exclusion) is retained. For illustration, the results from
FEYNHIGGS and CPH are also shown separately for the main CPV scenario
CPX (see Section~\ref{sect:mssmcpx}).

The limits obtained for the different benchmark sets are summarized in
Table~\ref{tab:mssmmhmalimits}.

%
%

\subsection{CPC benchmark scenarios}\label{sect:cpcons}

Of the eight CPC benchmark sets examined in this paper, sets 1, 3 and
6 have been used in the past. Scenarios 4 and 5 are motivated by
experimental constraints on the branching ratio of the inclusive decay
b$\ra$s$\gamma$ and recent measurements of the muon anomalous magnetic
moment $(g-2)_{\mu}$.  Benchmark sets 2, 7 and 8 are motivated by the
fact that the planned Higgs searches at the LHC may have low
sensitivity to detect Higgs bosons in these situations.  The choice of
parameters is summarized in Table~\ref{tab:mssmmhmalimits}.

In most cases, $\tanb$ is scanned between 0.4 and 40. For values below
0.4 the theoretical predictions become unreliable; for $\tanb$ larger
than 40 the decay width of the Higgs bosons may become comparable to
or larger than the experimental mass resolution, and the modelling of
the signal efficiencies may loose precision. The value of $\mA$ is
scanned between 0 and 1000~GeV. For values of $\mA<2$~GeV, the
branching ratios of the A become dominated by resonances and their
calculation is unstable.  However this area can be probed using direct
searches for the heavier $\ho$ boson, decay independent searches and
$\Gamma_{\mathrm{Z}}$ constraints.

In general there is good agreement between the data and the background
estimation, therefore limits on the MSSM parameters can be derived. In
the CPC scenarios, the largest observed excess of the data over the
background appears at $\mh=95$~GeV in the \mhmax{} benchmark set. At
this point the excess is $(1-\mathrm{CL}_b)=2.6\times10^{-3}$,
corresponding to a significance of $2.8\,\sigma$. It should be noted,
however, that there is a large statistical probability of such an
excess to appear somewhere in the parameter space under study.

\begin{figure}[p]
  \qquad\hspace{-1.9cm}{\centering{
      \begin{minipage}{1.15\textwidth}
        \epsfig{file=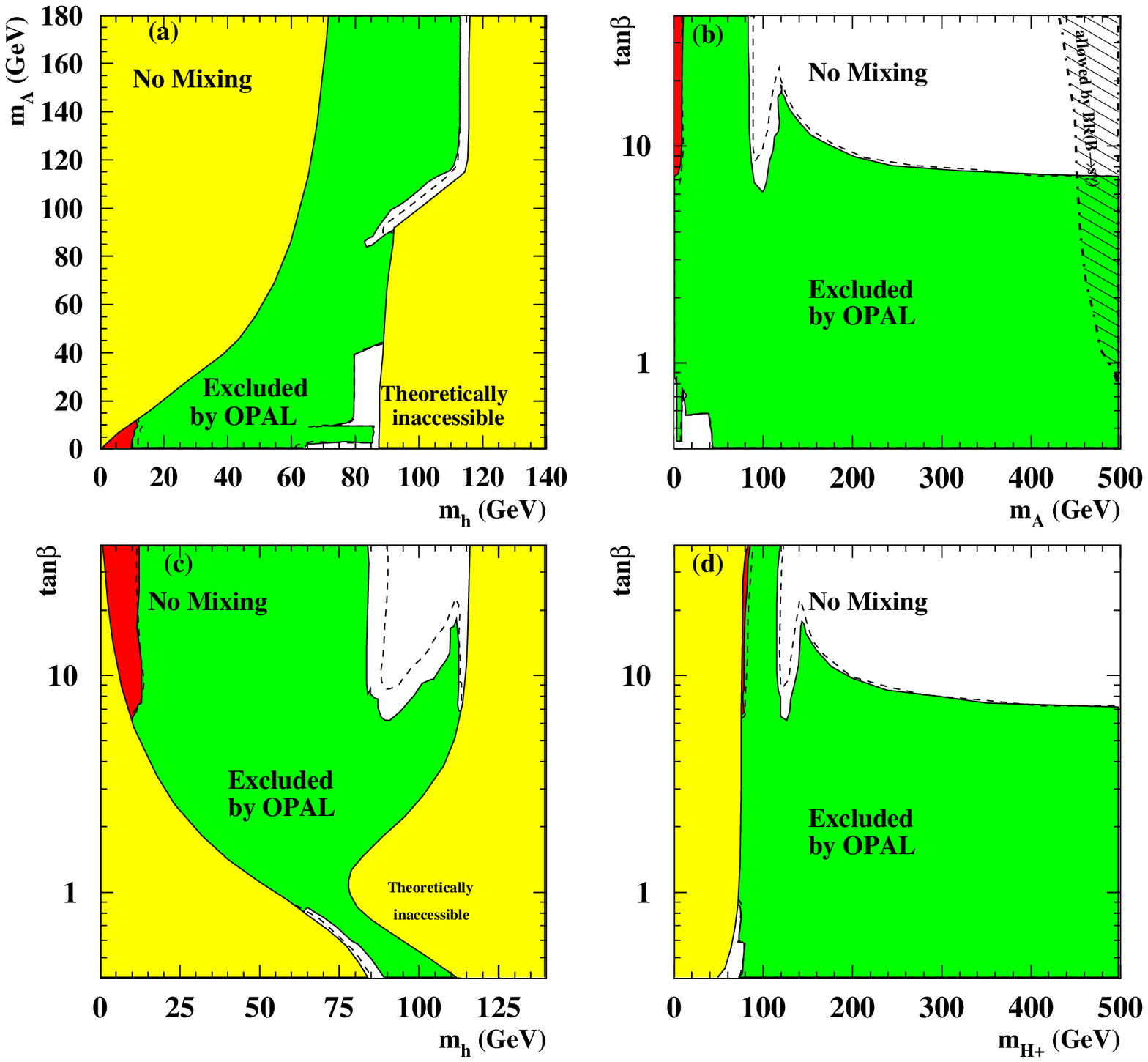, width=\textwidth}
      \end{minipage}}}
  \caption[]{\label{fig:nomix}
    \sl Results for the no mixing benchmark scenario.
    The figure shows the excluded regions in darker grey (green)
    and theoretically inaccessible regions in light grey (yellow) as functions of the MSSM
    parameters in four projections:
    (a) the (\mh,~\mA) plane,
    (b) the (\mA,~\tanb) plane,
    (c) the (\mh,~\tanb) plane and
    (d) the (\mHpm,~\tanb) plane.
    The dashed lines indicate the boundaries of the
    regions expected to be excluded at the 95\% CL if only SM background
    processes are present. The region excluded by Yukawa searches, $\Zo$-width constraints or 
    decay independent searches is shown in dark grey (red). In (b) the hatched area is still allowed by $\mathrm{BR(b\ra s\gamma)}$ searches.
    }
\end{figure}

\begin{figure}[p]
  \qquad\hspace{-1.9cm}{\centering{
      \begin{minipage}{1.15\textwidth}
        \epsfig{file=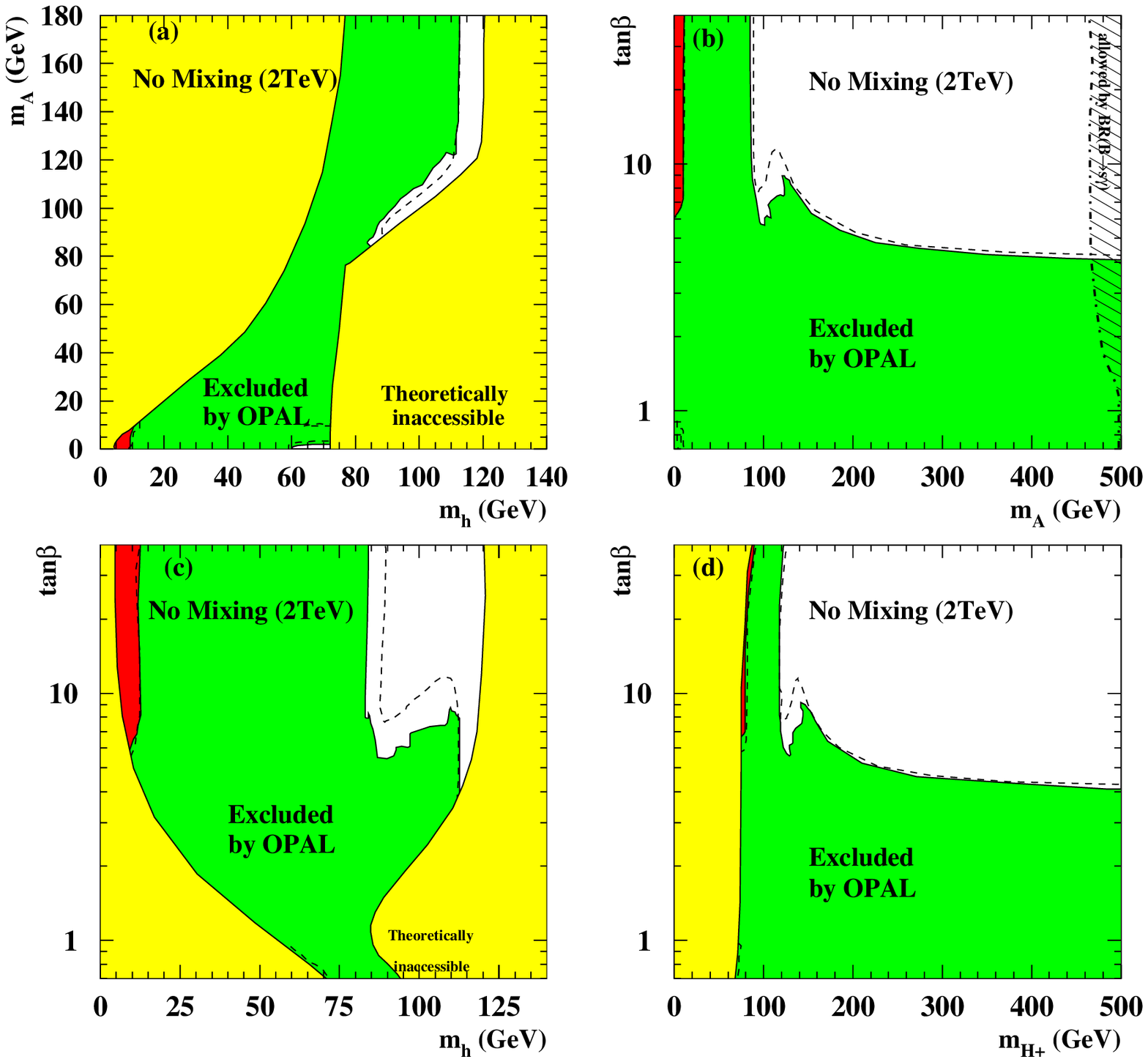, width=\textwidth}
      \end{minipage}}}
  \caption[]{\label{fig:nomixing2tev}
    \sl Results for the no mixing (2 TeV) benchmark scenario
    described in the text
    of Section~\ref{sect:cpcons}.
    See Fig.~\ref{fig:nomix} for the notation.
    }
\end{figure}

\begin{figure}[p]
  \qquad\hspace{-1.9cm}{\centering{
      \begin{minipage}{1.15\textwidth}
        \epsfig{file=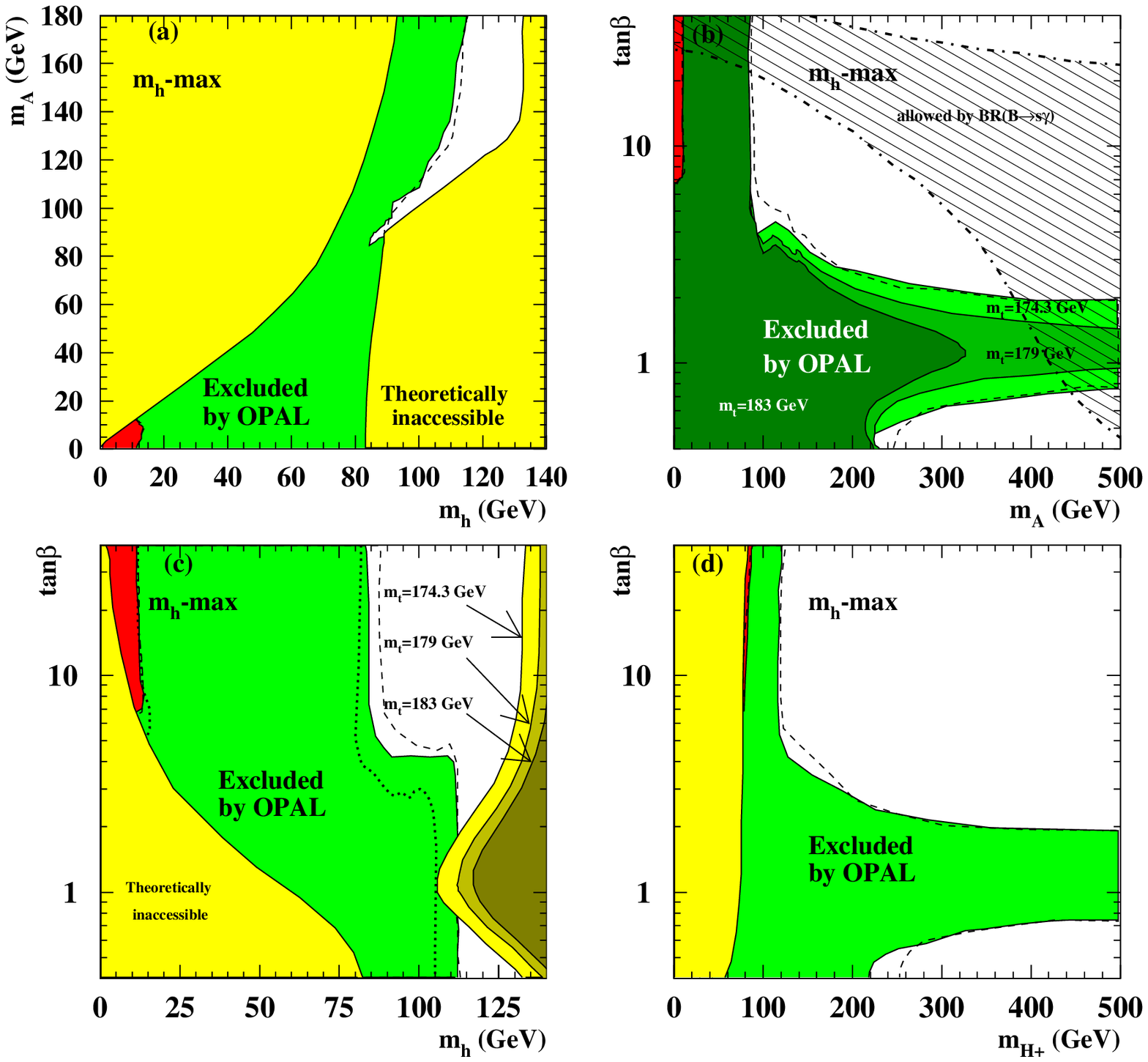, width=\textwidth}
      \end{minipage}}}
  \caption[]{\label{fig:maxmh}
    \sl Results for the \mhmax\ benchmark scenario.  See
    Fig.~\ref{fig:nomix} for the notation. The dotted line in (c)
    shows the observed 99.9\% confidence limit. The differently shaded
    regions in (b) show the exclusion for different values of
    $m_{\mathrm{top}}$, as written in the Plot. The upper limit on
    $\mh$ for different values of $m_{\mathrm{top}}$ is shown in (c),
    as expressed in the plot.}
\end{figure}

\begin{figure}[p]
  \qquad\hspace{-1.9cm}{\centering{
      \begin{minipage}{1.15\textwidth}
        \epsfig{file=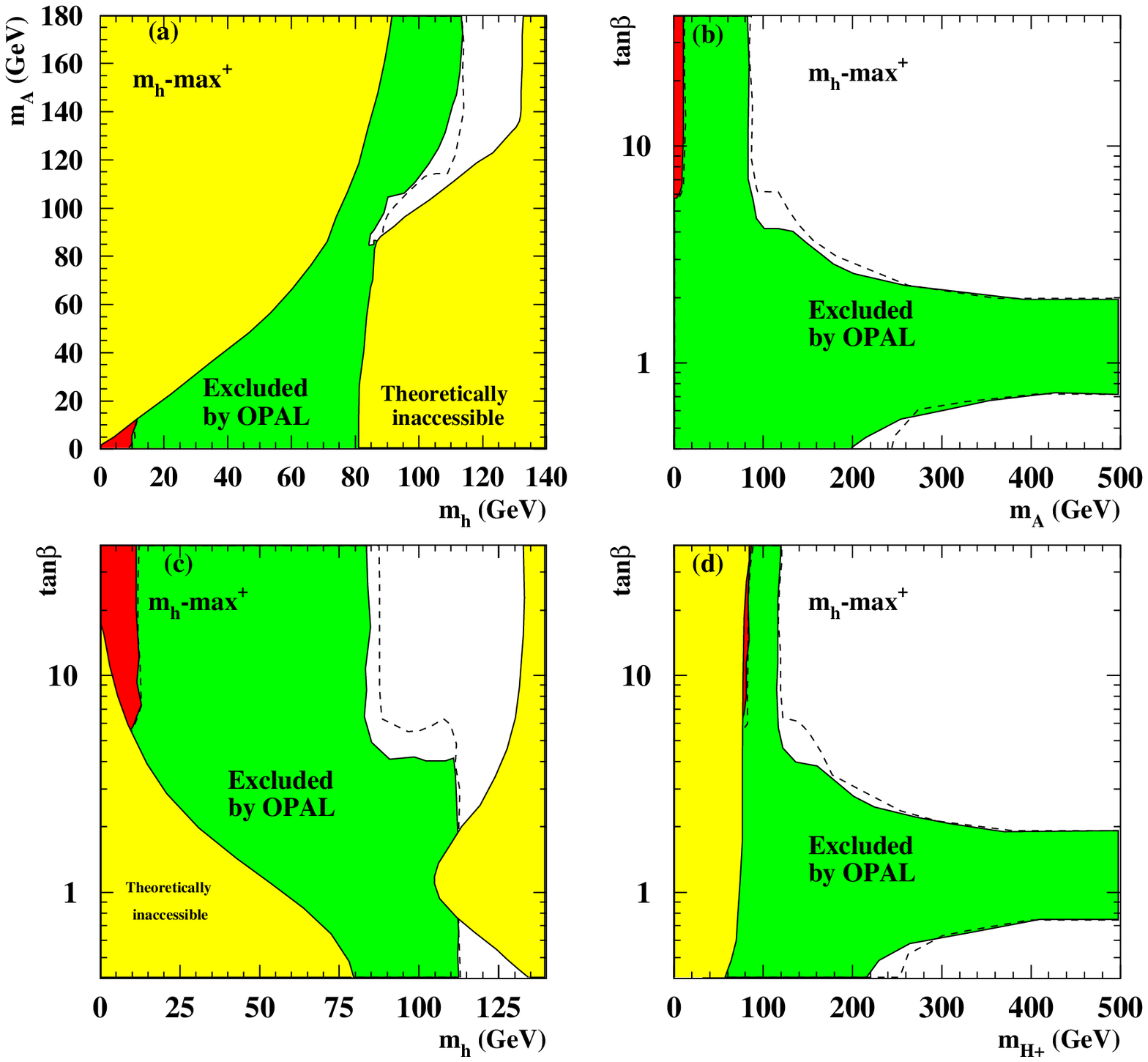, width=\textwidth}
      \end{minipage}}}
  \caption[]{\label{fig:mhmaxplus}
    \sl Results for the $\mhmax^+$ benchmark scenario described in the
    text of Section~\ref{sect:cpcons}.  See Fig.~\ref{fig:nomix} for
    the notation. This scenario is excluded by the BR(b$\ra$s$\gamma$)
    constraint for $\mA<600$~GeV.}
\end{figure}

\begin{figure}[p]
  \qquad\hspace{-1.9cm}{\centering{
      \begin{minipage}{1.15\textwidth}
        \epsfig{file=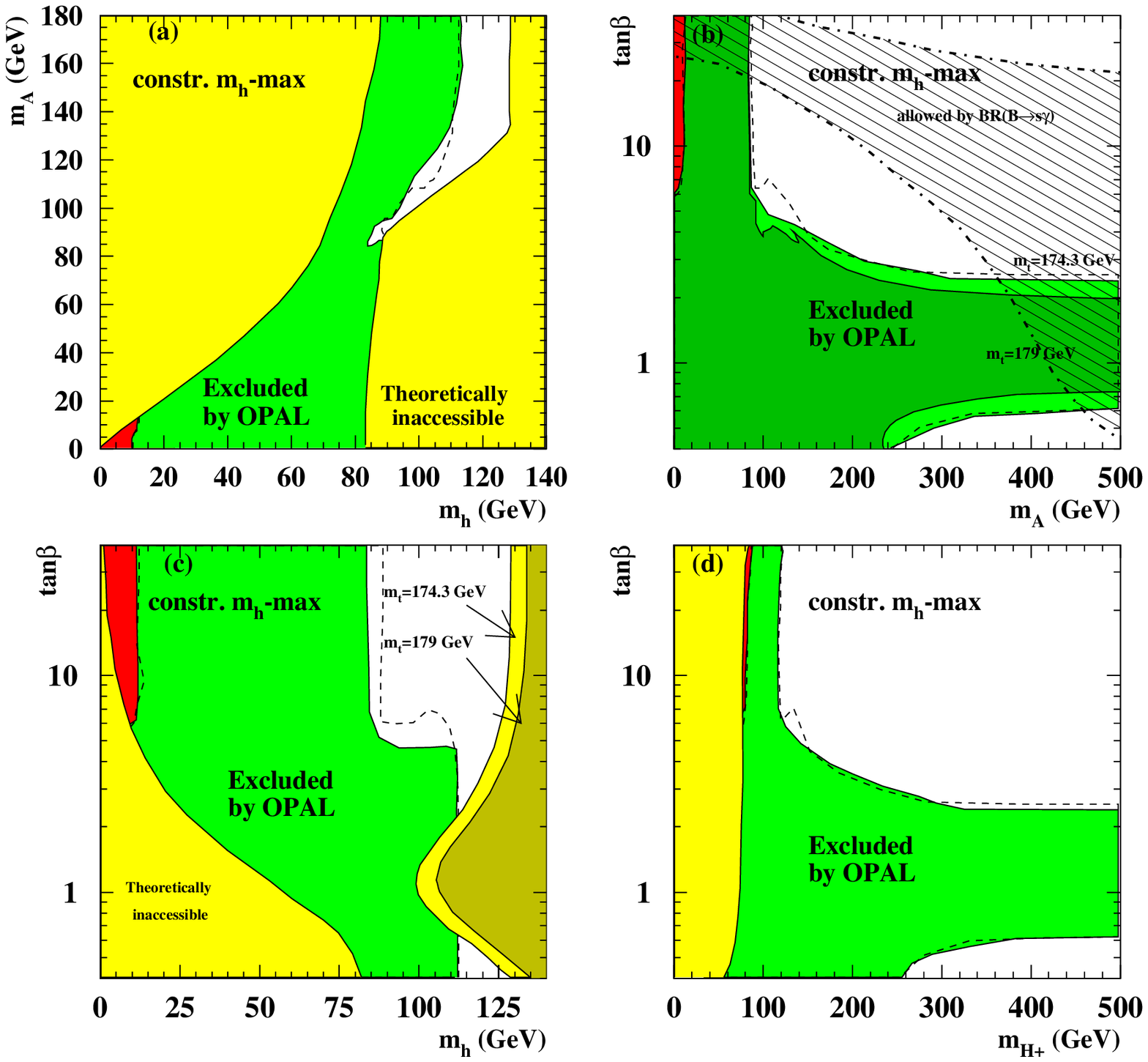, width=\textwidth}
      \end{minipage}}}
  \caption[]{\label{fig:constrmhmax}
    \sl Results for the constrained \mhmax\ benchmark scenario
    described in the text
    of Section~\ref{sect:cpcons}.
    See Fig.~\ref{fig:nomix} for the notation. The differently shaded
    regions in (b) show the exclusion for different values of
    $m_{\mathrm{top}}$, as written in the Plot. The upper limit on
    $\mh$ for different values of $m_{\mathrm{top}}$ is shown in (c),
    as also expressed in the plot.
    }
\end{figure}

\begin{enumerate}
  
\item In the no mixing benchmark set the stop mixing parameter
  $X_{\mathrm{t}}$ is put at zero.  The other parameters are fixed at
  the following values: $\msusy = 1$~TeV, $M_2 = 200$~GeV,
  $\mu=-200$~GeV.  The gluino mass \mg\ has little effect on the
  phenomenology of this scenario; its value is set to 800~\gevcs.
  
  The corresponding exclusion plots are shown in Fig.~\ref{fig:nomix}.
  The unexcluded region with $64<\mh<88$~GeV and $\mA<43$~GeV is due
  to the dominance of the cascade decay $\ho\ra\Ao\Ao$ for which the
  search sensitivity is lower than for the $\ho\ra\bb$ and $\tautau$
  channels.  One should note, however, that in this domain the charged
  Higgs boson mass $\mHpm$ is predicted to smaller than $81$~GeV. This
  area is probed by charged Higgs boson
  searches~\cite{OPALCHARGEDHiggs}, which will be further extended in
  the future (see Fig.~\ref{fig:nomix}~(d)).
  
  The region with $\mh>83$~GeV and $\mA>82$~GeV is still unexcluded.
  In this domain, either the cross-section for Higgsstrahlung
  $\ee\ra\Zh$ is small ($\sin^2(\beta-\alpha)$ is close to 0, see
  Eq.~\ref{equation:xsec_zh} in Section~\ref{sect:channels}) or the
  pair production process $\ee\ra\Ah$ is kinematically forbidden.
  
  Values of $\tanb$ are excluded from 0.8 to 6.2.  However, the
  $\tanb$ limit is strongly dependent on the top quark mass which was
  taken to be $m_{\mathrm{top}}=174.3$~GeV. For
  $m_{\mathrm{top}}=179$~GeV, the $\tanb$ exclusion is reduced to
  $0.8<\tanb<4.7$. If one disregards the unexcluded domain at low
  $\tanb$, the following lower bounds are obtained at the 95\%
  confidence level: $\mh>83$~GeV and $\mA>82$~GeV.
  
  The constraint from the measured value of BR(b$\ra$s$\gamma$) (see
  Section~\ref{sect:add_exp_constr}) is indicated in
  Fig.~\ref{fig:nomix}~(b).

\item The no mixing (2 TeV) benchmark set differs from the no mixing
  scenario in the flipped sign of $\mu$ (which is preferred by the
  current results on $(g-2)_{\mu}$) and by a larger SUSY mass scale
  $M_{SUSY} = 2\;\rm TeV$.  The value of $\tanb$ is scanned only from
  0.7 to 40 due to numerical instabilities in the diagonalisation of
  the mass matrix for very low $\tanb$. Therefore the largest part of
  the unexcluded region of the no mixing case at low $\tanb$ is not
  probed in this scenario.
  
  The corresponding exclusion region is shown in
  Fig.~\ref{fig:nomixing2tev}. For $\mA>2$~GeV, i.e. above the region
  of resonant Higgs boson decays, absolute limits can be set for the
  Higgs boson masses and on $\tanb$, which are $\mh>83.3$~GeV,
  $\mA>84.3$~GeV and $\tanb>4.2$. If the unexcluded area at
  $\mA<2$~GeV is also regarded, the exclusion in $\tanb$ is
  $0.9<\tanb<4.2$. The reduced $\tanb$ exclusion with respect to the
  no mixing case reflects the increased value of $M_{SUSY}$.  This
  limit is further weakened to $\tanb>3.2$ for
  $m_{\mathrm{top}}=179$~GeV.
  
  The measurements of BR(b$\ra$s$\gamma$) exclude the no mixing (2
  TeV) scenario for $\mA<450$~GeV, as can be seen in
  Fig.~\ref{fig:nomixing2tev}~(b).
  
\item The \mhmax\ benchmark set is designed to yield the largest range
  of \mh\ for a given ($\mA,\tanb$).  This scenario is therefore the
  most conservative in terms of exclusion in $\tanb$.  The other
  parameters are fixed as in the no mixing scenario, with the
  exception of the stop mixing parameter $X_t = \sqrt{6}$~TeV. 
  
  The exclusion plots for this benchmark set are shown in
  Fig.~\ref{fig:maxmh}.  The following absolute limits are obtained at
  the 95\% confidence level: $\mh>84.5$~GeV and $\mA>85.0$~GeV.
  Furthermore, values of $\tanb$ between 0.7 and 1.9 are excluded.
  For $m_{\mathrm{top}}=179$~GeV this exclusion shrinks to the domain
  $1.0<\tanb<1.3$. Since the $\mhmax$ benchmark set yields the most
  conservative exclusion in $\tanb$, also $m_{\mathrm{top}}=183$~GeV
  as the anticipated $1\,\sigma$ upper bound of an increased world
  average of $m_{\mathrm{top}}=179$~GeV was tested. This is
  illustrated in Fig~\ref{fig:maxmh}~(b) and in
  Fig~\ref{fig:maxmh}~(c), where the exclusion in the ($\tanb,\mA$)
  plane respectively the theoretical upper bounds on $\mh$ for
  $m_{\mathrm{top}}=179$~GeV and $m_{\mathrm{top}}=183$~GeV are also
  shown.  Should the world average of the top quark mass move beyond
  179.5~GeV, the exclusion in $\tanb$ would vanish completely.
  
  The supplementary constraint from the measured value of the
  BR(b$\ra$s$\gamma$) is shown in Fig.~\ref{fig:maxmh}~(b).
  
\item The \mhmax$^{+}$ benchmark set differs from the \mhmax{} case
  only by the flipped sign of $\mu$.  This choice is favored by the
  presently available results on $(g-2)_{\mu}$~\cite{bib:g-2results}.
  Since the Higgs boson properties depend only weakly on the sign of
  $\mu$, the accessible Higgs mass range as well as the excluded
  domains are very similar to those of the \mhmax{} scenario; they are
  shown in Fig.~\ref{fig:mhmaxplus}.
  
  The limits on the Higgs masses are $\mh>84.5$~GeV and
  $\mA>84.0$~GeV. The excluded range in $\tanb$ is $0.7<\tanb<1.9$,
  which decreases to $0.96<\tanb<1.4$ for $m_{\mathrm{top}}=179$~GeV.
  
  The \mhmax$^{+}$ scenario is excluded for $\mA<600$~GeV by
  BR(b$\ra$s$\gamma$) measurements for all values of $\tanb$
  considered (between 0.4 and 40).  This means that only the
  decoupling limit with $\mh$ at its maximum value for a given $\tanb$
  is still allowed by BR(b$\ra$s$\gamma$).

\item The constrained $m_h$-max benchmark set differs from the
  \mhmax$^{+}$ set by the flipped sign of $X_{\mathrm{t}}$, which
  yields better agreement with
  $\mathrm{BR}(\mathrm{b}\rightarrow\mathrm{s}\gamma)$ constraints.
  One observes that the maximum value of the Higgs boson mass at a
  given $\tan \beta$ is lowered by about $5\;\rm GeV$.
  
  The excluded areas for this scenario (see
  Fig.~\ref{fig:constrmhmax}) show similar features as the $\mhmax$
  and $\mhmax^+$ scenarios. The limits on the Higgs masses are
  $\mh>84.0$~GeV and $\mA>85.0$~GeV. The excluded range in $\tanb$ is
  $0.6<\tanb<2.2$, which shrinks to $0.8<\tanb<1.8$ for
  $m_{\mathrm{top}}=179$~GeV. This is illustrated in
  Fig.~\ref{fig:constrmhmax})~(b) and (c).
  
  The supplementary constraint from the measured value of the
  b$\ra$s$\gamma$ branching ratio is shown in
  Fig.~\ref{fig:constrmhmax}~(b) as the band delimited by the two
  dash-dotted lines.
  
\item The large $\mu$ benchmark set is designed to illustrate choices
  of parameters for which the detection of the Higgs bosons is
  believed to be a priori difficult at LEP. The parameters are set to
  the following values: \msusy=$400$~\gevcs, $\mu=1$~\tevcs,
  $M_2=400$~\gevcs, $\mg=200$~\gevcs, $X_t = -300$~\gevcs. It is
  scanned from $\tanb=0.7-40$ and $\mA=0-400$.
 
  For this set of parameters, the $\ho$ boson is always kinematically
  accessible ($\mh < 108$~GeV) but its decay to $\bb$, on which most
  of the searches are based, is suppressed.  For many of the scan
  points the decay $\ho\ra\tautau$ is also suppressed.  The dominant
  decay modes are thus $\ho\ra\cc,\glgl$ or $\WW$, and the detection
  of Higgs bosons has to rely more heavily on flavour-independent
  searches.
    
  In some of the scan points the Higgsstrahlung process $\ee\ra\Zh$ is
  suppressed all-together ($\sin^2(\beta-\alpha)$ small). However, the
  heavy neutral scalar is relatively light in such cases ($\mH <
  109$~GeV) and the cross-section for the process $\ee\ra\ZH$, being
  proportional to $\cos^2(\beta-\alpha)$, is large.
    
  The exclusions for this benchmark scenario are given in
  Fig.~\ref{fig:largemu}.  They show that the parameter space is
  essentially excluded even in this difficult scenario, with the
  exception of a few isolated ``islands''.  Those may slightly
  increase for higher values of the top quark mass.  The origin of the
  islands can best be explained using Fig.~\ref{fig:largemu}~(b).  The
  large diagonal island at $\mA>100$~GeV is due to the fact that
  BR($\ho\ra\bb$) goes to 0 there. The two thin vertical islands
  around $\mA>100$~GeV are due to an overlap between $\ee\ra\ho\Zo$
  and $\ee\ra\Ho\Zo$ production. Both are kinematically accessible,
  but either one or the other can be used in the interpretation.
  
  The supplementary constraint from the measured value of the
  b$\ra$s$\gamma$ branching ratio is shown in
  Fig.~\ref{fig:largemu}~(b).

\item The gluophobic benchmark set is constructed such that the Higgs
  coupling to gluons is suppressed due to a cancellation between the
  top and the stop loops at the $\ho\glgl$ vertex. Since at the LHC
  the searches will rely heavily on the production of the Higgs boson
  by gluon-gluon fusion, such a scenario may be difficult to
  investigate there.  The parameters chosen are : \msusy = 350~\gevcs,
  $M_2 = 300$~\gevcs, $\mu=300$~\gevcs, $X_t =-750$~\gevcs, $0.4<\tanb
  < 40$, 0~\gevcs~$<\mA < 1$~\tevcs\ and \mg=500~\gevcs.
  
  The exclusion for this benchmark set is shown in
  Fig.~\ref{fig:gluophobic}. It is excluded to a large extend.  The
  limits on the Higgs masses are $\mh>82$~GeV and $\mA>87.5$~GeV. The
  excluded range in $\tanb$ is $\tanb<6.0$. The excluded range is
  reduced to $\tanb<3.5$ for $m_{\mathrm{top}}=179$~GeV.
  
  The supplementary constraint from the measured value of the
  b$\ra$s$\gamma$ branching ratio is shown in
  Fig.~\ref{fig:gluophobic}~(b).
  
\item In the small $\alpha_{\mathrm{eff}}$ benchmark set the Higgs
  boson decay channels $\ho\rightarrow\bb$ and $\ho\rightarrow
  \tautau$ are suppressed with respect to their Standard Model
  coupling by the additional factor $-\sin \alpha_{\mathrm{eff}}/ \cos
  \beta$, stemming from corrections from $\tilde{\mathrm{b}} -
  \tilde{\mathrm{g}}$ loops.  This scenario may also be difficult to
  investigate by the LHC experiments.  Similarly to the large-$\mu$
  scenario, such suppressions occur for large $\tan \beta$ and not too
  large $m_A$. The parameters chosen are: $\msusy = 800$~\gevcs, $M_2
  = 500$~\gevcs, $\mu=2$~\tevcs, $X_t =-1100$~\gevcs, $0.4<\tanb <
  40$, and \mg\,=500~\gevcs.
  
  The exclusion for this benchmark set is shown in
  Fig.~\ref{fig:smallalphaeff}.  The limits on the Higgs masses are
  $\mh>79.0$~GeV and $\mA>90.0$~GeV. The excluded range in $\tanb$ is
  $0.4<\tanb<3.6$, which is reduced to $0.5<\tanb<2.9$ for
  $m_{\mathrm{top}}=179$~GeV.  It appears that effects of suppression
  of the decays $\ho\ra\bb$ and $\ho\ra\tautau$ do not play a role in
  the region kinematically accessible at LEP.
  
  The constraint from the measured value of BR(b$\ra$s$\gamma$) is shown in
  Fig.~\ref{fig:smallalphaeff}~(b).

\end{enumerate}

%
%

\subsection{CPV benchmark scenarios}
\label{sect:mssmcpx}

In the MSSM, the Higgs potential is invariant under CP transformations
at tree level.  However it is possible to explicitly or spontaneously
break CP symmetry by radiative corrections~\cite{Carena:2000yi}.  In
particular, the phases of $A_{\mathrm{t,b}}$ and
$m_{\tilde{\mathrm{g}}}$ introduce CP violation into the Higgs
potential via loop effects, leading to sizeable off-diagonal
contributions to the Higgs boson mass matrix (see
Section~\ref{sect:channels}).

\begin{figure}[ht]
  \begin{center}
    \epsfig{file=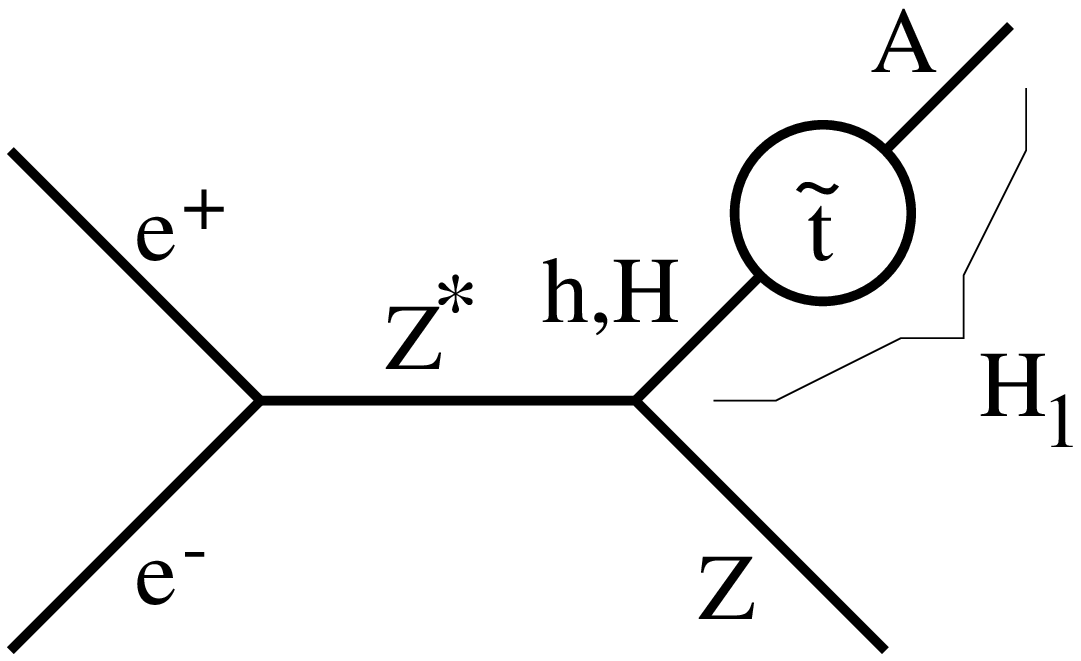, width=0.4\textwidth}
  \end{center}
  \caption{\sl{Diagram illustrating the effective coupling of a Higgs mass eigenstate 
      to the $\Zo$ in Higgsstrahlung. The complete mass eigenstate
      $\genHone$ is composed of admixtures of $\mathrm{h},\mathrm{H}$ and
      $\mathrm{A}$. Here the $\mathrm{h},\mathrm{H}$ and $\mathrm{A}$ denote the CP-even and CP-odd weak eigenstates, respectively. 
      Only the CP-even admixtures $\mathrm{h}$ and
      $\mathrm{H}$ couple to the $\Zo$, while the CP-odd $\mathrm{A}$
      does not. Therefore the coupling of the mass eigenstate is
      reduced with respect to a CPC scenario.
      }}\label{fig:ha-mixing}
\end{figure}

As a consequence the Higgs mass eigenstates $\Hone,\Htwo$ and
$\Hthree$ are not the CP eigenstates. This influences predominantly
the couplings in the Higgs sector.  Fig.~\ref{fig:ha-mixing} shows the
coupling of a mixed mass eigenstate $\Hone$ consisting of admixtures
from the CP eigenstates $\mathrm{h},\mathrm{H}$ and $\mathrm{A}$.
Since only the CP-even field component couples to the $\Zo$ boson, the
individual coupling of the mass eigenstates are reduced in the CPV
case with respect to a CPC case.

%
%

The size of the CPV off-diagonal elements, ${\cal M}^2_{ij}$, of the
Higgs boson mass matrix and hence the size of CPV effect scales
qualitatively \cite{Carena:2000ks} as
\begin{equation}
  {\cal M}^2_{ij} \propto \frac{m^4_{\mathrm{top}}}{v^2}\frac{\mathrm{Im}(\mu A_t)}
  {32\pi^2m^2_{\mathrm{SUSY}}}.\label{for:offdiag}
\end{equation}
Large CPV effects, and thus scenarios dissimilar from the CPC case,
are therefore obtained if the SUSY breaking scale $m_{\mathrm{SUSY}}$
is small and the imaginary contribution to $\mu A_t$ large. Also large
values of $m_{\mathrm{top}}$ increase the CPV effects.

When choosing the parameters, experimental constraints
\cite{Commins:gv,Harris:jx} from electric dipole moment (EDM)
measurements of the neutron and the electron have to be fulfilled.
However, cancellations among different contributions to the EDM may
naturally emerge \cite{Carena:2000yi}; hence those measurements
provide no universal exclusion in the MSSM parameter space, while
direct searches at LEP provide a good testing ground for a CPV MSSM.

The basic CPV MSSM benchmark set is CPX. Its parameters are chosen
such as to approximately fulfill the EDM constraints and to provide
features that are the most dissimilar from a CPC scenario.  The choice
of parameters~\cite{Carena:2000ks} is given in
Table~\ref{tab:benchmark_sum} (last column).  In the definition of the
CPX scenario~\cite{Carena:2000ks} the relations
$\mu=4m_{\mathrm{SUSY}}$ and
$|A_{\mathrm{t,b}}|=|m_{\tilde\mathrm{g}}|=2m_{\mathrm{SUSY}}$ are
fixed.  Here, $m_{\mathrm{SUSY}}=500$~GeV is chosen. The parameter
$m_2$ is set to 200~GeV.  Additionally the complex phases of
$A_{\mathrm{t,b}}$ and $m_{\tilde{g}}$ are fixed at $90^{\circ}$
degrees.  Variants of the CPX scenario are investigated to check the
stability of the CPX results with respect to the choice of its
parameters. The phases of $A_{\mathrm{t,b}}$ and $m_{\tilde{g}}$
varied from from $0^{\circ}$ to $180^{\circ}$, $\mu$ in between 500
and 4000~GeV.  The scenario with $\arg(A_{\mathrm{t,b}}) = 90^{\circ}$
has very different features from a CPC case and therefore has good
properties for a CPV benchmark scenario.

The benchmark scan databases, containing masses, cross-sections and
branching ratios for all three neutral Higgs bosons for a variety of
different input parameters, are generated using both
CPH~\cite{Carena:2000ks}, a modified version of SUBHPOLE, and
FEYNHIGGS~2.0~\cite{Heinemeyer:CFeyn}. They are implemented in a
modified version of HZHA~\cite{Janot:1996hzha}.  Initial-state
radiation and interference between Higgsstrahlung and boson fusion
processes are taken into account by HZHA. The parameter $\tanb$ is
scanned from $0.6$ to $40$, and $\mHpm$ is scanned from 4 to $1000$.
In this region both $\Hone$ and $\Htwo$ have a width below 1~GeV,
negligible with respect to the experimental resolution of several GeV.

Also in the CPV scenarios there is good agreement between the data and
the background estimation.  The largest observed excess of the data
over the background appears at $\mHone=40$~GeV and $\mHtwo=105$~GeV in
the CPX benchmark set. The excess has
$(1-\mathrm{CL}_b)=1.2\times10^{-3}$, corresponding to a significance
of $3.0\,\sigma$. Also here it should be noted that there is a large
statistical probability of such an excess to appear somewhere in the
parameter space under study. Limits on the MSSM parameter space are
derived.

\begin{figure}[pt]
  \quad\\[-2cm]
  \epsfig{file=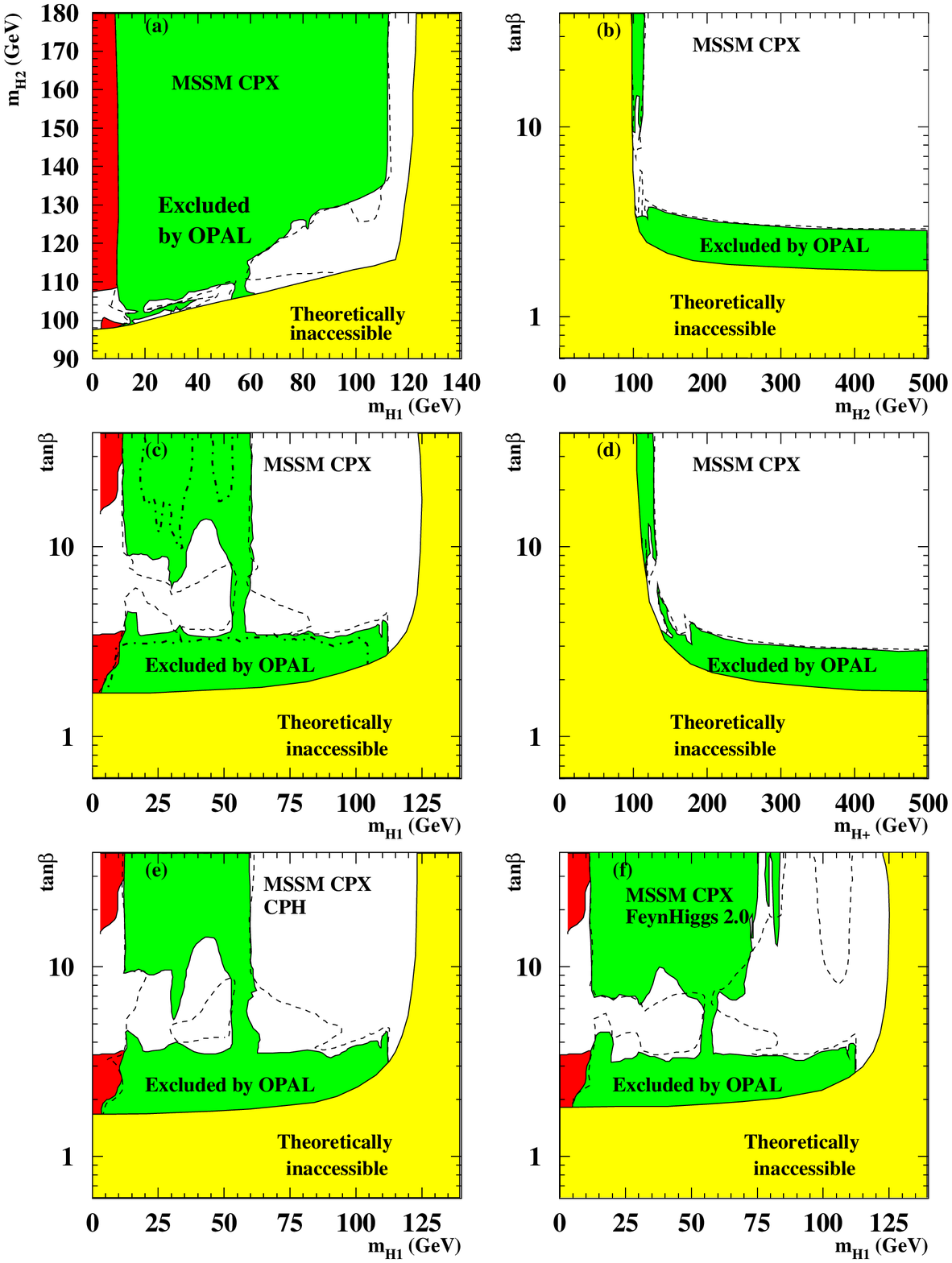, width=\textwidth}
  \caption{\sl{The CPX MSSM $95\,\%\,\mathrm{CL}$ exclusion areas. Excluded regions are
      shown for (a) the ($m_{\Hone}$,$m_{\Htwo}$) plane, (b) the ($m_{\Htwo}$,$\tanb$) plane,
      (c) the ($m_{\Hone}$,$\tanb$) plane and (d) the ($m_{\mathrm{H}^{\pm}}$,$\tanb$) plane. 
      Figure (e) shows the ($m_{\Hone}$,$\tanb$) of the CPH calculation alone, (f) shows 
      the same projection of the FEYNHIGGS~2.0 calculation.
      See Fig.~\ref{fig:nomix} for the notation. The dash-dotted line in (c) 
      shows the area excluded on the 99.9\% confidence level.
      In (b) and (d) the area excluded by $\Zo$ width constraints or by decay 
      independent searches is too small to be displayed.}}\label{fig:CPX90_excl}
\end{figure}

Fig.~\ref{fig:CPX90_excl} shows the combined exclusion result for the
CPX scenario with all phases equal to $90^{\circ}$,
$m_{\mathrm{SUSY}}=500$~GeV and $\mu=2$~TeV.
Fig.~\ref{fig:CPX90_excl}~(a) shows both the expected and observed
95\% CL exclusion areas in the plane of $m_{\Hone}$ and $m_{\Htwo}$.
For heavy $m_{\Htwo}$, $\Hone$ resembles the SM Higgs boson (almost
completely CP-even) with very little effect from CP violation.  The
limit on the allowed mass of $\Hone$ for large $m_{\Htwo}$ is found to
be $m_{\Hone}>112$~GeV.  In the region below $m_{\Htwo}\approx
130$~GeV CPV effects play a major role.
 
Fig.~\ref{fig:CPX90_excl}~(b) shows the 95\% CL exclusion areas in the
parameter space of $\tanb$ and $m_{\Htwo}$. One can see that
$\tanb<2.8$ is excluded. The band at $\tanb<2.8$ is excluded by
searches for the SM-like $\Hone$, while the band at $\tanb>10$ and
$m_{\Htwo}<120$~GeV is excluded by searches for $\Zo\Htwo$ and
$\Hone\Htwo$ topologies.

Fig.~\ref{fig:CPX90_excl}~(c) displays the parameter space of $\tanb$
and $m_{\Hone}$. The range $\tanb<2.8$ is excluded, and a lower limit
of $\tanb>3.2$ exists if $m_{\Hone}$ is below 112~GeV.  For
$4<\tanb<10$, $\Zo\Htwo$ production is dominant.  The large difference
between the expected and observed exclusion regions in the area of
$4<\tanb<10$ is mainly due to a less than $2\sigma$ excess in the data
between $m_{\mathrm{h}}\approx 95$~GeV and $m_{\mathrm{h}}\approx
110$~GeV~\cite{OPALSMPAPER}, which corresponds to the mass of $\Htwo$
in this region. For $m_{\Hone}<50$~GeV there are also unexcluded
regions in the expected exclusion, which is due to dominant
$\Zo\Htwo\ra\Zo\Hone\Hone$ production with relatively large
$m_{\Hone}$, yielding broad mass resolutions and therefore reduced
sensitivity.

In Fig.~\ref{fig:CPX90_excl}~(d) the exclusion area is shown in the
parameter space of the theoretical input parameters $\tanb$ and
$m_{\mathrm{H}^{\pm}}$, which are varied during the scan.  Since the
CPX scenario yields $m_{\Htwo}\approx m_{\mathrm{H}^{\pm}}$ for most
of the scan points, this is very similar to
Fig.~\ref{fig:CPX90_excl}~(b).

The uncertainty inherent to the two theoretical approaches, CPH and
FEYNHIGGS, is illustrated in parts (e) and (f) of
Fig.~\ref{fig:CPX90_excl}.  The largest discrepancy occurs for large
values of $\tanb$, where the FEYNHIGGS calculation (part (f)) predicts
a higher cross-section for Higgsstrahlung, and hence a better search
sensitivity than the CPH prediction (part(e)).

\begin{figure}[pt]
  \qquad\hspace{-1.9cm}{\centering{
      \begin{minipage}{1.15\textwidth}
        \epsfig{file=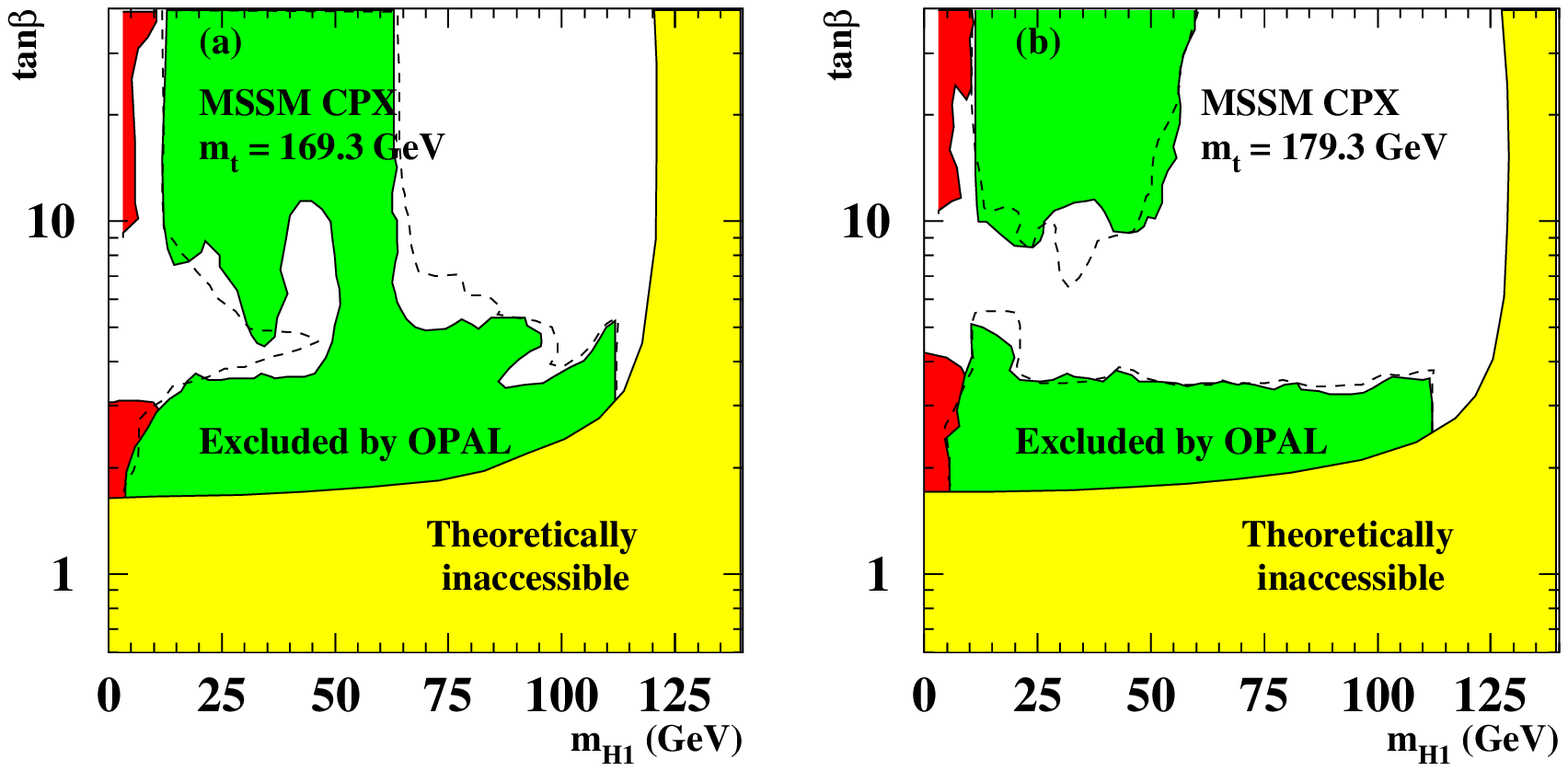, width=\textwidth}
      \end{minipage}}}
  \caption{\sl{The CPX MSSM $95\,\%\,\mathrm{CL}$ exclusion areas in the 
      ($m_{\Hone}$,$\tanb$) plane, using scans with (a) $m_{\mathrm{t}}=179.3$~$\mathrm{GeV}$
      and (b) $m_{\mathrm{t}}=169.3$~$\mathrm{GeV}$. Due to the change in the top
      masses a strong difference is observed compared to Fig.~\ref{fig:CPX90_excl}~(c). 
      See Fig.~\ref{fig:nomix} for the notation.
      }}\label{fig:CPX90syst}
\end{figure}

The large impact of the value of the top quark mass on the exclusion
limits is shown in Fig.~\ref{fig:CPX90syst}. For
$m_{\mathrm{top}}=179.3$~GeV, the excluded range in $\tanb$ shrinks to
$\tanb<2.4$.

\begin{figure}[pt]
  \qquad\hspace{-1.9cm}{\centering{
      \begin{minipage}{1.15\textwidth}
        \epsfig{file=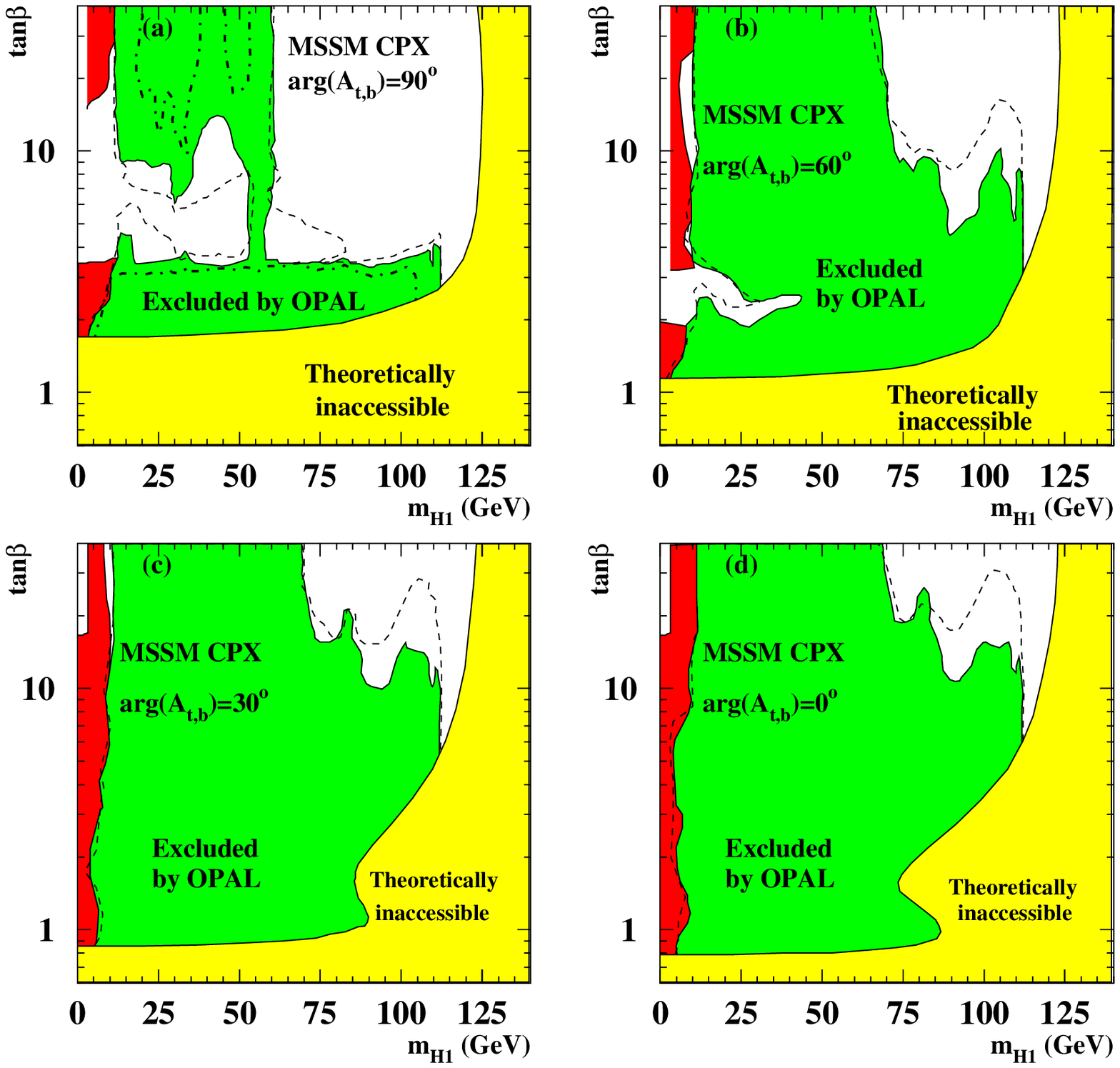, width=\textwidth}
      \end{minipage}}}
  \caption{\sl{The CPX MSSM $95\,\%\,\mathrm{CL}$ exclusion areas in the 
      ($m_{\Hone}$,$\tanb$) plane, using scans with (a) $\arg (A_{\mathrm{t,b}}) = 
      \arg (m_{\tilde\mathrm{g}}) = 90^{\circ}$,  (b) $\arg (A_{\mathrm{t,b}}) = 
      \arg (m_{\tilde\mathrm{g}}) = 60^{\circ}$,  (c) $\arg (A_{\mathrm{t,b}}) = 
      \arg (m_{\tilde\mathrm{g}}) = 30^{\circ}$ and (d) $\arg (A_{\mathrm{t,b}}) = 
      \arg (m_{\tilde\mathrm{g}}) = 0^{\circ}$. While the CPV phases 
      decrease, effects from CP violation like the strong $\Htwo\ra\Hone\Hone$
      contribution vanish. 
      See Fig.~\ref{fig:nomix} for the notation.
      }}\label{fig:CPXangles1_excl}
\end{figure}

\begin{figure}[pt]
  \qquad\hspace{-1.9cm}{\centering{
      \begin{minipage}{1.15\textwidth}
        \epsfig{file=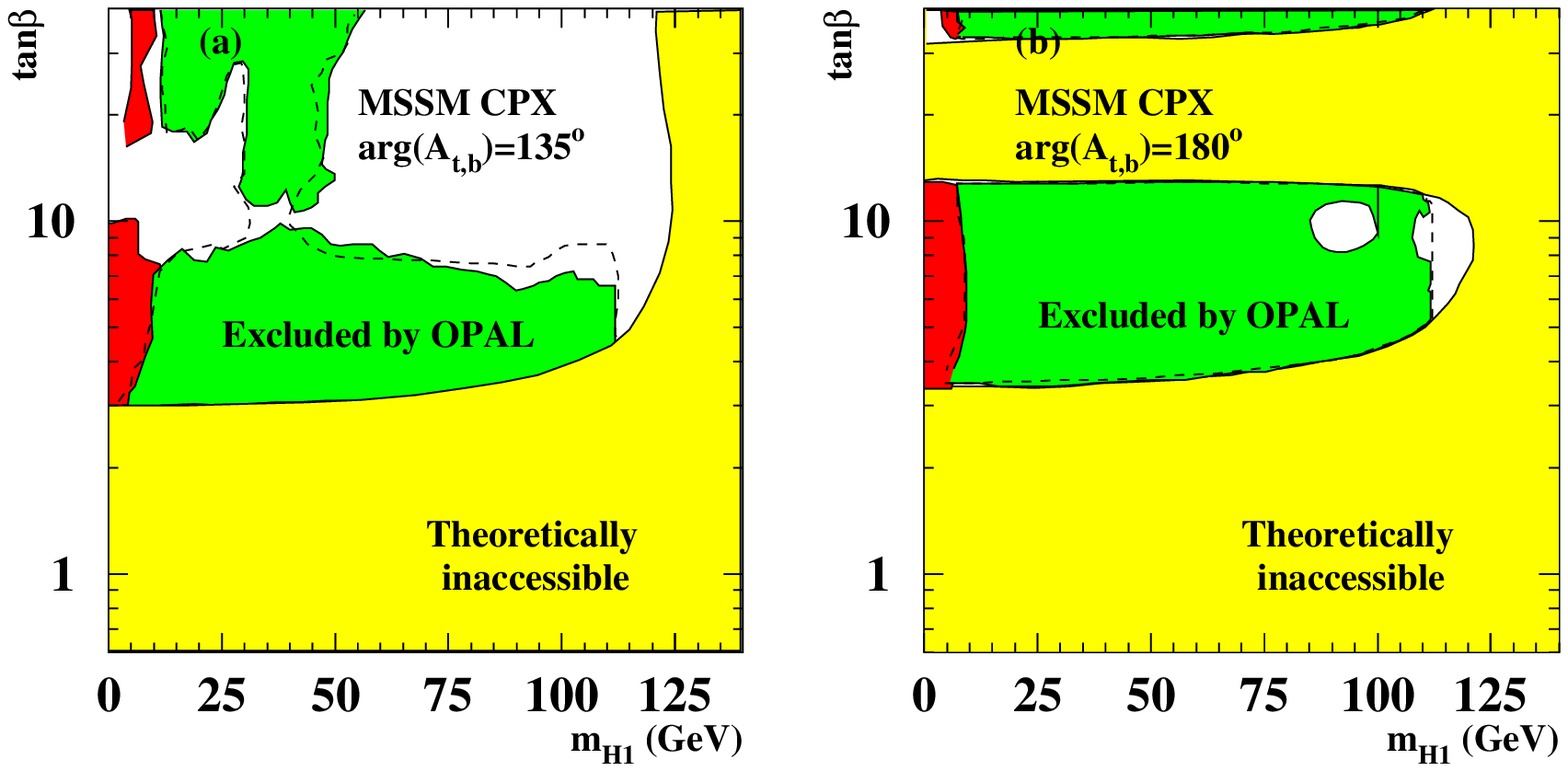, width=\textwidth}
      \end{minipage}}}
  \caption{\sl{The CPX MSSM $95\,\%\,\mathrm{CL}$ exclusion areas in the 
      ($m_{\Hone}$,$\tanb$) plane, using scans with (a) $\arg (A_{\mathrm{t,b}}) = 
      \arg (m_{\tilde\mathrm{g}}) = 135^{\circ}$ and  (b) $\arg (A_{\mathrm{t,b}}) = 
      \arg (m_{\tilde\mathrm{g}}) = 180^{\circ}$.
      See Fig.~\ref{fig:nomix} for the notation.
      }}\label{fig:CPXangles2_excl}
\end{figure}

The effect of different choices of the CPV phases is illustrated in
Figs.~\ref{fig:CPXangles1_excl} and \ref{fig:CPXangles2_excl}.  Values
of $\arg(A_{\mathrm{t,b}}) = \arg(m_{\tilde\mathrm{g}})$ from
$0^{\circ}$ to $180^{\circ}$ are displayed.
Fig.~\ref{fig:CPXangles1_excl} shows exclusion regions in the
parameter space of $\tanb$ and $m_{\Hone}$ for $\arg(A_{\mathrm{t,b}})
= \arg(m_{\tilde\mathrm{g}}) = 90^{\circ},60^{\circ},30^{\circ}$ and
$0 ^{\circ}$. The lower limit on $\tanb>1.9$ in the scenario with
phases of $60^{\circ}$ is the lowest limit on $\tanb$ in the CPV
scenarios. At $30^{\circ}$ and at $0^{\circ}$ all areas for low
$m_{\Hone}$ and low $\tanb$ are excluded.  The exclusion for the
maximally CPV scenario CPX with $90^{\circ}$ is very different from
the exclusion of a CPC scenario ($\arg(A_{\mathrm{t,b}}) =
\arg(m_{\tilde\mathrm{g}}) = 0^{\circ}$).  A variation of the second
main parameter governing the size of CPV effects, $m_{\mathrm{SUSY}}$,
has similar effects on the exclusion to those of a variation of
$\arg(A_{\mathrm{t,b}}) = \arg(m_{\tilde\mathrm{g}})$.

Fig.~\ref{fig:CPXangles2_excl} shows exclusion regions in the
parameter space of $\tanb$ and $m_{\Hone}$ for phases of (a)
$135^{\circ}$ and (b) $180^{\circ}$. The scenario in (a) is
phenomenologically still similar to the original CPX scenario. The
scenario in (b), which is in fact a CPC case, exhibits two allowed
regions, of which the lower one from $\tanb=3$ to $\tanb=13$ has a low
$\Hone\Zo$ coupling.  The unexcluded ``hole'' in the exclusion region
for $90<\mHone<100$~GeV is due to an excess of the background in the
SM-like channels.

\begin{figure}[pt]
  \qquad\hspace{-1.9cm}{\centering{
      \begin{minipage}{1.15\textwidth}
        \epsfig{file=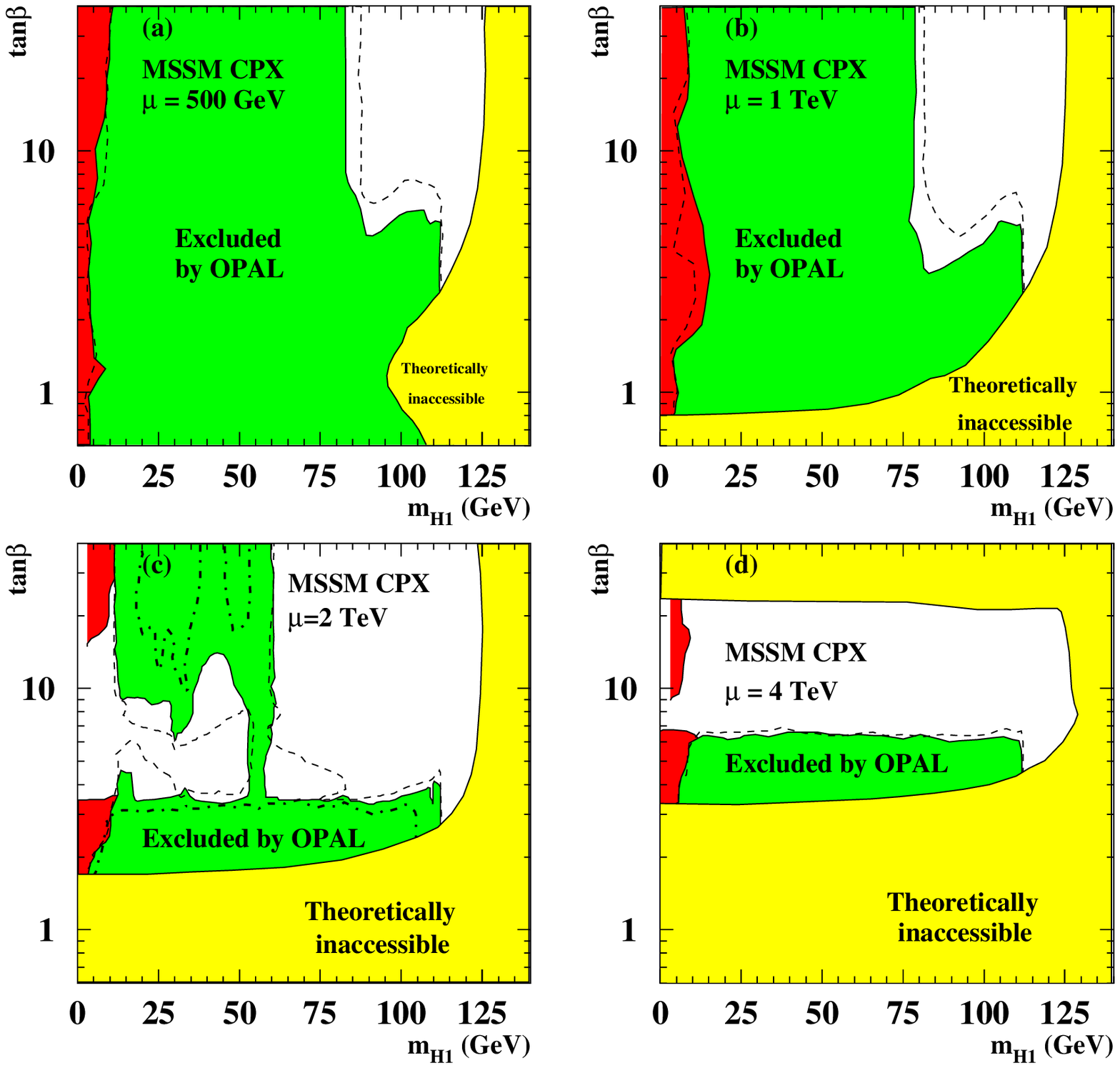, width=\textwidth}
      \end{minipage}}}
  \caption{\sl{The CPX MSSM $95\,\%\,\mathrm{CL}$ exclusion areas in the 
      ($m_{\Hone}$,$\tanb$) plane, using scans with (a) $\mu = 500$~GeV,  (b) $\mu=1000$~GeV, (c) $\mu=2000$~GeV (CPX)
      and (d) $\mu=4000$~GeV.
  See Fig.~\ref{fig:nomix} for the notation.
  }}\label{fig:CPXmu_excl}
\end{figure}

Since the CPX scenario has a relatively high value of $\mu=2$~TeV,
which influences the mixing of the CP eigenstates into the mass
eigenstates (see Eq.~(\ref{for:offdiag})), $\mu$ is varied from
$\mu=500$~GeV to $\mu=4$~TeV in Fig.~\ref{fig:CPXmu_excl}. For
$\mu=500$~GeV (Fig.~\ref{fig:CPXmu_excl}~(a)) and $\mu=1$~TeV
(Fig.~\ref{fig:CPXmu_excl}~(b)) the CPV effects are small. Therefore
no unexcluded regions occur at small $\mHone$. The scenario with
$\mu=4$~TeV (Fig.~\ref{fig:CPXmu_excl}~(d)) has strong mixing and a
suppression of pair production at large $\tanb$, resulting in an
exclusion area that is considerably smaller than in the CPX scenario
(Fig.~\ref{fig:CPXmu_excl}~(c)).

\begin{figure}[pt]
  \qquad\hspace{-1.9cm}{\centering{
      \begin{minipage}{1.15\textwidth}
        \epsfig{file=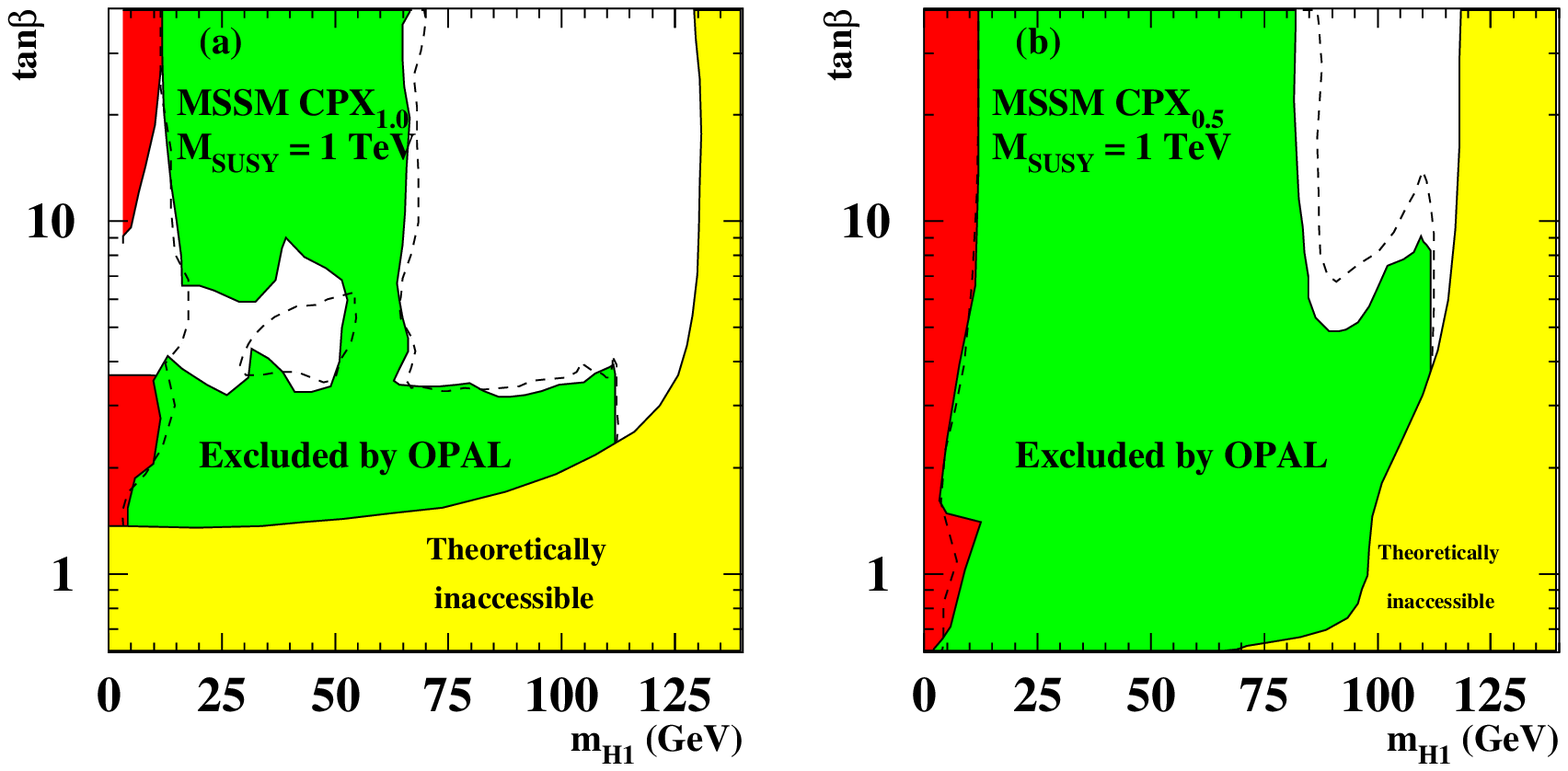, width=\textwidth}
      \end{minipage}}}
  \caption{\sl{The CPX MSSM $95\,\%\,\mathrm{CL}$ exclusion areas in the 
      ($m_{\Hone}$,$\tanb$) plane, using scans (a) preserving the CPX ratios of 
      $\mu,\, A_{\mathrm{b,t}}$ and  $m_{\mathrm{SUSY}}$, using 
      $m_{\mathrm{SUSY}} = 1$~TeV, and  (b) $m_{\mathrm{SUSY}}=1$~TeV keeping
      $\mu$ and $A_{\mathrm{b,t}}$ at their CPX values.
      See Fig.~\ref{fig:nomix} for the notation.
      }}\label{fig:CPXmsusy_excl}
\end{figure}

The proposal of the CPX scenario in~\cite{Carena:2000ks} leaves the
choice of $m_{\mathrm{SUSY}}$ open, as long as the relations
$|A_{\mathrm{t,b}}| = 2 m_{\mathrm{SUSY}}$, $|m_{\tilde{\mathrm{g}}}|
= 2 m_{\mathrm{SUSY}}$ and $\mu = 4 m_{\mathrm{SUSY}}$ are preserved.
In order to test the dependence on $m_{\mathrm{SUSY}}$, two scenarios
are tested: Fig.~\ref{fig:CPXmsusy_excl}~(a) shows the scenario
CPX$_{1.0}$, where the ratio between the parameters in the CPX
proposal is preserved, while $m_{\mathrm{SUSY}}$ is increased from
500~GeV to 1~TeV.  Only small differences with respect to the CPX
scenario with $m_{\mathrm{SUSY}}=500$~GeV can be seen.
Fig.~\ref{fig:CPXmsusy_excl}~(b) shows the CPX scenario as given in
Table~\ref{tab:benchmark_sum}, but with only $m_{\mathrm{SUSY}}$ set
to 1~TeV, while the values of $|A_{\mathrm{t,b}}|$,
$|m_{\tilde{\mathrm{g}}}|$ and $\mu$ are kept fixed. This results in a
decrease of the CPV effects and thus no unexcluded regions at small
$\mHone$ are observed.

%
%

\section{Summary}
\label{sect:conclusion}

The searches for neutral Higgs bosons described in this paper are
based on all data collected by the OPAL experiment, at energies in the
vicinity of the Z resonance (LEP1 phase) and between 130 and 209~GeV
(LEP2 phase).  The corresponding integrated luminosities are of about
$720\,\ipb$.  The searches addressing the Higgsstrahlung process
$\ee\ra{\cal H}Z$ and those for the pair production process
$\ee\ra{\cal H}_1{\cal H}_2$ are statistically combined. None of these
searches reveals a significant excess of events beyond the predicted
background level, which would indicate the production of Higgs bosons.

From these results, model-independent limits are derived for the
cross-section of a number of event topologies that could be associated
to Higgs boson production. These limits cover a wide range of Higgs
boson masses and are typically much lower than the largest
cross-sections predicted by the MSSM.

The search results are also used to test a number of ``benchmark
scenarios'' of the MSSM, with and without the inclusion of
CP-violating effects.

In the CP-conserving case, new benchmark situations are investigated
as compared to earlier publications. These are motivated either by new
measurements of the b$\ra$s$\gamma$ branching ratio and the muon
anomalous magnetic moment $(g-2)_{\mu}$, or in anticipation of the
forthcoming searches at the proton-proton collider LHC. In all these
scenarios the searches conducted by OPAL exclude sizeable domains of
the MSSM parameter space, even in those situations where the
sensitivity of the LHC experiments is expected to be low. An overview
of the results is given in Table~\ref{tab:mssmmhmalimits}.  In the
``$\mh$-max'' scenario which, among all scenarios predicts the widest
range of $\mh$ values, the following limits can be set at the 95\%
confidence level: $\mh > 84.5$~GeV and $\mA > 85.0$~GeV; furthermore,
if the top quark mass is fixed at the current experimental value of
174.3 GeV, the range $0.7 < \tanb < 1.9$~GeV can be excluded (this
range shrinks for higher values of $m_{\mathrm{top}}$).

For the first time, a number of CP-violating MSSM scenarios are
studied experimentally, where the CP-violating effects are introduced
in the Higgs potential by radiative corrections. The ``CPX'' benchmark
scenario is designed to maximize the phenomenological differences in
the Higgs sector with respect to the CP-conserving scenarios. In this
case the region $\tanb < 2.8$ is excluded at 95\% confidence level but
no universal limit is obtained for either of the Higgs boson masses.
However, for $\tanb < 3.3$, the limit $\mHone > 112$~GeV can be set
for the mass of the lightest neutral Higgs boson of the model.

\section*{Acknowledgments}

We particularly wish to thank Marcela Carena, Sven Heinemeyer, Gudrun Hiller, 
Steve Mrenna, Apostolos Pilaftsis and Georg Weiglein 
for their support.
We also thank the SL Division for the efficient operation
of the LEP accelerator at all energies
 and for their close cooperation with
our experimental group.  In addition to the support staff at our own
institutions we are pleased to acknowledge the  \\
Department of Energy, USA, \\
National Science Foundation, USA, \\
Particle Physics and Astronomy Research Council, UK, \\
Natural Sciences and Engineering Research Council, Canada, \\
Israel Science Foundation, administered by the Israel
Academy of Science and Humanities, \\
Benoziyo Center for High Energy Physics,\\
Japanese Ministry of Education, Culture, Sports, Science and
Technology (MEXT) and a grant under the MEXT International
Science Research Program,\\
Japanese Society for the Promotion of Science (JSPS),\\
German Israeli Bi-national Science Foundation (GIF), \\
Bundesministerium f\"ur Bildung und Forschung, Germany, \\
National Research Council of Canada, \\
Hungarian Foundation for Scientific Research, OTKA T-038240, 
and T-042864,\\
The NWO/NATO Fund for Scientific Research, the Netherlands.\\



\newcommand{\EPC}[3]  {Eur.\ Phys.\ J.\ C\ {\bf#1} (#2) #3}

\clearpage


\begin{table}
  \qquad\hspace{-2.05cm}{\centering{
      \begin{minipage}{1.15\textwidth}
  \begin{center}
    {\small
      \begin{tabular}{|l||c|c|c|c|}
        \hline  \multicolumn{5}{|c|}{Luminosity table for $\ee\ra\genH\Zo$ production}\\\hline   
        Channel Name  & Energies & Luminosity   & Mass range & Described \\
        $\genH\Zo\ra$ & (GeV)    & (pb$^{-1}$)  &    (GeV)   &  in \\
        \hline\hline
        \multicolumn{5}{|c|}{LEP 1 Channels}\\\hline\hline
        
        qq$\tau\tau$/$\tau\tau$qq&            91.2    &        46.3 &   $\genmH=0 - 70            $ & \cite{bib:opalhiggsold4,bib:opalhiggsold5}\\ 
        ($\genHone\genHone\ra\qq\qq$)$\nu\nu$  &            91.2    &        46.3 &   $\genmHtwo=10-75,\genmHone=0-35  $ & \cite{bib:opalhiggsold4,bib:opalhiggsold5}\\ 
        qq$\nu\nu$               &            91.2    &        46.3 &   $\genmH=0 - 70            $ & \cite{bib:opalhiggsold4,bib:opalhiggsold5}\\ 
        qqll                     &            91.2    &        46.3 &   $\genmH=20 - 70           $ & \cite{bib:opalhiggsold4,bib:opalhiggsold5}\\ 
        
        \hline\hline
        \multicolumn{5}{|c|}{LEP 2 Channels}\\\hline
        \hline

        bbqq                             &  161--172  & 20.4  &  $\genmH=40 - 80              $  & \cite{bib:opalhiggsold2,bib:opalhiggsold3} \\ 
        bb$\nu\nu$                       &  161--172  & 20.4  &  $\genmH=50 - 70            $  & \cite{bib:opalhiggsold2,bib:opalhiggsold3} \\ 
        $\tau\tau$qq, bb$\tau\tau$  &  161--172  & 20.4  &  $\genmH=30 - 95            $  & \cite{bib:opalhiggsold2,bib:opalhiggsold3} \\ 
        Xee                              &  161--172  & 20.4  &  $\genmH=35 - 80            $  & \cite{bib:opalhiggsold2,bib:opalhiggsold3} \\ 
        X$\mu\mu$                        &  161--172  & 20.4  &  $\genmH=35 - 80            $  & \cite{bib:opalhiggsold2,bib:opalhiggsold3} \\ 
        
        \hline   
        
        bbqq                                              & 183 & 54.1 &  $\genmH=40 - 95           $  & \cite{bib:opalhiggsold1} \\
        ($\genHone\genHone$$\ra$4b)qq                     & 183 & 54.1 &  $\genmHtwo=40 - 80,\genmHone=10.5-38            $  & \cite{bib:opalhiggsold1}\\
        bb$\nu\nu$/($\genHone\genHone$$\ra$4b)$\nu\nu$    & 183 & 53.9 &  $\genmH=50 - 95           $  & \cite{bib:opalhiggsold1}\\
        bb$\tau\tau$/                                     & 183 & 53.7 &  $\genmH=30 - 100          $  & \cite{bib:opalhiggsold1}\\
        \quad $\tau\tau$qq/($\genHone\genHone$$\ra$4b)$\tau\tau$                &     &       &                              &                         \\
        bbee,  bb$\mu\mu$                                  & 183 & 55.9 &  $\genmH=60 - 100          $  & \cite{bib:opalhiggsold1}\\
        
        \hline
        
        bbqq                                          &  189        & 172.1 &  $\genmH=40 - 100         $  & \cite{pr285} \\ 
        ($\genHone\genHone$$\ra$4b)qq                 &  189        & 172.1 &  $\genmHtwo=40 - 100, \genmHone=10.5-48         $  & \cite{pr285} \\ 
        bb$\nu\nu$/($\genHone\genHone$$\ra$4q)$\nu\nu$    &  189        & 171.4 &  $\genmH=50 - 100         $  & \cite{pr285} \\ 
        bb$\tau\tau$/                                     &  189        & 168.7 &  $\genmH=30 - 100         $  & \cite{pr285} \\ 
        \quad $\tau\tau$qq/($\genHone\genHone$$\ra$4b)$\tau\tau$               &     &       &                              &                         \\
        bbee, bb$\mu\mu$                                   &  189        & 170.0 &  $\genmH=70 - 100         $  & \cite{pr285} \\ 
        low \mA ($\genHone\genHone$)($\nunu$,ee,$\mu\mu$)                               &  189--192   & 201.7 &  $\genmHtwo=45-90 ,\genmHone=2-10.5  $  & \cite{lowma} \\ 
        
        
        \hline
        
        bbqq                          &           192--209       &      421.2 & $\genmH=80-120               $   & \cite{OPALSMPAPER}           \\     
        ($\genHone\genHone\ra$4b)qq   &           192--209       &      421.2 & $\genmHtwo=80-120,\genmHone=12-\genmHtwo/2 $   & \ref{sect:fourjet}           \\     
        bb$\nu\nu$                    &           192--209       & 419.9 & $\genmH=30-120               $   & \cite{OPALSMPAPER}            \\     
        bbbb$\nu\nu$                  &           199--209  &      207.2  & $\genmHtwo=100-110,\genmHone=12-\genmHtwo/2$  & \ref{sect:missing_energy}     \\     
        bb$\tau\tau$ / $\tau\tau$qq   &           192--209  &      417.4  & $\genmH=80-120               $   & \cite{OPALSMPAPER}           \\     
        bbee, bb$\mu\mu$               &           192--209  &      418.3 & $\genmH=40-120               $   & \cite{OPALSMPAPER}           \\     
        low \mA ($\genHone\genHone$)($\nunu$,ee,$\mu\mu$)           &           196--209  &      396.9 & $\genmHtwo=45-90 ,\genmHone=2-10.5      $   & \cite{lowma}     \\     
        
        \hline
        \hline
        \multicolumn{5}{|c|}{LEP 2 Flavour-Independent Channels}\\\hline
        \hline
        qqqq        &    189       &      174.1 &  $\genmH=60 - 100$    &   \cite{Abbiendi:2000ug}    \\ 
        qq$\nu\nu$  &    189       &      171.8 &  $\genmH=30 - 100$    &   \cite{Abbiendi:2000ug}     \\ 
        qq$\tau\tau$, $\tau\tau$qq &    189       &      168.7 &  $\genmH=30 - 100$    &   \cite{Abbiendi:2000ug}     \\ 
        qqee, qq$\mu\mu$  &    189       &      170.0 &  $\genmH=70 - 100$    &   \cite{Abbiendi:2000ug}     \\ 
        
        \hline
        qqqq                   &       192--209       &     424.2 & $\genmH=60 - 120$ &      \cite{2HDMFINAL}                  \\ 
        qq$\nu\nu$             &       192--209       &     414.5 & $\genmH=30 - 110$ &     \cite{2HDMFINAL}                   \\ 
        qq$\tau\tau$, $\tau\tau$qq          &       192--209       &     418.9  & $\genmH=60 - 115$ &      \cite{2HDMFINAL}                  \\ 
        qqee, qq$\mu\mu$          &       192--209       &     422.0 & $\genmH=60 - 120$ &      \cite{2HDMFINAL}                  \\

        \hline

      \end{tabular}
      }
  \end{center}
\end{minipage}}}
\caption{\sl List of the searches for the Higgsstrahlung process.
    The last column gives the reference or section where the search is 
    described.} \label{tab:channel_intro_hz}
\end{table}

\begin{table}
  \begin{center}
    {\small
      \begin{tabular}{|l||c|c|c|c|}
        \hline  \multicolumn{5}{|c|}{Luminosity table for $\ee\ra\genHone\genHtwo$ production}\\\hline   
        Channel Name           & Energies & Luminosity   & Mass range & Described \\
        $\genHone\genHtwo\ra$ & (GeV)    & (pb$^{-1}$)  &    (GeV)   &  in \\
        \hline\hline
        \multicolumn{5}{|c|}{LEP 1 Channels}\\\hline\hline
        
        6b                   &            91.2    &        27.6 &   $\genmHtwo=40-70, \genmHone=5-35 $ & \cite{bib:opalhiggsold4,bib:opalhiggsold5} \\ 
        qq$\tau\tau$, $\tau\tau$qq&            91.2    &        46.3 &   $\genmHtwo=12-75, \genmHone=10-78$ & \cite{bib:opalhiggsold4,bib:opalhiggsold5}\\ 
        6$\tau$, 4$\tau$2q, 2$\tau$4q  &            91.2    &        46.3 &   $\genmHtwo=30-75, \genmHone=4-30 $ & \cite{bib:opalhiggsold4,bib:opalhiggsold5}\\ 
        
        \hline\hline
        \multicolumn{5}{|c|}{LEP 1.5 Channels}\\\hline
        \hline
        
        4b          &                    130--136     &       5.2  & $\Sigma=80-130,\Delta=0-50$  & \cite{bib:opalhiggsold3} \\
        6b          &                    130--136     &       5.2  & $\genmHtwo=55-65,\genmHone>27.5 $  & \cite{bib:opalhiggsold3} \\

        \hline\hline
        \multicolumn{5}{|c|}{LEP 2 Channels}\\\hline
        \hline
        
        4b                               &   161  & 10.0  &  $\Sigma=80-130,\Delta=0-60      $  & \cite{bib:opalhiggsold2,bib:opalhiggsold3} \\ 
        6b                               &   161  & 10.0  &  $\genmHtwo=55-65,\genmHone>20.0       $  & \cite{bib:opalhiggsold2,bib:opalhiggsold3} \\ 
        bb$\tau\tau$, $\tau\tau$bb        &   161  & 10.0  &  $\genmHtwo=40-160,\genmHone=52-160$  & \cite{bib:opalhiggsold2,bib:opalhiggsold3} \\ 
        
        \hline
        
        4b                               &  172  & 10.4  &  $\Sigma=80-130,\Delta=0-60      $  & \cite{bib:opalhiggsold2,bib:opalhiggsold3} \\ 
        6b                               &  172  & 10.4  &  $\genmHtwo=55-65,\genmHone=25-35  $  & \cite{bib:opalhiggsold2,bib:opalhiggsold3} \\ 
        bb$\tau\tau$, $\tau\tau$bb        &  172  & 10.4  &  $\genmHtwo=37-160,\genmHone=28-160$  & \cite{bib:opalhiggsold2,bib:opalhiggsold3} \\ 
        
        \hline
        
        4b                                                & 183 & 54.1 &  $\Sigma=80-150,\Delta=0-60     $  & \cite{bib:opalhiggsold1}\\
        6b                                                & 183 & 54.1 &  $\genmHtwo=30-80,\genmHone=12-40 $  & \cite{bib:opalhiggsold1}\\
        bb$\tau\tau$, $\tau\tau$bb                         & 183 & 53.7 &  $\Sigma=70-170,\Delta=0-70     $  & \cite{bib:opalhiggsold1}\\
        
        \hline
        
        4b                                                &  189        & 172.1 &  $\Sigma=80-180,\Delta=0-70    $  & \cite{pr285} \\ 
        6b                                                &  189        & 172.1 &  $\genmHtwo=24-80,\genmHone=12-40$  & \cite{pr285} \\ 
        bb$\tau\tau$, $\tau\tau$bb                         &  189        & 168.7 &  $\Sigma=70-190,\Delta=0-90    $  & \cite{pr285} \\ 
        
        \hline
        
        4b                            &           192       &      28.9 & $\Sigma=83-183 ,\Delta=0-70       $   & \ref{sect:ahbb_eq}           \\     
        4b                            &           196       &      74.8 & $\Sigma=80-187 ,\Delta=0-70       $   & \ref{sect:ahbb_eq}         \\     
        4b                            &           200       &      77.2 & $\Sigma=80-191 ,\Delta=0-70       $   & \ref{sect:ahbb_eq}         \\     
        4b                            &           202       &      36.1 & $\Sigma=80-193 ,\Delta=0-70       $   & \ref{sect:ahbb_eq}         \\     
        high $\genmHone$ 4b           &           199--209  &      207.3 & $\Sigma=120-190 ,\Delta=0-70       $   & \ref{sect:ahbb_eq}         \\     
        low $\genmHone$ 4b            &           199--209  &      207.3 & $\Sigma=100-140 ,\Delta=60-100     $   & \ref{sect:ahbb_gg}         \\     
        6b                            &           199--209  &      207.3 & $\Sigma=90-200 ,\Delta=40-160     $   & \ref{sect:ah6b}         \\     
        bb$\tau\tau$, $\tau\tau$bb     &           192       &      28.7   & $\Sigma=10-174 ,\Delta=0-182      $   & \ref{btau}          \\     
        bb$\tau\tau$, $\tau\tau$bb     &           196       &      74.7   & $\Sigma=10-182 ,\Delta=0-191      $   & \ref{btau}         \\     
        bb$\tau\tau$, $\tau\tau$bb     &           200       &      74.8   & $\Sigma=10-182 ,\Delta=0-191      $   & \ref{btau}         \\     
        bb$\tau\tau$, $\tau\tau$bb     &           202       &      35.4   & $\Sigma=10-174 ,\Delta=0-182      $   & \ref{btau}         \\     
        bb$\tau\tau$, $\tau\tau$bb     &           199--209  &      203.6  & $\Sigma=70-190 ,\Delta=0-90       $   & \ref{btau}         \\     
        
        \hline

      \end{tabular}
      }
  \end{center}
  \caption{\sl List of the searches for pair production.
    The last column gives the reference or section where the search is 
    described. The symbols $\Sigma=\genmHone+\genmHtwo$ and $\Delta=\genmHtwo-\genmHone$ denote 
    the Higgs mass sum and difference.} \label{tab:channel_intro_ah}
\end{table}

\begin{table}[htbp]
\begin{center}
\begin{tabular}{|c|c||c|c|c|c|c|}
\hline
$m_{\genHtwo}$ & $m_{\genHone}$ & \multicolumn{5}{|c|}{Efficiency 
  for the process $\genHtwo\Zo\ra\bb\bb\qq$ at $\sqrt{s}$}\\
 (GeV)      & (GeV)       & 192 GeV & 196 GeV & 200 GeV & 202 GeV & 206 GeV \\
\hline\hline
100. & 12. &  0.689 & 0.684  & 0.717  & 0.733  &  0.693\\
100. & 20. &  0.651 & 0.639  & 0.653  & 0.659  &  0.586\\
100. & 30. &  0.460 & 0.461  & 0.461  & 0.470  &  0.480\\
100. & 40. &  0.270 & 0.260  & 0.283  & 0.315  &  0.323\\
100. & 48. &  0.328 & 0.325  & 0.361  & 0.392  &  0.400\\
\hline
105. & 12. &  0.538 & 0.658  & 0.702  & 0.709  &  0.701\\
105. & 20. &  0.562 & 0.618  & 0.697  & 0.658  &  0.681\\
105. & 30. &  0.490 & 0.525  & 0.509  & 0.536  &  0.497\\
105. & 40. &  0.407 & 0.306  & 0.309  & 0.316  &  0.319\\
105. & 50. &  0.433 & 0.368  & 0.355  & 0.359  &  0.370\\
\hline
110. & 12. &     &           & 0.637  & 0.682  &  0.720\\
110. & 20. &     &           & 0.625  & 0.646  &  0.532\\
110. & 30. &     &           & 0.556  & 0.549  &  0.565\\
110. & 40. &     &           & 0.380  & 0.328  &  0.343\\
110. & 53. &     &           & 0.395  & 0.341  &  0.358\\
\hline
\end{tabular}
\end{center}
\caption{\sl{
    Efficiencies of the standard $\epm\ra\Zo\mathrm{H}\ra\qq\,\bb$ 
    analysis~\cite{OPALSMPAPER} for the $\epm\ra\Zo\genHtwo\ra\Zo\genHone\genHone\ra\qq\,\bb\,\bb$ final state (see Section~\ref{sect:fourjet}). 
    The uncertainties from Monte Carlo statistics are of the order of $\pm0.015$.}}\label{tab:effbbbbqq}
\end{table}

\begin{table}[htbp] 
\begin{center} 
\begin{tabular}{|c||c|c|c|c|c|c|c|}\hline 
cut &   data    &   tot. bkg.   &   qq($\gamma$)   &   tot 4-f   & Eff. & Eff. & Eff. \\
    &         &     & & & \genHtwo\ra\genHone\genHone  &  \genHtwo\ra\genHone\genHone  &   \genHtwo\ra\bb   \\ 
    &         &     & & & \genmHtwo=105   &  \genmHtwo=105    &  \genmHtwo=105    \\ 
 &       &      &      &      &  \genmHone=20  &  \genmHone=40   & \\ \hline 
(1)--(5) &   503 &   424.77  &   123.94  &   297.99  &   0.75   & 0.74   &   0.59   \\ \hline \hline 
\multicolumn{8}{|c|}{2-jet topology subsample A}     \\ 
\hline 
(6) &   371 &   308.56  &   108.66  &   197.17  &   0.75   &   0.16   &   0.53   \\ 
(7) &   213 &   201.77  &   24.70   &   177.07  &   0.70   &   0.15   &   0.50   \\ 
(8) &   135 &   126.29  &   22.85   &   103.43  &   0.68   &  0.14   &   0.49   \\ 
\hline 
ANN$_{\mathrm{A}}$ &   11   &   10.0   &  2.63   &  7.39 &   0.59   & 0.11 &   0.40   \\ \hline \hline 
\multicolumn{8}{|c|}{4-jet topology subsample B}     \\ 
\hline 
(6) &   118 &   112.32  &   14.42   &   97.80   &   0.008    & 0.57  &   0.06    \\ 
\hline 
ANN$_{\mathrm{B}}$ &   8   &    7.20   &  2.83   &  4.37 &   0.003    & 0.55   &   0.05    \\ \hline \hline 
\multicolumn{8}{|c|}{Total (2+4 jets) }     \\ 
\hline 
Sum &   19  &   $17.2\,\pm\,0.6$   &   5.46   &   11.8  &   0.59   &   0.66   &   0.45   \\ 
\hline 
\end{tabular} 
\end{center} 
\caption{\sl{Cut flow in the missing energy analysis for
    $\genHtwo\nunu\ra\bb\nunu$ and  $\genHtwo\nunu\ra\genHone\genHone\nunu\ra\bb\bb\nunu$ for $100\le\genmHtwo\le110$~GeV 
    for data taken at $\sqrts=199$ to 209~GeV (see Section~\ref{sect:missing_energy}). The uncertainty quoted for the total background is from Monte Carlo statistics only.}}\label{tab:cutbbbbnn}
\end{table}

\begin{table}[htbp]
\begin{center}
\begin{tabular}{|c|c||c|c|}
\hline
\multicolumn{4}{|c||}{Efficiency A and B for $\genHtwo\nunu\ra\bb\nunu$ decays}\\
\hline
\multicolumn{2}{|c||}{$m_{\genHtwo}$ (GeV)} & Efficiency of selection A & Efficiency of selection B \\
\hline
\multicolumn{2}{|c||}{100.} & 0.382 & 0.044 \\
\multicolumn{2}{|c||}{105.} & 0.396 & 0.050 \\
\multicolumn{2}{|c||}{110.} & 0.379 & 0.050 \\
\hline
\hline
\multicolumn{4}{|c|}{Efficiency A and B for $\genHtwo\nunu\ra\genHone\genHone\nunu\ra\bb\bb\nunu$ decays}\\
\hline
$m_{\genHtwo}$ (GeV) & $m_{\genHone}$ (GeV) & Efficiency of selection A & Efficiency of selection B \\
\hline
100.  & 12.  &  0.686  & 0.0    \\
100.  & 20.  &  0.561  & 0.001  \\
100.  & 30.  &  0.280  & 0.301  \\
100.  & 40.  &  0.090  & 0.522  \\
100.  & 48.  &  0.195  & 0.436  \\
\hline                   
105.  & 12.  &  0.707  & 0.0    \\
105.  & 20.  &  0.587  & 0.0    \\
105.  & 30.  &  0.349  & 0.254  \\
105.  & 40.  &  0.113  & 0.550  \\
105.  & 50.  &  0.179  & 0.487  \\
\hline                   
110.  & 12.  &  0.677  & 0.0    \\
110.  & 20.  &  0.585  & 0.001  \\
110.  & 30.  &  0.402  & 0.189  \\
110.  & 40.  &  0.109  & 0.555  \\
110.  & 50.  &  0.131  & 0.537  \\
110.  & 53.  &  0.186  & 0.495  \\
\hline
\end{tabular}
\end{center}

\caption{\sl{Efficiencies of the selections A and B of the missing energy analysis
    for $\genHtwo\ra\bb\nunu$ and  $\genHtwo\ra\genHone\genHone\ra\bb\bb\nunu$ for $100\le\genmHtwo\le110$~GeV 
    for data taken at $\sqrts=199$ to $209$~GeV  is used (see Section~\ref{sect:missing_energy}).
    The uncertainty from Monte Carlo statistics is of the order of $\pm0.010$. 
    }}\label{tab:zhqa_eff}
\end{table}

\begin{table}[htbp]
  \begin{center}
    \begin{tabular}{|c||r||r||r|r||c|} \hline
      Cut & Data & Total bkg. & q\=q($\gamma$) & 4-fermi. & Efficiency (\%)\\
      &      &            &              &     &  $\genmHone = \genmHtwo =90$ GeV \\\hline\hline

      \multicolumn{6}{|c|}{$\genHtwo\genHone\ra$4b Channel} \\\hline
      (1) &  39367& 39375.6&30958.7&8325.6&$99.8$\\
      (2) &  13792& 13895.2& 8914.4&4976.0&$98.4$\\
      (3) &   4682&  4509.6& 1110.5&3397.9&$88.3$\\
      (4) &   3997&  3994.5&  707.7&3285.7&$86.4$\\
      (5) &   3474&  3431.0&  566.3&2863.6&$85.6$\\
      (6) &   3331&  3271.5&  520.4&2749.9&$83.7$\\
      \hline 
      ${\cal L}^{\genHone\genHtwo}>0.95$ & 22& 19.9 $\pm$  0.3&  6.5&13.4&$49.4$  \\\hline\hline
      \multicolumn{6}{|c|}{$\genHtwo\genHone\ra\bb\tautau$ Channel } \\ \hline
      Pre-sel                 &   336.0 & 354.6  & 96.0 & 258.5 & $53.8$\\
\hline
      ${\cal L}^{\genHtwo\genHone}>0.64$     &      13&  $13.2\pm0.4$  & 0.7 & 12.4& $42.5$\\
      \hline
    \end{tabular}
\end{center}
\caption{\label{tab:ah_cutflow}\sl
  Cut flow in the \genHone\genHtwo\ channels for high $\genmHone$ (see Section~\ref{fourb}) and for all data taken 
  at $\sqrts=192$ to $209$~GeV:
  effect of the cuts on the data and the
  background, normalised to the integrated luminosity of the data. 
  The two-photon background, not shown separately, is included in the total
  background.
  The signal efficiencies are given in the last column for
  \genmHone=\genmHtwo=90~GeV.
  }
\end{table}

\begin{table}[htbp]
\begin{center}
\begin{tabular}{|c||c|c|c|c|c|c|c|}
\hline
 & \multicolumn{7}{|c|}{Efficiency for the process  $\genHtwo\genHone\ra\bb\bb$ (high $\genmHone$) at  }      \\
 & \multicolumn{7}{|c|}{$\sqrt{s}=206$~GeV }  \\
\hline\hline
$m_{\genHone}$ (GeV) & 30.0 & 40.0 & 50.0 & 60.0 & 70.0 & 80.0 & 90.0 \\
\hline\hline
$m_{\genHtwo}$ (GeV) & \multicolumn{7}{|c|}{ } \\
\hline
 30.0 & 0.001  &       &       &       &       &        &  \\
 40.0 & 0.0008 & 0.004 &       &       &       &        &  \\
 50.0 &        & 0.110 & 0.215 &       &       &        &  \\
 60.0 & 0.103  &       & 0.274 & 0.364 &       &        &  \\
 70.0 &        & 0.254 &       & 0.381 & 0.388 &        &  \\
 80.0 & 0.200  &       & 0.384 &       & 0.425 &  0.470 &  \\
 90.0 &        & 0.319 &       & 0.388 &       &  0.472 & 0.479 \\  
100.0 &        &       & 0.374 &       & 0.432 &        & 0.435 \\
110.0 &        & 0.341 &       & 0.371 &       &        &  \\
120.0 & 0.261  &       & 0.349 &       & 0.399 &        &  \\
130.0 &        & 0.253 &       & 0.368 &       &        &  \\
140.0 & 0.231  &       & 0.290 &       &       &        &  \\
150.0 &        & 0.177 &       &       &       &        &  \\
160.0 & 0.116  &       &       &       &       &        &  \\
\hline
\end{tabular}
\end{center}
\caption{\sl{Efficiencies of the $\genHone\genHtwo\ra\bb\bb$ analysis for high $\genmHone$ (see Section~\ref{sect:ahbb_eq}). 
    The uncertainty from Monte Carlo statistics is typically of the order of $\pm0.015$. The table is showing the Monte Carlo points produced. 
    }}\label{tab:ahbb_eff}
\end{table}

\begin{table}[htbp]
  {\small 
    \qquad\hspace{-2.05cm}{\centering{
        \begin{minipage}{1.15\textwidth}
          \begin{center}
            \begin{tabular}{|l||c|c||c|c||c|c|}\hline
              \multicolumn{7}{|c|}{Systematic uncertainties 
                at \sqrts = 206~\G}    \\\hline
              & \multicolumn{2}{|c|}{$\genHtwo\genHone\ra\bb\bb$ (high $\genmHone$)} & \multicolumn{2}{|c|}{$\genHtwo\genHone\ra\bb\bb$ (low $\genmHone$)}
              & \multicolumn{2}{|c|}{$\genHtwo\genHone\ra\bb\tautau,\tautau\bb$}\\\hline
              Source & Signal eff. & Background & Signal eff. & Background & Signal eff. & Background \\\hline\hline
              Detector modelling         &   0.9\%    & 8.0\%     &   1.1\%    & 8.6\%       &   1.0\%    & 1.0\%  \\ 
              B-had. Decay Mult.         &   0.9\%    & 1.2\%     &   2.1\%    & 1.9\%       &   1.0\%    & 3.0\%  \\
              B-had. Fragment.           &   1.8\%    & 1.5\%     &   2.7\%    & 1.5\%       &   1.8\%    & 1.5\%  \\
              C-had. Fragment.           &   0.0\%    & 0.5\%     &   0.0\%    & 0.0\%       &   0.0\%    & 0.0\%  \\
              4f-cross-section           &   0.0\%    & 1.6\%     &   0.0\%    & 1.0\%       &   0.0\%    & 1.9\%  \\
              MC-Generators              &   0.0\%    & 2.6\%     &   0.0\%    & 2.6\%       &   0.0\%    & 0.0\%  \\
              $\tau$ identification      &   0.0\%    & 0.0\%     &   0.0\%    & 0.0\%       &   0.9\%    & 15.0\%  \\
              LH modelling               &   0.9\%    & 2.3\%     &   1.0\%    & 2.1\%       &   0.0\%    & 0.0\%  \\
              MC statistics              &   2.0\%    & 5.0\%     &   3.0\%    & 5.2\%       &   1.0\%    & 3.0\%  \\
              \hline
              Combined                   &   3.1\%    & 10.3\%    &   4.7\%    & 10.9\%        &   2.6\%    & 15.8\% \\
              \hline
            \end{tabular}
          \end{center}
      \end{minipage}}}}
  \caption[]{\sl Systematic uncertainties on the signal efficiency and background at \sqrts
    = 206~GeV for the processes $\genHone\genHtwo\ra\bb\bb$ with high $\genmHone$, low $\genmHone$ and for $\genHtwo\genHone\ra\bb\tautau,\tautau\bb$
    (see Sections~\ref{sect:ahbb_eq}, \ref{sect:ahbb_gg} and \ref{btau}).}
  \label{ah:allsyst}
\end{table}

\begin{table}[htbp]
  \begin{center}
    \begin{tabular}{|c||r||r||r|r||c|} \hline
      Cut & Data & Total bkg. & q\=q($\gamma$) & 4-fermi. & Efficiency (\%)\\
      &      &            &              &     & $m_{\genHone} = 30,\,m_{\genHtwo} =100$ GeV \\
      \hline\hline
      \multicolumn{6}{|c|}{$\genHone\genHtwo\ra$4b Channel for low $\genmHone$ } \\
      \hline
      (1) &  18519& $17802.0$&13705.2&4096.9&$ 99.6$\\
      (2) &   6538& $ 6427.8$& 3971.6&2456.3&$ 96.2$\\
      (3) &   4215& $ 4048.0$& 2082.4&1965.6&$ 94.2$\\
      (4) &   3618& $ 3497.5$& 1546.8&1950.6&$ 93.6$\\
      (5) &   2712& $ 2625.9$& 1188.5&1437.4&$ 90.5$\\
      (6) &   2477& $ 2389.4$& 1060.0&1329.5&$ 83.5$\\
      \hline 
      ${\cal L}^{\genHone\genHtwo}>0.98$ &  8& 10.4 $\pm$ 0.1&  6.1& 4.3& 36.9 \\
      \hline
    \end{tabular}
\end{center}
\caption{\label{tab:ahflow200-209_offdiag}\sl
  Cut flow in the \genHone\genHtwo\ channel for low $\genmHone$ (see Section~\ref{sect:ahbb_gg}) and for all data taken at $\sqrts=199$ to $209$~GeV:
  effect of the cuts on the data and the
  background, normalized to the integrated luminosity of the data. 
  The two-photon background, not shown separately, is included in the total
  background.
  The signal efficiencies are given in the last column for
  \genmHone=30~GeV and \genmHtwo=100~GeV in the $\genHone\genHtwo\ra\bb\bb$ channel. 
  }
\end{table}

\begin{table}[htbp]
\begin{center}
\begin{tabular}{|c||c|c|c|}
\hline
\multicolumn{4}{|c|}{Efficiency for the process $\genHtwo\genHone\ra\bb\bb$ (low $\genmHone$) at   }     \\
\multicolumn{4}{|c|}{$\sqrt{s}=206$~GeV  } \\
\hline\hline
$m_{\genHone}$ (GeV) &  \qquad 12.0 \qquad  &  \qquad 20.0 \qquad  &  \qquad 30.0 \qquad  \\
\hline\hline
$m_{\genHtwo}$ (GeV) & \multicolumn{3}{|c|}{ } \\
\hline
  90. & \qquad 0.269 \qquad  & \qquad 0.330 \qquad & \qquad 0.370 \qquad \\
  95. & \qquad 0.286 \qquad  & \qquad 0.341 \qquad & \qquad 0.384 \qquad \\
 100. & \qquad 0.305 \qquad  & \qquad 0.366 \qquad & \qquad 0.369 \qquad \\
 105. & \qquad 0.310 \qquad  & \qquad 0.358 \qquad & \qquad 0.369 \qquad \\
 110. & \qquad 0.298 \qquad  & \qquad 0.351 \qquad & \qquad 0.366 \qquad \\
\hline
\end{tabular}
\end{center}
\caption{\sl{Efficiencies of the $\genHone\genHtwo\ra\bb\bb$ analysis for low $\genmHone$ (see Section~\ref{sect:ahbb_gg}). 
    The uncertainty from Monte Carlo statistics is of the order of $\pm0.011$.
    }}\label{tab:ahbb_eff_offdiag}
\end{table}

\begin{table}[htbp]
\begin{center}
\begin{tabular}{|c||c|c|c|c|c|c|c|c|}
\hline
 & \multicolumn{8}{|c|}{Efficiency for \genHtwo\genHone\ra\bb\bb\bb\ at }\\
 & \multicolumn{8}{|c|}{$\sqrt{s}=206$~GeV     }  \\
\hline\hline
$m_{\genHone}$ (GeV) & 12.0 & 20.0 & 30.0 & 40.0 & 45.0 & 50.0 & 60.0 & 70.0 \\
\hline\hline
$m_{\genHtwo}$ (GeV) & \multicolumn{8}{|c|}{ } \\
\hline
  80. &  0.002  &  0.188 &  0.390 &  0.465 &        &        &        & \\
  90. &  0.002  &  0.263 &  0.447 &  0.595 &  0.569 &        &        & \\
 100. &  0.001  &  0.283 &  0.486 &  0.594 &  0.639 &  0.629 &        & \\
 110. &  0.001  &  0.300 &  0.552 &  0.627 &  0.662 &  0.659 &        & \\
 120. &  0.002  &  0.214 &  0.512 &  0.671 &  0.664 &  0.650 &  0.695 & \\
 130. &  0.002  &  0.292 &  0.519 &  0.635 &  0.680 &  0.670 &  0.657 & \\
 140. &  0.000  &  0.255 &  0.536 &  0.636 &  0.646 &  0.670 &  0.649 &  0.382 \\
\hline
\end{tabular}
\end{center}
\caption{\sl{Efficiencies of the $\genHone\genHtwo\ra\genHone\genHone\genHone\ra\bb\bb\bb$ analysis (see Section~\ref{sect:ah6b}). 
    The uncertainty from Monte Carlo statistics is $\pm 0.010$.  
    }}\label{tab:effbbbbbb}
\end{table}

\begin{table}[htbp]
\begin{center}
\begin{tabular}{|c||c|c|c|c|c|c|c|}
\hline
 & \multicolumn{7}{|c|}{Efficiency for $\genHtwo\genHone\ra\bb\tautau$ at }\\
 & \multicolumn{7}{|c|}{$\sqrt{s}=206$~GeV     }  \\
\hline\hline
$m_{\genHone}$ (GeV) &  30. & 40. & 50. & 60. & 70. & 80. & 90. \\
\hline\hline
$m_{\genHtwo}$ (GeV) & \multicolumn{7}{|c|}{ } \\
\hline
  30. & 0.0818 &       &       &      &        &       &        \\ 
  40. & 0.1016  & 0.2024 &       &      &        &       &        \\ 
  50. &        & 0.3296 & 0.3690 &      &        &       &        \\ 
  60. & 0.3152  &       & 0.4086 & 0.3668&        &       &        \\ 
  70. &        & 0.3802 &       & 0.3936&  0.4631 &       &        \\ 
  80. & 0.4265  &       & 0.3856 &      &  0.4343 & 0.4247 &        \\ 
  90. &        &  0.3599&       & 0.4285&        & 0.3925 &  0.4305 \\ 
 100. &  0.3854 &       & 0.4367 &      &  0.3951 & 0.4298 &        \\ 
 110. &        &  0.3628&       & 0.4227&        &       &        \\ 
 120. & 0.3631  &       & 0.3609 &      &  0.4020 &       &        \\ 
 130. &        & 0.2996 &       & 0.3180&        &       &        \\ 
 140. &        &       & 0.3049 &      &        &       &        \\ 
 150. &        & 0.2278 &       &      &        &       &        \\ 
 160. &        &       &       &      &        &       &        \\ 
 170. &        &       &       &      &        &       &        \\  
\hline
\end{tabular}
\end{center}
\caption{\sl{Efficiencies of the $\genHone\genHtwo\ra\bb\tautau$ analysis (see Section~\ref{btau}). 
    The uncertainty from Monte Carlo statistics is of the order of $\pm0.02$.  }}\label{tab:bbtautaueff}
\end{table}

\begin{table}[hp] 
\begin{center} 
\begin{tabular}{|l||c|c|c|}
\hline 
Channel & Total & Data & Section \\ 
 & Background & & \\  
\hline
\hline

$\genHtwo\Zo\ra\bb(\bb)\qq$     & 135.4 & 140 & \ref{sect:fourjet} \\
$\genHtwo\nunu\ra\bb(\bb)\nunu$ & 39.5  & 36  & \ref{sect:missing_energy} \\ 
$\genHtwo\genHone\ra\bb\bb(\bb)$                   & 19.9  & 22 & \ref{sect:ahbb_eq},  \ref{sect:ahbb_gg}, \ref{sect:ah6b}  \\
$\genHtwo\genHone\ra\bb\tautau$                    & 13.2  & 13 & \ref{btau} \\
\hline 
\end{tabular} 
\end{center} 
  \caption{\sl 
    Typical numbers of background and data for the searches described in this publication.}\label{tab:channel_sum}
\end{table}

\begin{table}[hp] 
  \qquad\hspace{-2.05cm}{\centering{
      \begin{minipage}{1.15\textwidth}
        \begin{center}
{\footnotesize
\begin{tabular}{|l||c|c|c|c|c||c|}
\hline 

%
%

Parameter & no mixing           & $m_{\mathrm{h}}$-max          & large-$\mu$  & gluophobic & small $\mathrm{\alpha_{\mathrm{eff}}}$ & CPX \\
          & /(no-mixing (2 TeV))&/($m_{\mathrm{h}}$-max$^{+}$)  &  & & & \\
          &                  &/(C $\mathrm{m_h}$-max)        &  & & & \\
\hline
\hline
\multicolumn{7}{|c|}{Parameters varied in the scan} \\
\hline
$\tanb$                      & 0.4--40& 0.4--40  & 0.7--40 &  0.4--40 & 0.4--40 & 0.6--40 \\
                             & /(0.7--40) &      &       &        &         &    \\
$m_{\mathrm{A}}$ (GeV)       & 0--1000& 0--1000  & 0--400&  0--1000 & 0--1000 & -- \\
$m_{\mathrm{H}^{\pm}}$ (GeV) &  --    & --       &  --    & --      &  --     & 4--1000 \\
\hline
\multicolumn{7}{|c|}{Fixed parameters} \\
\hline
$m_{\mathrm{t}}$  (GeV)      & 174.3   & 174.3   & 174.3  & 174.3   & 174.3 & 174.3  \\
$m_{\mathrm{SUSY}}$ (GeV)    & 1000    & 1000    & 400    & 350     & 800   & 500   \\
                             & /(2000) &         &        &         &       &       \\
$M_2$ (GeV)                  & 200     & 200     & 400  &300 & 500 & 200   \\
$\mu$ (GeV)                  & -200    & -200    & 1000   & 300 & 2000 & 2000  \\
                             & /(200)  & /(200)  & & & & \\
                             &         & /(200)  & & & & \\
 
$m_{\tilde\mathrm{g}}$ (GeV) & 800  & 800                             & 200 & 500 & 500   & 1000  \\
$X_{\mathrm{t}}$  (GeV)      & 0    & $\sqrt{6}\,m_{\mathrm{SUSY}}$   & -300  & -750 & -1100 & $A_{\mathrm{t}}-\mu\cot\beta$ \\

&  & /($\sqrt{6}\,m_{\mathrm{SUSY}}$) & & & & \\
&  &  /($-\sqrt{6}\,m_{\mathrm{SUSY}})$ & & & & \\

$A_{\mathrm{t,b}}$ (GeV)     & $X_{\mathrm{t}}+\mu\cot\beta$ & $X_{\mathrm{t}}+\mu\cot\beta$  & $X_{\mathrm{t}}+\mu\cot\beta$ & $X_{\mathrm{t}}+\mu\cot\beta$  & $X_{\mathrm{t}}+\mu\cot\beta$& 1000 \\

$\arg(A_{\mathrm{t,b}})$      & 0 & 0  & 0 &  0 & 0 & $90^{\circ}$ \\  
$\arg(m_{\tilde\mathrm{g}})$  & 0 & 0 & 0 & 0 & 0 & $90^{\circ}$ \\  

\hline 
\end{tabular} }
\end{center} 
\end{minipage}}}
\caption{\sl Parameters of  benchmark scenarios considered.
Note that the values for $X_{\mathrm{t}}$ and $A_{\mathrm{t,b}}$ are given for the 
    $\overline{\mathrm{MS}}$-renormalization
    scheme. For a description of the choice of parameters see Section~\ref{sect:benchmark}.
    Columns 2 to 6 refer to the CPC benchmark sets and the last column refers to the
    basic CPV benchmark set CPX.}\label{tab:benchmark_sum}
\end{table}

\begin{table}[hp]
  \qquad\hspace{-2.05cm}{\centering{
      \begin{minipage}{1.15\textwidth}
        \begin{center}
          \begin{footnotesize}
            \begin{tabular}{|l|c|c|c|} \hline
              \multicolumn{4}{|c|}{Limits on the MSSM scenarios}\\
              \hline
              Benchmark set & Lower limit on \mh\ (\gevcs ) & Lower limit on \mA\ (\gevcs ) & Excluded $\tan\beta$  \\
              \hline
              no mixing          & 64.0 (60.0) &    --       & $0.8<\tan\beta<6.2$ ($0.9<\tan\beta<7.2$) \\
              no mixing (2 TeV)  & 83.3 (88.0) & 84.3 (88.8) & $0.9<\tanb<4.2$ ($0.9<\tanb<4.3$)\\
              \mhmax\            & 84.5 (88.5) & 85.0 (89.0) & $0.7<\tan\beta<1.9$ $(0.7<\tan\beta<1.9)$ \\
              $\mhmax^+$         & 84.5 (88.0) & 84.0 (89.5) & $0.7<\tanb<1.9$ $(0.7<\tanb<1.9)$ \\ 
              constr. \mhmax\    & 84.0 (88.0) & 85.0 (89.0) & $0.6<\tanb<2.2$ $(0.6<\tanb<2.2)$ \\ 
              gluophobic         & 82.0 (87.0) & 87.5 (90.5) & $\tan\beta<6.0$  $(\tan\beta<8.0)$ \\
              small $\alpha_{\mathrm{eff}}$ & 79.0 (83.0) & 90.0 (95.0) & $0.4<\tanb<3.6$  $(0.4<\tanb<3.6)$ \\
              CPX                & -- & -- & $\tanb<2.8$ $(\tanb<2.8)$\\
              \hline\hline
              \multicolumn{4}{|c|}{Allowed regions in the ``large $\mu$'' scenario}\\
              \hline
              large $\mu$        & $80.0<\mh<107.0$ & $87.0<\mA$ & $\tanb>6$ \\
                                 & $(81.0<\mh<107.0)$ & $(87.0<\mA)$ & $(\tanb>12)$ \\
              \hline
            \end{tabular}
          \end{footnotesize}
        \end{center}
      \end{minipage}}}
  \vspace{0.5cm}
  \caption{\label{tab:mssmmhmalimits} 
    \sl Limits on \mh\, \mA\  and $\tanb$ for the various  
    benchmark sets.  The 
    median expected limits in an ensemble
    of SM background-only experiments are listed in parentheses. The lower limits on
    \mh\ and \mA\ in the no mixing (2 TeV) scenario are only valid for $\mA>2$~GeV.
    }
\end{table}

\clearpage

%
%


\clearpage
\newpage

\clearpage


\begin{figure}[p]
  \qquad\hspace{-1.9cm}{\centering{
      \begin{minipage}{1.15\textwidth}
        \epsfig{file=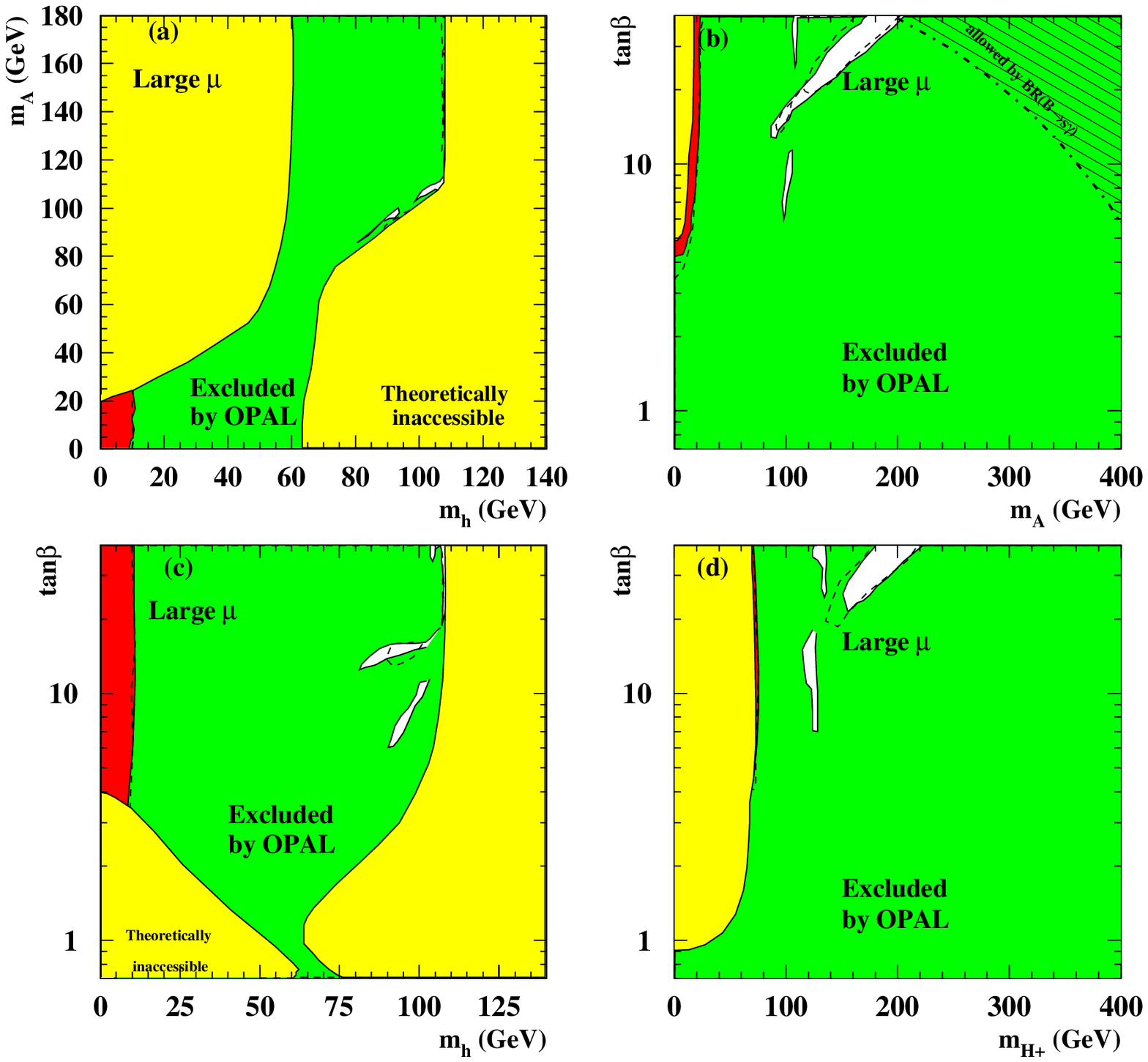, width=\textwidth}
      \end{minipage}}}
  \caption[]{\label{fig:largemu}
    \sl Results for the large $\mu$ benchmark scenario
    described in the text
    of Section~\ref{sect:cpcons}.  
    See Fig.~\ref{fig:nomix} for the notation.
    }
\end{figure}

\begin{figure}[p]
  \qquad\hspace{-1.9cm}{\centering{
      \begin{minipage}{1.15\textwidth}
        \epsfig{file=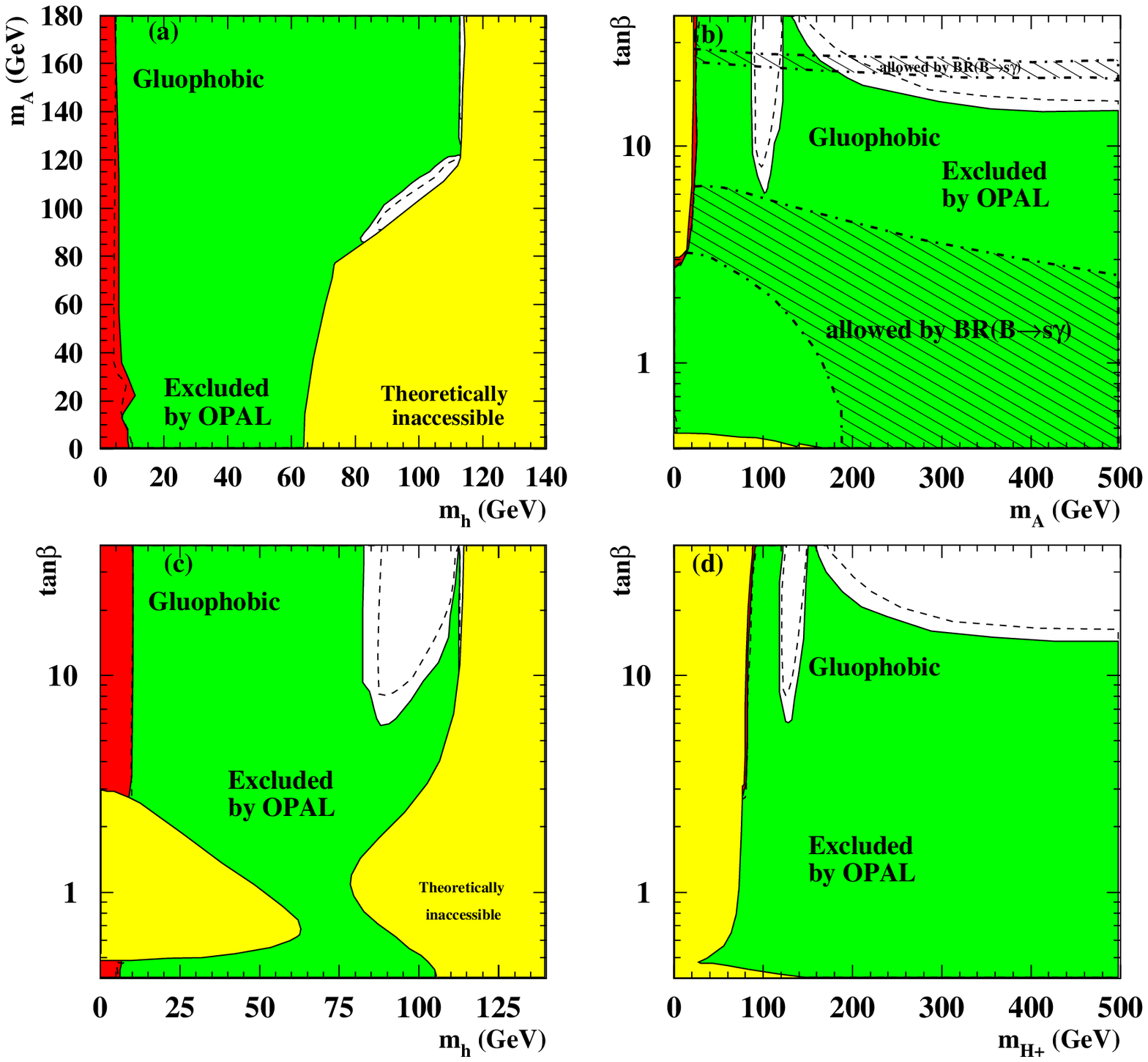, width=\textwidth}
      \end{minipage}}}
  \caption[]{\label{fig:gluophobic}
    \sl Results for the gluophobic benchmark scenario
    described in the text
    of Section~\ref{sect:cpcons}.
    See Fig.~\ref{fig:nomix} for the notation. The hatched area in (c) is allowed by 
    the BR(b$\ra$s$\gamma$) constraint.
    }
\end{figure}

\begin{figure}[p]
  \qquad\hspace{-1.9cm}{\centering{
      \begin{minipage}{1.15\textwidth}
        \epsfig{file=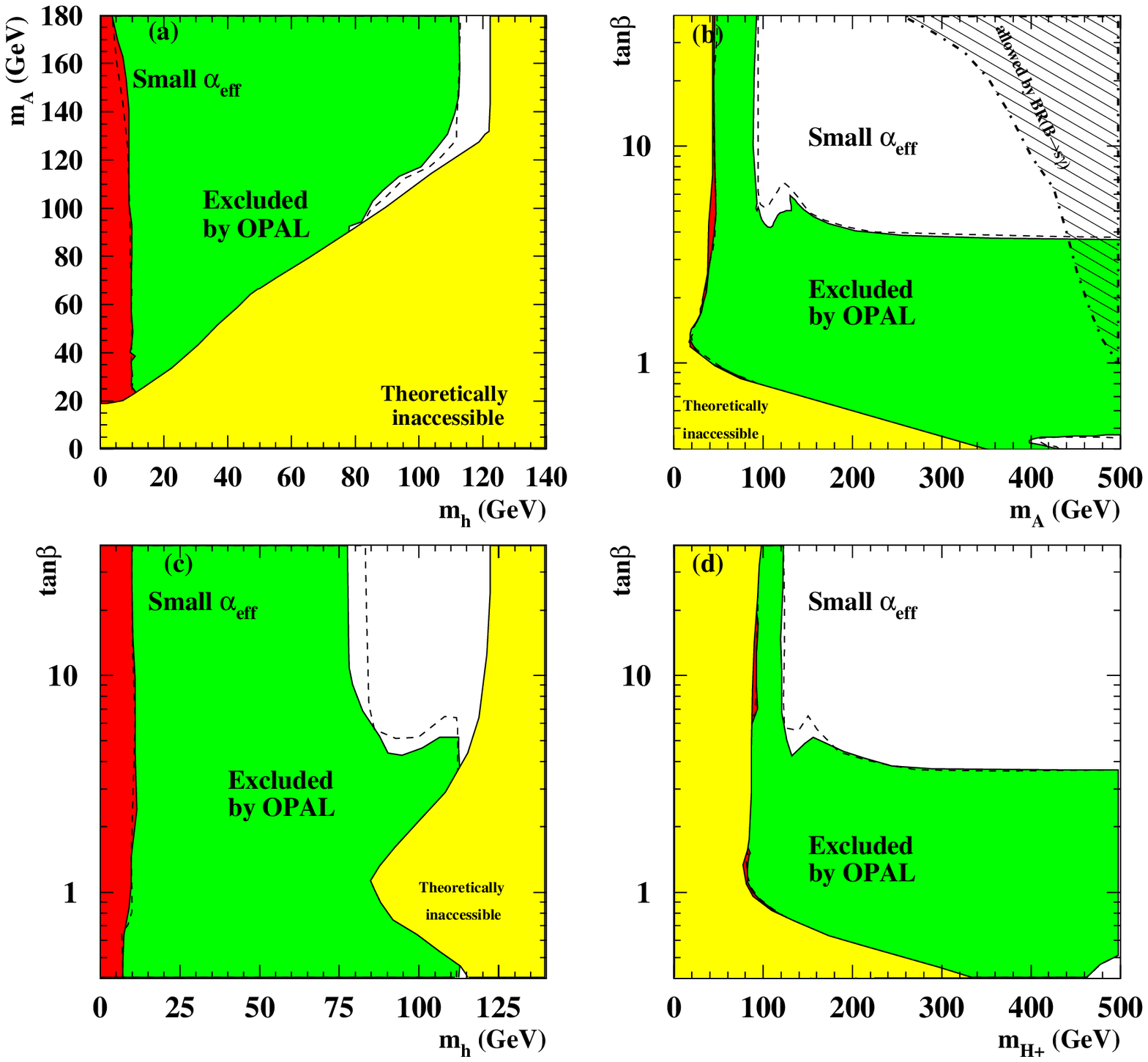, width=\textwidth}
      \end{minipage}}}
  \caption[]{\label{fig:smallalphaeff}
    \sl Results for the small $\alpha_{\mathrm{eff}}$ benchmark scenario
    described in the text
    of Section~\ref{sect:cpcons}.
    See Fig.~\ref{fig:nomix} for the notation.
    }
\end{figure}



\begin{thebibliography}{99}
%

\bibitem{Higgs:ia}
P.~W.~Higgs,
Phys.\ Lett.\  {\bf 12} (1964) 132;\\
F.~Englert and R.~Brout,
Phys.\ Rev.\ Lett.\  {\bf 13} (1964) 321.


\bibitem{Barate:2003sz}
R.~Barate {\it et al.}  [ALEPH, DELPHI, L3, OPAL Collaborations],
Phys.\ Lett.\ B {\bf 565} (2003) 61.

\bibitem{Degrassi:2002fi}
A.~Brignole, G.~Degrassi, P.~Slavich and F.~Zwirner,
Nucl.\ Phys.\ B {\bf 631} (2002) 195;\\
G.~Degrassi, S.~Heinemeyer, W.~Hollik, P.~Slavich and G.~Weiglein,
Eur.\ Phys.\ J.\ C\ {\bf28} (2003) 133.


\bibitem{Pilaftsis:1999qt}
A.~Pilaftsis and C.~E.~Wagner,
Nucl.\ Phys.\ B {\bf 553} (1999) 3.

\bibitem{Carena:2000id}
M.~Carena {\it et al.}, 
Nucl.\ Phys.\ B {\bf 599} (2001) 158.

\bibitem{sven_hAA} 
S.~Heinemeyer and W.~Hollik,
Nucl.\ Phys.\ B {\bf 474} (1996) 32.


\bibitem{OPALSMPAPER}  G.Abbiendi \etal{} [OPAL Collaboration], 
 \EPC{26}{2003}{479}.

\bibitem{pr285} G.~Abbiendi {\it et. al.} [OPAL Collaboration],
  Eur.~Phys.~J.~C~{\bf12} (2000) 567.

\bibitem{bib:opalhiggsold1} G.~Abbiendi {\it et. al.} [OPAL Collaboration],
  Eur.~Phys.~J.~C~{\bf7} (1999) 407.

\bibitem{bib:opalhiggsold2} K.~Ackerstaff {\it et. al.} [OPAL Collaboration],
  Eur.~Phys.~J.~C~{\bf1} (1998) 425.

\bibitem{bib:opalhiggsold3} K.~Ackerstaff {\it et. al.} [OPAL Collaboration],
  Eur.~Phys.~J.~C~{\bf5} (1998) 19.

\bibitem{bib:opalhiggsold4} G.~Alexander {\it et. al.} [OPAL Collaboration],
  Z.~Phys.~C~{\bf73} (1997) 189.

\bibitem{bib:opalhiggsold5} R.~Akers {\it et. al.} [OPAL Collaboration],
  Z.~Phys.~C~{\bf64} (1994) 1.


\bibitem{bib:otherlep} 
A.~Heister {\it et al.}  [ALEPH Collaboration],
Phys.\ Lett.\ B ~{\bf526} (2002) 191.\\
J.~Abdallah {\it et al.}  [DELPHI Collaboration],
Eur.\ Phys.\ J.\ C {\bf 32} (2004) 145;
P.~Achard {\it et al.}  [L3 Collaboration],
Phys.\ Lett.\ B {\bf 545} (2002) 30.

\bibitem{bib:tevatronhiggs} 
T.~Affolder {\it et al.}  [CDF Collaboration],
Phys.\ Rev.\ Lett.\  {\bf 86} (2001) 4472.

\bibitem{lhc_benchmarks} M.~Carena, S.~Heinemeyer, C.~E.~M.~Wagner and G.~Weiglein,
hep-ph/0202167.

\bibitem{detector}
K.~Ahmet \etal{}  [OPAL Collaboration], Nucl.~Instr.~and Meth.~A~{\bf305} (1991) 275.

\bibitem{simvtx}
S. Anderson \etal, Nucl.~Instr.~and~Meth.~A~{\bf403} (1998) 326.


\bibitem{Janot:1996hzha}
P.~Janot, \emph{Physics at LEP2}, 
CERN 96-01 Vol.~2 309.

\bibitem{kk2f} S.~Jadach, B.~F.~Ward and Z.~W\'as, Comput.\ Phys.\
Commun.\ {\bf 130} (2000) 260.

\bibitem{grc4f}
J. Fujimoto \etal, Comp. Phys. Comm. {\bf 100} (1997) 128;\\
J.~Fujimoto \etal, {\it Physics at LEP2}, CERN 96-01, Vol.2, 30.

\bibitem{bhwide}
S.~Jadach, W.~P{\l}aczek, and B.F.L.~Ward, {\it Physics at LEP2},
CERN 96-01, Vol.2, 286; Phys.~Lett.~B~{\bf390} (1997), 298. 

\bibitem{phojet}
E. Budinov \etal, {\it Physics at LEP2}, 
CERN 96-01, Vol.2, 216;\\
R.~Engel and J.~Ranft, Phys.~Rev.~D~{\bf54} (1996) 4244.

\bibitem{herwig}
G.~Marchesini \etal, Comp.~Phys.~Comm.~{\bf 67} (1992) 465;
G.~Corcella {\it et al.}, JHEP 0101 (2001) 10.

\bibitem{vermaseren}
J.A.M.~Vermaseren, Nucl.~Phys.~B~{\bf229} (1983) 347.

\bibitem{pythia}
T. Sj\"ostrand, Comp. Phys. Comm. {\bf 82} (1994) 74;\\
T. Sj\"ostrand, LU TP 95-20.


\bibitem{gopal}
J. Allison \etal, Nucl.~Instr.~and Meth.~A~{\bf317} (1992) 47.


\bibitem{Carena:2000ks}
M.~Carena, J.~R.~Ellis, A.~Pilaftsis and C.~E.~Wagner,
Phys.\ Lett.\ B {\bf 495} (2000) 155.

\bibitem{lowma}  G.~Abbiendi \etal{} [OPAL Collaboration], Eur.~Phys.~J. {\bf C27} (2003) 483.  

\bibitem{Abbiendi:2000ug}
G.~Abbiendi \etal{}  [OPAL Collaboration],
Eur.\ Phys.\ J.\ C {\bf 18} (2001) 425.

\bibitem{2HDMFINAL} 
G.~Abbiendi \etal{} [OPAL Collaboration],
hep-ex/0312042, Submitted to Phys. Lett. B.

\bibitem{durham_jetfinder}
N.~Brown and W.~J.~Stirling, Phys.~Lett.~B~{\bf252} (1990) 657.


\bibitem{l2mh}
K.~Ackerstaff \etal{} [OPAL Collaboration], Eur.~Phys.~J.~C~{\bf2} (1998) 441.

\bibitem{Parisi:1978eg}
G.~Parisi,
Phys.\ Lett.\ B {\bf 74} (1978) 65.\\
J.~F.~Donoghue, F.~E.~Low and S.~Y.~Pi,
Phys.\ Rev.\ D {\bf 20} (1979) 2759.

\bibitem{opal_eflow}
K. Ackerstaff \emph{et al.} [OPAL Collaboration], Eur.~Phys.~J.~C~{\bf 2} (1998) 213.

\bibitem{Zwidth:2000se}
 The LEP Collaborations ALEPH, DELPHI, L3, OPAL, the LEP Electroweak
Working Group, and the SLD Heavy Flavour and Electroweak Groups,
hep-ex/0212036.



\bibitem{Bardin:1999yd}
D.~Y.~Bardin, P.~Christova, M.~Jack, L.~Kalinovskaya, A.~Olchevski, S.~Riemann and T.~Riemann,
Comput.\ Phys.\ Commun.\  {\bf 133} (2001) 229.

\bibitem{DECAYMODEINDEP} G.Abbiendi \etal{} [OPAL Collaboration], Eur.~Phys.~J.~C~{\bf27}
(2002) 311.

\bibitem{bib:yukawa}
G.~Abbiendi {\it et al.} [OPAL Collaboration],
Eur.\ Phys.\ J.\ C {\bf 23} (2002) 397.

\bibitem{bsg_calc} P.~Cho, M.~Misiak and D.~Wyler,
              {Phys. Rev. D} {\bf 54}, 3329 (1996);\\
              A.~Kagan and M.~Neubert,
              {Eur. Phys. J. C} {\bf 7} (1999) 5;\\
              K.~Chetyrkin, M.~Misiak and M.~Munz,
              {Phys. Lett. B} {\bf 400}, (1997) 206,
              [Erratum-ibid.\ {\bf 425} (1998) 414];\\
              P.~Gambino and M.~Misiak,
              {Nucl. Phys. B} {\bf 611} (2001) 338;\\
              J.~R.~Ellis, T.~Falk, G.~Ganis, K.~A.~Olive and M.~Srednicki,
              Phys.\ Lett.\ B {\bf 510} (2001) 236.

\bibitem{bsg_measurement}
              R.~Barate et al.\ [ALEPH Collaboration],
              {Phys. Lett. B} {\bf 429} (1998) 169;\\
              S.~Chen et al.\ [CLEO Collaboration],
              {Phys. Rev. Lett.} {\bf 87} (2001) 251807;\\
              K.~Abe et al.\ [Belle Collaboration],
              {Phys. Lett. B} {\bf 511} (2001) 151;\\
              B.~Aubert et al.\ [BABAR Collaboration],
              hep-ex/0207074;
              hep-ex/0207076.


\bibitem{bib:kleissgen}
F.~A.~Berends and R.~Kleiss,
Nucl.\ Phys.\ B {\bf 260} (1985) 32.


\bibitem{RPP2000} D.E. Groom {\it et al}, Eur. Phys. J. C {\bf15} (2000) 1,
available on the PDG WWW pages \href{http://pdg.lbl.gov/}{\tt http://pdg.lbl.gov/}.

\bibitem{newbenchmarks} M.~Carena, S.~Heinemeyer, C.~E.~M.~Wagner and G.~Weiglein,
hep-ph/9912223.

\bibitem{feynhiggs} S.~Heinemeyer, W.~Hollik and~G. Weiglein, 
       Comp.~Phys.~Comm.~{\bf 124} (2000) 76;\\ Also see \href{http://www.feynhiggs.de}{\tt http://www.feynhiggs.de}.


\bibitem{Heinemeyer:CFeyn} 
                    M.~Frank, S.~Heinemeyer, W.~Hollik and G.~Weiglein,
                    hep-ph/0212037\\
                    Also see \href{http://www.feynhiggs.de}{\tt http://www.feynhiggs.de}.

\bibitem{MSSMMHBOUND7} S.~Heinemeyer, W.~Hollik and G.~Weiglein, Eur. Phys. Jour. C {\bf9} (1999) 343.

\bibitem{weigheiholl}  
S.~Heinemeyer, W.~Hollik and G.~Weiglein, Phys.~Rev.~D~{\bf58} (1998) 091701, 
Phys. Lett. B {\bf440} (1998) 296, hep-ph/9807423 and 
JHEP~0006 (2000) 009.

\bibitem{MSSMMHBOUND5} M.~Carena, M.~Quir\'os and C.E.M.~Wagner, Nucl. Phys. B {\bf461} (1996) 407.

\bibitem{carenamrennawagner} M.~Carena, S.~Mrenna and C.~Wagner, Phys. Rev. D {\bf60} (1999) 075010.

\bibitem{reconciliation} M.~Carena, H.~E.~Haber, S.~Heinemeyer, W.~Hollik, C.~E.~M.~Wagner
and G.~Weiglein, 
Nucl. Phys. B {\bf580} (2000) 29.

\bibitem{espinosareconciliation} J.~R.~Espinosa and R.-J.~Zhang, JHEP 0003 (2000) 026.

\bibitem{OPALCHARGEDHiggs}
P.~Bock {\it et al.}  [ALEPH, DELPHI, L3 and OPAL Collaborations],
CERN-EP-2000-055


\bibitem{bib:g-2results}
G.~W.~Bennett {\it et al.}  [Muon g-2 Collaboration],
Phys.\ Rev.\ Lett.\  {\bf 89} (2002) 101804
[Erratum-ibid.\  {\bf 89} (2002) 129903].

\bibitem{Carena:2000yi}
M.~Carena, J.~R.~Ellis, A.~Pilaftsis and C.~E.~Wagner,
Nucl.\ Phys.\ B {\bf 586} (2000) 92.

\bibitem{Commins:gv}
E.~D.~Commins, S.~B.~Ross, D.~DeMille and B.~C.~Regan,
Phys.\ Rev.\ A {\bf 50} (1994) 2960.

\bibitem{Harris:jx}
P.~G.~Harris {\it et al.},
Phys.\ Rev.\ Lett.\  {\bf 82} (1999) 904.



\end{thebibliography}
\end{document}